\documentclass{iopart}
\usepackage{epsf}
\usepackage{latexsym}
\begin{document}   

\newcommand{\tbox}[1]{\mbox{\tiny #1}} 
\newcommand{\mboxs}[1]{\mbox{\small #1}} 
\newcommand{\mbf}[1]{{\mathbf #1}}	
\newcommand{\half}{\mbox{\small $\frac{1}{2}$}} 
\newcommand{\const}{\mbox{const}} 
\newcommand{\mpg}[2]{\begin{minipage}[t]{#1cm}{#2}\end{minipage}}	
\newcommand{\mpb}[2]{\begin{minipage}[b]{#1cm}{#2}\end{minipage}}
\newcommand{\mpc}[2]{\begin{minipage}[c]{#1cm}{#2}\end{minipage}}

{\bf\Large 
Chaos and Energy Spreading for Time-Dependent 
Hamiltonians, and the various Regimes in the} 
\title{Theory of Quantum Dissipation}  
 
\author{
\rm Doron Cohen \\ 
{\it Department of Physics, Harvard University}. \\ 
(December 1999)
}


\begin{abstract}
We make the first steps towards a generic theory for 
energy spreading and quantum dissipation.  
The {\em Wall formula} for the calculation of friction in 
nuclear physics and the {\em Drude formula} for the calculation 
of conductivity in mesoscopic physics can be regarded 
as two special results of the general formulation. 
We assume a time-dependent Hamiltonian $H(Q,P;x(t))$ 
with $x(t)=Vt$, where $V$ is slow in a classical sense. 
The rate-of-change $V$ is not necessarily slow in the 
quantum-mechanical sense. 
The dynamical variables $(Q,P)$ may represent some `bath' 
which is being parametrically driven by $x$.  
This `bath' may consist of just few degrees-of-freedom, 
but it is assumed to be classically chaotic. 
In case of either the Wall or Drude formula, the 
dynamical variables $(Q,P)$ may represent a single particle. 
In any case, dissipation means an irreversible systematic 
growth of the (average) energy. It is associated with the 
stochastic spreading of energy across levels.  
The latter can be characterized by a transition 
probability kernel $P_t(n|m)$ where $n$ and $m$ 
are level indices.  This kernel is the main 
object of the present study.  
In the classical limit, due to the (assumed) chaotic 
nature of the dynamics, the second moment of $P_t(n|m)$ 
exhibits a crossover from ballistic to diffusive behavior. 
In order to capture this crossover 
within quantum-mechanics, a proper theory 
for the quantal $P_t(n|m)$ should be constructed. 
We define the $V$ regimes where either 
perturbation theory or semiclassical considerations 
are applicable in order to establish this crossover.   
In the limit $\hbar\rightarrow 0$ perturbation theory 
does not apply but semiclassical considerations can be used 
in order to argue that there is detailed correspondence, 
during the crossover time, between the quantal and the 
classical $P_t(n|m)$. In the perturbative regime  
there is a lack of such correspondence. Namely, 
$P_t(n|m)$ is characterized by a perturbative core-tail 
structure that persists during the crossover time.  
In spite of this lack of (detailed) correspondence 
there may be still a restricted correspondence as far as the 
second-moment is concerned. Such restricted 
correspondence is essential in order to establish 
the universal fluctuation-dissipation relation.
\end{abstract}

\section{Introduction}

\subsection{Definition of the problem} 

\begin{figure}
\begin{center}
\leavevmode
\epsfysize=2.2in 
\epsffile{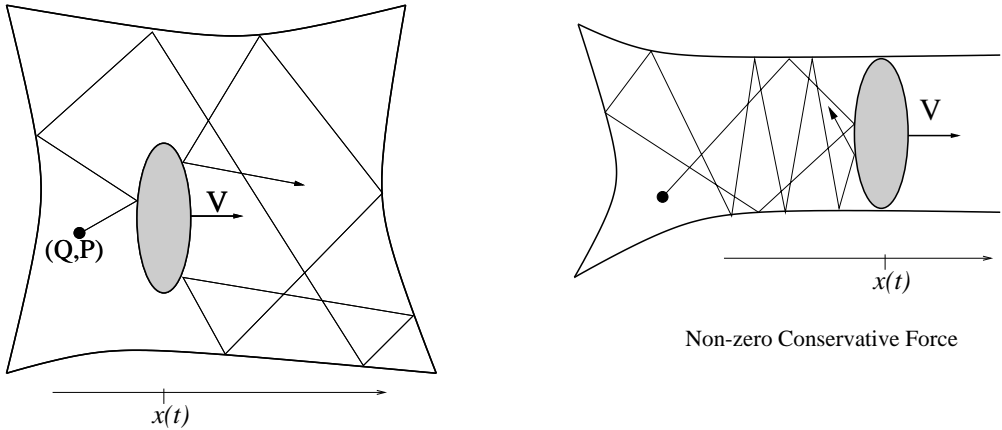}
\end{center}
\caption{\protect\footnotesize 
The `piston' example: The slow degree of freedom 
is the `piston', and the `bath' consists of one gas 
particle. In this paper the position of the `piston' 
$x(t)=Vt$ is treated as a classical parameter. 
Dissipation means a systematic growth of the  
bath-energy. For simplicity we assume throughout 
the paper that the conservative work is zero. 
This is not the case in the right illustration. } 
\label{f_pistons}
\end{figure}

We consider in this paper a system that is described by 
an Hamiltonian ${\cal H}(Q,P;x)$ where $(Q,P)$ are canonical 
variables and $x$ is a parameter. It is assumed that 
${\cal H}(Q,P;x)$ with $x=\const$ generates classically 
chaotic motion. We are mainly interested in the case of 
time dependent $x(t)$. However, it is assumed that $\dot{x}=V$ 
is a classically small velocity. The notion of classical 
slowness will be defined in Sec.\ref{s_flc}. The theory that 
we are going to present is quite general. In some particular 
applications $x(t)$ may represent, for example, 
a time-dependent electric field. However, the theory is best  
illustrated by considering the `piston' example: In this 
example $x$ represent the position of a small rigid body that 
is translated inside a large cavity, and $(Q,P)$ are the 
coordinates of a tiny gas particle. See Fig.\ref{f_pistons}.

It is assumed that initially the system is characterized by  
some energy distribution $\rho(E)$. In particular we may 
assume a microcanonical preparation. For $V=0$ energy is 
a constant of the motion, and therefore the energy distribution 
$\rho(E)$ will not change as a function of time. 
On the other hand, for $V\ne 0$ the energy will be re-distributed 
and $\rho(E)$ will become time dependent. In this paper 
we are interested in the study of this time dependence. 
Of particular interest is the time dependence of the 
{\em first} and of the {\em second} moments. 
A systematic increase of the {\em average} energy has, 
by definition, the meaning of dissipation. 
In case of the `piston' example, dissipation means that 
the gas particle is being `heated up'.

\subsection{Restrictive sense of `Quantum dissipation'}

The subject of this paper is the quantum-mechanical (QM) 
theory of energy-spreading and dissipation, as defined 
in the previous subsection. In short we 
may say that we are interested in the theory of 
{\em Quantum Dissipation}. However, it is important to realize 
that we are using the term `Quantum Dissipation' in 
a quite {\em restrictive sense}.   
This is because of mainly two reasons: 
(a) We assume a classical driving force; 
(b) We are not considering a many-body bath. 
Note that an infinite number of 
degrees-of-freedom is not important for having 
stochastic behavior: this is the main idea behind 
the term `chaos' when applied to dynamical systems. 
We can have dissipation even if $(Q,P)$ represent  
a few degrees-of-freedom `bath'.

The interest in Quantum Dissipation is very old 
\cite{FV,ZG,textbook,wall,koonin,MS,CL}. 
However, in most of the literature, the term 
`Quantum Dissipation' is used in a more {\em general sense}. 
Namely, $x$ becomes a dynamical variable, and one looks 
for its reduced dynamics. Thus, in most of the 
literature, dissipation-of-energy becomes only one aspect 
of a much more complicated problem. The `grand problem' 
includes, besides `dissipation', other issues such 
as `dephasing' and `thermalization'.  It also should 
be noticed that the standard literature usually adopts 
an effective-bath approach (see subsection 1.4) or 
other effective formulations \cite{textbook} that 
do not necessarily reflect the actual dynamics 
of the bath degrees-of-freedoms. Important exceptions are 
works such as \cite{kolovsky} and \cite{srednicki}.

Of particular interest is the `piston' model. 
If the `piston' is treated as a dynamical object, than its 
reduced dynamics is called `quantal Brownian motion' (QBM). 
According to our (restricted) definition, `dissipation' 
means systematic irreversible growth of the bath-energy. 
In case of an un-driven Brownian particle, the `dissipation' 
is balanced eventually by 'noise' leading to `thermalization'. 
In the QM case the issue of `irreversibility' is more 
complicated because we may have `recurrences'. The relevant 
time scale for these recurrences is the Heisenberg time for 
the combined BrownianParticle-GasParticle system. This 
latter time scale may be extremely large if the Brownian 
particle has a large mass.

In this paper $x$ is not a dynamical degree of freedom, 
and therefore the `recurrences' that have been mentioned in 
the previous paragraph are not an issue. (It is as if we assume 
that the `piston' has an infinite mass, hence the  
frequent use of the term `moving walls').  For $V=0$ 
the Hamiltonian is time-independent, and we will have 
recurrences that are associated with the dynamics of 
the GasParticle (alone). The remnant of this latter type 
of recurrences is QM-adiabaticity, which we are going 
to discuss soon. Another type of `recurrences' are   
associated with {\em periodic} driving and are discussed 
in Sec.\ref{s8}. It should become clear from the above discussion,  
and subsection 1.6 below, that `recurrences' are not an 
important issue in this paper.

\subsection{The classical theory of dissipation}

The classical understanding of the dissipation-process 
is based mainly on the works of 
\cite{wall,koonin,ott,wilk1,jar} and followers. 
We are going to sketch briefly the main idea  
of the classical theory, and the associated 
derivation of the fluctuation-dissipation (${\cal F\!D}$) relation. 
In the {\em time-independent case} ($V=0$) the motion 
of $(Q(t),P(t))$ is irregular due to the chaotic 
nature of the dynamics. We shall denote the ergodic 
time by $t_{\tbox{erg}}$.  We can define a fluctuating quantity   
${\cal F}(t) = -(\partial {\cal H} / \partial x)$
that has stochastic features. The intensity of these 
fluctuations will be denoted by $\nu$.   
In the classical case ${\cal F}(t)$ is essentially 
like noise whose correlation time $\tau_{\tbox{cl}}$ 
is smaller than or equal to $t_{\tbox{erg}}$.  
In the {\em time-dependent case} ($V\ne0$) energy 
is not a constant of the motion and consequently 
the energy distribution $\rho(E)$ becomes time 
dependent. It is argued that for $t\gg t_{\tbox{erg}}$ 
the energy distribution satisfy a diffusion equation. 
The energy-dependent diffusion coefficient will be 
denoted by $D_{\tbox{E}}$. 
It turns out that quite generally $D_{\tbox{E}} = \half \nu V^2$. 
Associated with this diffusion is a systematic growth 
of the average energy. This systematic growth of energy 
is due to the $E$-dependence of the diffusion process. 
The rate of energy growth will be denoted by $\dot{{\cal Q}}$. 
It can be written as $\dot{{\cal Q}}=\mu V^2$.

The considerations above lead to the conclusion that 
in the classical case the dissipation is of 
ohmic nature ($\dot{{\cal Q}}\propto V^2$). 
The dissipation coefficient is denoted by $\mu$. 
It is implied that the fluctuating quantity ${\cal F}(t)$ 
has a non-zero average, namely 
$\langle {\cal F} \rangle = -\mu V$. 
In the 'piston' example the latter represents the  
`friction' force that is experienced by the moving object.   
The considerations above also imply that 
{\em the analysis of dissipation 
is reduced to the study of energy spreading}.
The difficult issue is to establish a stochastic 
energy spreading with a coefficient $D_{\tbox{E}} = \half \nu V^2$. 
Then, the ${\cal F\!D}$ relation between $\mu$ and the 
noise intensity $\nu$ follows as an immediate 
consequence. If $\rho(E)$ is a canonical distribution 
(which is not necessarily the case) 
then the ${\cal F\!D}$ relation reduces to the familiar form  
$\mu=\nu/(2k_{\tbox{B}}T)$ where $T$ is the temperature.

\begin{table}
\begin{center}
\leavevmode  
\setlength{\baselineskip}{0.5cm}
\begin{tabular}{|l|}
\hline
\ \\
{\bf Generic classical parameters} $(\tau_{\tbox{cl}}, \nu)$ \\ 
$\tau_{\tbox{cl}} \ = \ $ classical correlation time. \\
$\nu \ = \ $ intensity of fluctuations \\
\ \\
{\bf Generic quantal parameters} $(\Delta,b,\sigma,\hbar)$ \\ 
$\Delta \ = \ $ mean level spacing of the eigen-energies $\{E_n\}$ \\
$b \ = \ $ Dimensionless bandwidth of the matrix 
$(\partial {\cal H} / \partial x)_{nm}$ \\ 
$\sigma \ = \ $ Root-mean-square of in-band matrix elements of
$(\partial {\cal H} / \partial x)_{nm}$ \\     
\ \\
{\bf Semiclassical relations} \\ 
$\tau_{\tbox{cl}} \ = \ 2\pi\hbar/(b\Delta)$ \\
$\nu = (2\pi\hbar/\Delta)\ \sigma^2$ \\
\ \\
{\bf Linear response theory} \\
$D_{\tbox{E}} \ = \ \half \nu V^2 \ = \ 
(\pi\hbar/\Delta)\ \sigma^2 V^2$ \\
\ \\
{\bf Ohmic dissipation} \\
$d\langle {\cal H} \rangle / dt \ = \ \mu V^2$ \\
$\mu \ = \ {\cal F\!D} [\nu]$ \\
\ \\
{\bf Primary dimensionless parameters} $(b,v_{\tbox{PR}})$ \\ \ \\ 
$v_{\tbox{PR}} \ = \ (1/\hbar) \ 
\sqrt{\nu\tau_{\tbox{cl}}^3 \ } \ V \ = \ 
b^{\tbox{-3/2}} \ (2\pi\hbar/\Delta)^2 \ (\sigma/\hbar) \ V$ \\
\ \\
\hline
\end{tabular}
\end{center}
\caption{\protect\rm\footnotesize 
Overview of the common theory for dissipation.  
Two generic parameters should be specified for 
the classical theory, while four are required 
for the QM theory. Note that $V=\dot{x}$  always 
appears in the combination $\nu V^2$ or $\sigma V$, 
and therefore it should not be counted as an 
additional (independent) parameter.   
The two classical parameters 
can be expressed in terms of the QM parameters 
via semiclassical relations. In the absence of 
well defined classical limit (as in the case of 
RMT models) this relations can be regarded as 
definitions. The so called Kubo-Greenwood result 
of linear response theory can be obtained using FGR picture, 
and it coincides  with the classical expression. 
General considerations lead to a fluctuation-dissipation (${\cal F\!D}$) 
relation between $\mu$ and $\nu$. An important observation 
of this paper is that the validity of the 
linear-response approach is controlled by the 
dimensionless parameter $v_{\tbox{PR}}$. 
}
\end{table}

\subsection{The effective-bath approach to Quantum Dissipation}

The most popular approach to `Quantum Dissipation' 
is the {\em effective-bath approach} \cite{FV,ZG,MS,CL}. 
When applied to `our' problem (as defined in the first subsection) 
it means that the chaotic $(Q,P)$ degrees-of-freedom are replace 
by an effective-bath that has the same {\em spectral-properties}. 
This may be either harmonic-bath 
(with infinitely many oscillators) or 
random-matrix-theory (RMT) bath \cite{rmt}. 

It turns out that quantal-classical correspondence (QCC)  
is a natural consequence of this procedure: The dissipation 
coefficient $\mu$ turns out to be the same classically 
and quantum-mechanically. In order to explain this point 
let us use the Caldeira-Leggett notations \cite{CL}.  
The distribution of the frequencies of the bath-oscillators 
is characterized by an ohmic spectral-function 
$J(\omega)=\eta\omega$.  The classical analysis leads 
to a friction force with a coefficient $\mu=\eta$, 
and white noise whose intensity is 
$\nu=2\eta k_{\tbox{B}}T$.   
Using Feynman-Vernon \cite{FV} formalism one obtains the 
same value $\mu=\eta$ in the QM case. The quantal 
noise is characterized by an $\hbar$-dependent 
power-spectrum, but the noise intensity $\nu$ is defined  
as the $\omega=0$ component, and it is still equal to 
$2\eta k_{\tbox{B}}T$.  Hence the classical ${\cal F\!D}$ relation   
$\mu=\nu/(2k_{\tbox{B}}T)$ holds also in the QM case.

The effective-bath approach will not be adopted in this letter 
since its applicability is a matter of {\em conjecture}.  
In this paper we want to have a direct understanding 
of quantum-dissipation.

\subsection{The QM theory of Dissipation}
 
Quantum-mechanics introduces additional energy scales, 
as well as additional parametric scales into the problem 
(See Table 2). Consequently there are few $V$ regimes 
in the QM theory (See Table 3). 
The QM-adiabatic regime \cite{wilk1} is quite well 
understood. We shall discuss this regime only briefly 
since it is not related to the main concern of this 
paper. The further distinction between the QM-slow regime 
and the QM-fast regime is the main issue of this paper.

Let us assume that initially the energy is concentrated 
in one particular level.  For extremely slow velocities 
($v_{\tbox{LZ}}\ll 1$) and relatively long time the energy will 
remain mainly concentrated in the initial level. This is the 
QM-adiabatic approximation. The term  `QM-adiabaticity' 
is a beat confusing, because it actually does not 
correspond (in the $\hbar\rightarrow 0$ limit) 
to adiabaticity in the classical sense. Maybe 
a better term would be `perturbative localization'. 
In the QM-adiabatic regime Landau-Zener transitions between 
neighboring levels constitute the predominant mechanism
for energy spreading. This mechanism does not correspond 
to the classical mechanism of energy-spreading. 
The QM-adiabatic regime is a genuine quantal regime.

For higher velocities ($v_{\tbox{LZ}}\gg 1$) it is 
essential to take into account transitions between 
non-neighboring levels. The contribution of 
near-neighbor transitions to the energy spreading 
becomes negligible rather than predominant.  
An obvious approach for the study of energy spreading would 
be to adopt a Fermi-golden-rule (FGR) picture. 
FGR is one possible picture of {\em perturbation theory}. 
The same results for $D_{\tbox{E}}$ and $\mu$ can 
be derived by using other, equivalent formulations 
of perturbation theory. The most popular variation is 
known as `linear response theory' or as `Kubo-Greenwood formalism'.  
Whatever version of perturbation theory is being used 
the standard result is always the same (See Table 1). 
It should be realized that the standard result is 
in complete correspondence with the classical result, and it 
becomes identical with the classical result upon taking 
the formal limit $\hbar\rightarrow 0$.

\begin{table}
\begin{center}
\leavevmode  
\setlength{\baselineskip}{0.5cm}
\begin{tabular}{|lll|}
\hline
\ & \ & \\
\multicolumn{3}{|l|}{\bf Energy Scales:} \\ 
$\Delta$  &$\propto$&  $\hbar^{d} \ = \ $ 
mean level spacing of the eigen-energies $\{E_n\}$. \\
$\Delta_b$  &=&  $b\Delta \ = \ 2\pi\hbar/\tau_{\tbox{cl}} \ = \ $ 
bandwidth of the matrix 
$(\partial {\cal H} / \partial x)_{nm}$ \\
$\Delta_{\tbox{SC}}$  &$\propto$& $\hbar^{2/3} \ = \ $
semiclassical width of Wigner function. \\
\ & \ & \\
\multicolumn{3}{|l|}{\bf Parametric scales:} \\ 
$\delta x_c^{\tbox{cl}}$  &=& 
parametric correlation scale of the $x$-dependent Hamiltonian \\ 
$\delta x_c^{\tbox{qm}}$  &=& 
$(\Delta/\sigma) \ \propto \ \hbar^{(1{+}d)/2}\ = \ $
The $\delta x$ required to mix neighboring levels. \\
$\delta x_{\tbox{prt}}$   &=& 
$\sqrt{b}(\Delta/\sigma) \ \propto \ \hbar \ = \ $ 
The $\delta x$ to mix all the levels within the bandwidth. \\
$\delta x_{\tbox{SC}}$  &$\propto$& $\hbar^{2/3} \ = \ $ 
The $\delta x$  required to get detailed QCC. \\
\ & \ & \\
\multicolumn{3}{|l|}{\bf Temporal scales:} \\ 
$\tau_{\tbox{cl}}$   &=& 
Classical correlation time of ${\cal F}(t)$. \\
$t_{\tbox{erg}}$   &=& 
Ergodic time of the classical chaotic motion. \\
$t_{\tbox{frc}}$   &=& $\nu/(\mu V)^2 \ = \ $
Breaktime of the classical adiabatic approximation. \\
$\tau_c^{\tbox{qm}}$   &=& $\delta x_c^{\tbox{qm}}/V \ = \ $
The time it takes to mix neighboring levels . \\
$t_{\tbox{prt}}$   &=& 
Ultimate breaktime of the QM perturbation theory. \\
$t_{\tbox{sdn}}$   &=& 
Breaktime of the QM sudden approximation. \\
$t_{\tbox{H}}$  &=& $2\pi\hbar/\Delta \ = \ $
Time needed to resolve individual levels (Heisenberg time). \\
\ & \ & \\
\hline
\end{tabular}
\end{center}
\caption{\protect\rm\footnotesize 
Various scales in the theory of energy spreading. 
The generic $\hbar$ dependence is indicated in most cases.
The parametric scales and the temporal scales are associated 
with the kernels $P(n|m)$ and $P_t(n|m)$ respectively. 
The determination of $t_{\tbox{prt}}$ and $t_{\tbox{sdn}}$ 
is an important issue of this paper. Their dependence  
on $V$ is illustrated in Fig.\ref{f_regimes}. It should be 
realized that $\tau_{\tbox{cl}}$ can be defined, from a 
purely QM point of view, as the time which is required in 
order to resolve the energy scale $\Delta_b$. Similarly 
$t_{\tbox{sdn}}$ is defined as the time which is required in 
order to resolve the spreading profile. The time $t_{\tbox{sdn}}$ 
can be either equal or shorter than $\tau_{\tbox{cl}}$.  
}
\end{table}

\subsection{Specific motivation for the preset study}

Reading some of the early literature one  
gets the impression that quantum dissipation is 
conceptually well-understood. Specifically, it 
looks as if the perturbative methods are effective 
for the purpose of constructing a general theory. 
However, this is a wrong impression.
{\em A general theory of energy spreading is 
still lacking,  a-fortiori there is no 
general theory of quantum dissipation}.  
This point becomes most evident once we read 
the work by Wilkinson and Austin (W\&A) \cite{wilk2}.  
Their observations constitute the original motivation 
for the present study \cite{crs}.

W\&A \cite{wilk2} have defined two important dimensionless 
parameters that are associated with the velocity $V$. 
These are, (using our notations), 
the scaled velocity $v_{\tbox{LZ}}$, 
and the scaled velocity  $v_{\tbox{RMT}}$.  
The QM-adiabatic regime is distinguished 
by the condition $v_{\tbox{LZ}} \ll 1$, where 
we have the relatively simple picture of 
spreading due to Landau-Zener transitions. 
At higher velocities ($v_{\tbox{LZ}} \gg 1$) 
the QM-adiabatic nature of the dynamics 
is lost, and the Landau-Zener picture no longer apply. 
In order to extend the perturbative treatment  
to such higher velocities W\&A have suggested 
to adopt an innocent-looking RMT assumption.  
As long as the velocity is sufficiently slow
($v_{\tbox{RMT}}\ll 1$, but still $v_{\tbox{LZ}} \gg 1$) 
a classical-like result for $D_{\tbox{E}}$ 
is obtained. On the other hand, once $v_{\tbox{RMT}}\gg 1$,   
the classical-like expression for $D_{\tbox{E}}$
no longer holds. It is modified in such a way that 
correspondence with the classical result is being lost!

Obviously, W\&A have realized that the above conclusion 
is inconceivable. We will have to understand what is wrong 
with their innocent-looking RMT assumption. 
We shall argue that $v_{\tbox{RMT}} \sim 1$ does not mark 
a crossover to a non-classical regime.  
Rather, we shall find out that there is a {\em different}  
dimensionless parameter ($v_{\tbox{PR}}$) that controls 
the route towards quantal-classical correspondence (QCC).

\begin{table}
\begin{center}
\leavevmode  
\setlength{\baselineskip}{0.5cm}
\begin{tabular}{|ll|}
\hline
\ & \ \\
{\bf Classical slowness conditions:} & \ \\
$V\tau_{\tbox{cl}} \ \ \ll \ \ \delta x_c^{\tbox{cl}}$ &  Trivial condition \\
$\ \tau_{\tbox{cl}} \ \ \ \ll \ \ t_{\tbox{frc}}$  & Non-trivial Condition \\
\ & \ \\
{\bf Quantal regimes:} & \ \\
$Vt_{\tbox{H}}\ \ \ll \ \ \delta x_c^{\tbox{qm}}$ &  
QM-adiabaticity (extremely slow velocities) \\
$V\tau_{\tbox{cl}}\ \ \ll \ \ \delta x_{\tbox{prt}}$ &  
QM-slow velocity (linear response regime)\\
$V\tau_{\tbox{cl}}\ \ \gg\ \ \delta x_{\tbox{SC}}$ &  
QM-fast velocity (semiclassical regime) \\
\ & \ \\
{\bf Scaled velocities:} & \ \\
$v_{\tbox{SC}} \ \ \ \ = \ \  
\sqrt{2D_{\tbox{E}} \ \tau_{\tbox{cl}} } 
\ / \ \Delta_{\tbox{SC}}$ &  
$= \ \ V \ / \ (\delta x_{\tbox{SC}}/\tau_{\tbox{cl}})$ \\
$v_{\tbox{PR}}  \ \ \ \ = \ \ 
\sqrt{2D_{\tbox{E}} \ \tau_{\tbox{cl}}} \ / \ \Delta_b$ & 
$= \ \ V \ / \ (\delta x_{\tbox{prt}}/\tau_{\tbox{cl}})$ \\
$v_{\tbox{RMT}} \ \  = \ \ b^{\tbox{1/2}} \ v_{\tbox{PR}}  
\ \ \ = \ \ \  \tau_{\tbox{cl}} \ / \ \tau_c^{\tbox{qm}}$  & 
$= \ \ V \ / \ (\delta x_c^{\tbox{qm}}/\tau_{\tbox{cl}})$ \\
$v_{\tbox{LZ}} \ \ \ \ = \ \  b^{\tbox{3/2}} \ v_{\tbox{PR}}    
\ \ \ = \ \ \ t_{\tbox{H}} \ / \ \tau_c^{\tbox{qm}}$  & 
$= \ \ V \ / \ (\delta x_c^{\tbox{qm}}/t_{\tbox{H}})$ \\
\ & \ \\
\hline
\end{tabular}
\end{center}
\caption{\protect\rm\footnotesize 
Definitions of the various $V$ regimes in the theory of 
energy spreading and dissipation. The classical 
slowness condition is always assumed to be satisfied. 
In the QM case we distinguish between the regimes of 
QM-adiabaticity (extremely slow velocities), 
QM-slow velocities, and QM-fast velocities.  
Dimensionless (scaled) velocities can be defined in 
order to distinguish between the various regimes. 
For reasonably small $\hbar$ we have 
$v_{\tbox{LZ}} \gg v_{\tbox{RMT}} \gg v_{\tbox{PR}} \gg v_{\tbox{SC}}$. 
In the classical limit all of them $\gg 1$.
The condition $v_{\tbox{LZ}}\ll 1$ defines the 
QM-adiabatic regime. The condition $v_{\tbox{PR}}\ll 1$ 
define the regime of QM-slow velocities. 
The condition $v_{\tbox{SC}}\gg 1$ defines the 
regime of QM-fast velocities. The parameter 
$v_{\tbox{RMT}}$ has been introduced in \cite{wilk2}, 
and we are going to explain that it determines the 
limitation of an over-simplified RMT approach.}
\end{table}

\subsection{Main claims of this paper}

The purpose of this paper is to describe the time-evolution 
of the energy spreading in the various velocity regimes 
(See Table 3). The various `scenarios' are graphically 
illustrated in Fig.\ref{f_regimes}.  The main 
claims of this paper are implied by this 
illustration. Disregarding the QM-adiabatic regime 
we are motivated by the following two questions: 
\begin{itemize} 
\setlength{\itemsep}{0cm}
\item What is the regime where FGR/RMT picture is valid? 
\item What is the regime where QCC considerations are valid? 
\end{itemize}
In particular we would like to know whether 
there is a `clash' between FGR/RMT considerations 
on the one hand, and QCC considerations on the other hand.

The main object of this paper is the transition 
probability kernel $P_t(n|m)$. 
The variable $m$ denotes the initial energy-state 
of  the system, and $n$ stands for one of the 
instantaneous energy-states at a later time $t$. 
This kernel is well defined quantum-mechanically 
as well as classically. The energy distribution $\rho_t(E)$
can be obtained by operating with the kernel $P_t(n|m)$ 
on the initial microcanonical preparation $\rho_{t{=}0}(E)$, 
and making a simple change of variables $n \mapsto E$.  
An important distinction in this paper is between 
{\em restricted} QCC and {\em detailed} QCC. 
Detailed QCC implies that the quantal $P_t(n|m)$ is similar to the 
classical $P_t(n|m)$. We shall see that detailed QCC 
can be established in an intermediate time regime 
provided the velocity $V$ is large enough. 
In the absence of detailed QCC we still may have 
restricted QCC. The latter implies that only 
the first and the second moments of the corresponding 
distributions (quantal versus classical) are similar. 
Restricted QCC is sufficient in order to guarantee 
QCC as far as the diffusion coefficient $D_{\tbox{E}}$ 
and the dissipation coefficient $\mu$  are concerned. 
Our main statements are: 
\begin{itemize} 
\setlength{\itemsep}{0cm}
\item The FGR picture implies restricted rather than detailed QCC. 
\item The FGR picture is valid in the regime $v_{\tbox{PR}} \ll 1$.
\item Detailed QCC considerations are valid in the regime $v_{\tbox{SC}} \gg 1$.
\end{itemize}
It should be realized that the FGR picture is not {\em valid} 
in the regime $v_{\tbox{PR}} \gg 1$. However, this does not  
necessarily imply that the standard Kubo-Greenwood result 
is not {\em correct} there. On the contrary: In the the limit 
$\hbar\rightarrow 0$ we have detailed QCC ($v_{\tbox{SC}} \gg 1$), 
and at the same time Kubo-Greenwood result simply 
coincides with the classical result. Thus we may say that 
for $v_{\tbox{SC}} \gg 1$, the standard Kubo-Greenwood result 
is not valid but correct. 
The distinction between `valid' and `correct' is crucial here:  
A correct result sometimes follows from using wrong assumptions. 
In the intermediate regime ($v_{\tbox{PR}} \gg 1$ but $v_{\tbox{SC}} \ll 1$) 
neither FGR nor QCC consideration apply and we may have qualitatively 
different results for $\mu$. It is suspected \cite{wbr}, 
but not yet proved in the present context, that some artificial RMT models, 
that does not possess a well defined classical limit,  
may exhibit for $v_{\tbox{PR}} \gg 1$ a significantly different 
behavior compared with the expected FGR or classical result.  
This latter observation is in the spirit of \cite{wilk2}, 
but it is quite different as far as details are concerned.

It is important to understand what is the origin of the 
FGR picture validity condition  $v_{\tbox{PR}}\ll 1$, 
and why it is different from the condition $v_{\tbox{RMT}}\ll 1$
that has been suggested in \cite{wilk2}.  
Again we assume that initially the energy is concentrated 
in one particular level. We shall argue that in order to 
determine $D_{\tbox{E}}$ it is important to estimate how many 
levels are mixed non-perturbatively at the time $t\sim\tau{\tbox{cl}}$. 
If the related parametric change $\delta x=Vt$, does not 
mix neighboring levels, then we are on ``safe ground'' of 
standard first-order perturbation theory (FOPT), 
and we can trust completely the FGR picture. 
Such circumstances are guaranteed 
by the condition $v_{\tbox{RMT}}\ll 1$. 
If $v_{\tbox{RMT}} \gg 1$ we have a breakdown of 
the standard FOPT picture, but this does 
not imply that the FGR picture becomes non-valid. 
It turns out that the FGR result for transition rate between 
levels is valid on ``large'' energy scales, even if there 
is non-perturbative mixing of levels on ``small'' energy scales. 
This is true as long as the ``small'' scale is much 
smaller compared with the bandwidth $\Delta_b$ of 
first-order transitions. Such circumstances are guaranteed 
by the condition $v_{\tbox{PR}}\ll 1$.

\subsection{The `piston' example - The wall formula} 

We are using throughout this paper the `piston' model 
of Fig.\ref{f_pistons} as an illustrative example.  
It should be noticed that for simplicity of presentation 
we picture the `piston' as a small moving obstacle 
whose motion is constrained to be in one space direction.   
From purely linguistic point of view 'piston' implies 
also hermetic closure along the margins. We do not assume 
such a closure. 

Application of the ${\cal F\!D}$ relation in order to get an expression 
for the dissipation coefficient $\mu$ leads to the `wall' 
formula. This formula has been originally derived 
using kinetic considerations \cite{wall}, 
and only later using other approaches \cite{koonin,jar}, 
including the ${\cal F\!D}$ approach that we are using here.

In the {\em proper classical limit} 
(taking $\hbar{\rightarrow}0$, while all the other parameters 
are held fixed) the walls of the `piston' always become `soft', 
meaning that De-Broglie wavelength becomes much smaller than 
the penetration distance.  
The hard walls limit (meaning that the penetration distance 
is taken to be zero, while $\hbar$ is being kept fixed),   
is non-generic. It is important to 
understand the consequences of taking this limit. 
In the hard wall limit $(\partial {\cal H} / \partial x)_{nm}$ 
is not a banded matrix, and the (generic) problem of having  
a non-perturbative regime for $\hbar\rightarrow 0$ 
is being avoided. The non-generic features 
of the hard wall limit are possibly responsible for some 
prevailing miss-conceptions, and in particular to 
the {\em illusion} that perturbative techniques can be 
used in order to get a {\em general} theory for 
`quantum dissipation'. This is the reason for the inclusion 
of a quite detailed discussion of the `piston' example. 
The consequences of taking the hard-wall limit, as well as 
other non-generic features of the 'piston' example are further 
discussed in the concluding section and in \cite{wls}.

\subsection{The `mesoscopic' example - Drude formula} 

Another physical examples that can be treated by 
the general theory of dissipation is taken from 
the realm of mesoscopic physics. 
Consider the case where $x$ is the magnetic flux via a ring. 
The velocity $V=\dot{x}$ has then the meaning of electro-motive-force.  
Let us assume that the ring contains one charged particle $(Q,P)$ 
that performs diffusive motion. Ohmic dissipation 
($\dot{{\cal Q}}=\mu V^2$) means that the charged particle 
gains kinetic energy, where the dissipation coefficient $\mu$ 
is just the conductivity of the ring. Equivalently, 
having $\langle {\cal F} \rangle = -\mu V$ just means 
that the drift velocity along the ring is proportional 
to the electro-motive-force. It is a trivial exercise to 
get Drude formula from the general ${\cal F\!D}$ relation. The advantage 
of this procedure is that the derivation can easily be extended 
to the case where the motion of the charged particle is chaotic rather 
then diffusive.  It should be noted that in actual circumstances 
the charged-particle is an electron, and its (increasing) kinetic energy 
is eventually transfered to the vibrational modes (phonons) 
of the ring.  The latter process (that leads to Joule heating) 
is `on top' of the generic dissipation problem that we are going 
to analyze.

\subsection{Overview of the paper}

This paper divides roughly into three parts.  
The appendixes (A-J) should be considered an integral component 
in the reading of the main text. The reason for transforming some 
of the sections into appendixes was the desire to maintain a simple 
logical flow. We turn now to give a brief description of the paper.     
The first part of this paper ({\bf sections 2-7}),  
in a superficial glance, looks like a review.  
However, inspite of its textbook style, it is not a review. 
It gives the necessary introduction for the later QM analysis,  
and in particular it contains a careful examination of the 
various assumptions that are involved in the common approaches to 
the theory of dissipation. (It turns out that a satisfactory  
presentation is lacking in the existing literature). 
The main items of the first part are: 
\begin{itemize} 
\setlength{\itemsep}{0cm}
\item The crossover from ballistic to diffusive energy spreading.
\item Precise formulation of the classical slowness conditions. 
\item Brief description of the derivation of the ${\cal F\!D}$ relation. 
\item Critical discussion of the QM linear-response theory.  
\item Critical discussion of the standard FGR picture.
\item The wall formula generalized to arbitrary dimensionality ($d=2,3...$).  
\end{itemize}
The second part of this paper ({\bf sections 8-10})  
contains the precise formulation of the theory and an 
overview of the general picture. 
The main items of the second part are:
\begin{itemize}
\setlength{\itemsep}{0cm} 
\item Definitions of the kernels $P(n|m)$ and $P_t(n|m)$.  
\item The stochastic description of the energy spreading process.   
\item Restricted QCC versus detailed QCC, and the classical approximation.   
\item Overview of the dynamical scenarios in the different $V$ regimes.  
\end{itemize}
The third part of this paper ({\bf sections 11-20}) gives a detailed 
presentation of perturbation theory and RMT considerations. 
The main items are:
\begin{itemize} 
\setlength{\itemsep}{0cm}
\item The Schroedinger equation in the $x$-dependent basis.    
\item The QM-sudden approximation and parametric evolution. 
\item The over-simplified RMT picture.   
\item An improved version of perturbation theory. 
\item The core-tail structure of the spreading kernel.   
\end{itemize}
The paper is concluded ({\bf section~21}) by pointing out 
some important questions that have been left open.  
Some future directions for research are indicated.

\subsection{The need for a generic theory}

Our main interest in this paper is to construct 
a {\em generic} theory for energy spreading and 
quantum dissipation.  In particular we want 
to define the conditions for getting ohmic dissipation, 
to establish the associated ${\cal F\!D}$ relation and 
to explore the validity limits of the QCC principle. 

One may wonder what is the practical gain in 
achieving the above mentioned goals. Is it just 
a matter of doing `mathematics' properly?  
A similar type of question is frequently asked 
with regard to the efforts to re-derive well known 
RMT results using semiclassical methods. 
The answer to such questions should be clear: It is 
not possible to analyze non-generic 
(or non-universal) features unless
one possess a thorough understanding of 
a generic theory along with its limitations. 
In the future, our intention is to analyze 
circumstances that go beyond these limitations, 
and to look for genuine QM effects \cite{rsp}.  

We are using the term `generic' frequently, and it 
is now appropriate to define what do we mean by that. 
The answer is as follows: In order to understand a phenomena 
(energy spreading and dissipation in the present case) 
it is a common practice to make the maximum simplifications 
possible. Then we get a theory with a minimal number 
of parameters. It turns out that the theory of energy spreading 
involves the minimal number of {\em two} dimensionless parameters.  
We are going to consider any additional (dimensionless) parameter  
as non-generic.  Finally, it should be clear that 
some of our predictions concerning energy spreading 
are completely non-trivial.  The kernel $P_t(n|m)$, 
as well as its parametric version $P(n|m)$, are accessible to 
numerical studies \cite{prm,lds} as well as to real experiments.

\newpage
\section{Energy surfaces and eigenstates}
\label{s_surfaces}

We consider a system that is described by an Hamiltonian
${\cal H}(Q,P;x)$ where $(Q,P)$ are canonical variables and 
$x$ is a parameter. The phase space volume that corresponds 
to a the energy surface ${\cal H}(Q,P;x)=E$ is 
\begin{eqnarray} \label{e_1} 
n \ = \ \Omega(E;x) \ = \ \int  \frac{dQdP}{(2\pi\hbar)^d} 
\ \Theta(E-{\cal H}(Q,P;x))
\end{eqnarray}
where $d$ is the number of degrees of freedom. 
Measuring phase-space volume in units of $(2\pi\hbar)^d$ 
is insignificant classically, but very convenient upon 
quantization. The density of phase space cells will be denoted 
by $g(E)=\partial_{\tbox{E}} \Omega(E;x)$. 
The energy surface that corresponds to 
a phase space volume $n$ will be denoted 
by $|n(x)\rangle$ and its energy 
will be denoted by $E_n(x)$. 
We assume a simple phase space topology such that for 
given $n$ and $x$ corresponds a unique energy surface. Thus 
\begin{eqnarray} \label{e_2} 
 |n(x)\rangle \ = \ 
\{(Q,P) | \ \ {\cal H}(Q,P;x)=E_n(x) \ \} 
\end{eqnarray}
The microcanonical distribution which is 
supported by $|n(x)\rangle$ is 
\begin{eqnarray} \label{e_3}  
\hspace*{-2cm}
\rho_{n,x}(Q,P) \ = \ 
\frac{1}{g(E)}\delta({\cal H}(Q,P;x)-E_n(x)) \ = \
\delta(\Omega({\cal H}(Q,P;x)) - n)
\end{eqnarray}
In the QM case the energy becomes quantized, and the mean level density 
is related to the classical density of phase space cells.  
By Weyl law we have:  
\begin{eqnarray} \label{e_4}  
\frac{1}{\Delta} \ \  \equiv \ \ 
\overline{\sum_n \delta(E-E_n)} \ \  = \ \ g(E)
\end{eqnarray}
Thus, upon quantization, the variable $n$ 
becomes a level-index, and $\rho_{n,x}(Q,P)$ 
should be interpreted as the Wigner function 
that corresponds to the eigenstate $|n(x)\rangle$. 
With these definitions we will be able to address 
the QM theory and the classical theory simultaneously. 
We shall use from now on an admixture of classical 
and quantum-mechanical jargon. This should not 
cause any confusion. The QM discussion however is 
postponed to later sections. 
The following discussion is purely classical.

Let us consider a set of {\em parametrically related} 
energy surfaces $|n(x)\rangle$ that enclose  
the same phase space volume $n$. By differentiation 
of  the expression $\Omega(E(x);x) = n$ with respect 
to the parameter $x$ one obtains:
\begin{eqnarray} \label{e_5} 
\delta E = -F(x) \delta x
\ , \ \ \ \ \ \ \ 
F(x) \ \equiv \ \left\langle 
- \frac{\partial {\cal H}}{\partial x}
\right\rangle_{\tbox{E}} 
\end{eqnarray}
The angular brackets denote microcanonical average
over all the phase space points that satisfy ${\cal H}(Q,P;x)=E$.
Later we shall see that the quantity $F(x)$ 
has the meaning of a generalized (conservative) force. 
Having $F(x)=0$ for any $x$ is equivalent to 
having $\Omega(E;x)$ which is independent of $x$. 
Such is the case for a gas particle which is affected by 
collisions with a small rigid body that is being translated 
inside a large cavity. We shall refer to the latter 
example as the `piston' example. See Fig.\ref{f_pistons} 
and Sec.\ref{s_piston} for more details.  
In order to simplify notations we shall assume, 
with almost no loss of generality, that indeed $\Omega(E;x)$ 
is independent of $x$.  It is also useful to define 
\begin{eqnarray} \label{e_6} 
{\cal F}(Q,P;x) \ \equiv \  
\left( -\frac{\partial {\cal H}}{\partial x} \right)
-\left\langle -\frac{\partial {\cal H}}{\partial x} 
\right\rangle_{\tbox{E}} 
\end{eqnarray}
We can define a parametric 
correlation scale $\delta x_c^{\tbox{cl}}$   
that is associated with the function ${\cal F}(Q,P;x)$. 
For the `piston' example (to be discussed in Sec.\ref{s_piston}) 
it is just the penetration distance into the `piston' upon 
collision (the effective `thickness' of the wall). 
If either $(Q,P)$ or $x$ become time-dependent, then 
${\cal F}$ becomes a fluctuating quantity. The nature 
of these fluctuations is discussed in the next section.

\section{Fluctuations}
\label{s_flc}

For a given $x=x(t)$ and initial conditions $(Q(0),P(0))$
we may find the time-history $(Q(t) P(t))$. 
We shall also use the notations 
\begin{eqnarray} \nonumber
{\cal E}(t) \ & \equiv & \ {\cal H}(Q(t),P(t);x(t)) \\ \nonumber
{\cal F}(t) \ & \equiv & \ {\cal F}(Q(t),P(t);x(t))
\end{eqnarray}
The correlator of the fluctuating force is 
\begin{eqnarray}  \label{e_7}
C(t,\tau) \ \equiv \ \langle {\cal F}(t) {\cal F}(t+\tau) \rangle
\end{eqnarray}
The angular brackets denote microcanonical average 
over the initial ($t{=}0$) phase-space point $(Q(0),P(0))$. 
In the next two paragraphs we are going to discuss the 
statistical properties of of the fluctuating 
force ${\cal F}(t)$. First we consider the  
time independent case ($x=\mbox{\it const}$), and then 
the time dependent case. 
 
If $x=\mbox{\it const}$, then  ${\cal E}(t) = E$ is a 
constant of the motion, and $C(t,\tau) \equiv C_{\tbox{E}}(\tau)$
is independent of $t$. It is assumed that the 
dynamics is {\em chaotic}, and consequently the 
the stochastic force looks like white noise.   
The fluctuations spectrum  
$\tilde{C}_{\tbox{E}}(\omega)$ is defined as the Fourier transform 
of $C_{\tbox{E}}(\tau)$.  The intensity of the fluctuations 
is characterized by the parameter 
\begin{eqnarray} \label{e_8} 
\nu_{\tbox{E}} \ \equiv \ 
\int_{-\infty}^{\infty} C_{\tbox{E}}(\tau) d\tau 
\ \equiv \  \tilde{C}_{\tbox{E}}(\omega{=}0)
\end{eqnarray}
The fluctuations are also characterized by a short correlation time 
$\tau_{\tbox{cl}} = \tilde{C}_{\tbox{E}}(0)/ C_{\tbox{E}}(0)$. 
For generic Hamiltonian 
system it is natural to identify $\tau_{\tbox{cl}}$ with 
the ergodic time $t_{\tbox{erg}}$. However, in specific applications 
we may have $\tau_{\tbox{cl}}<t_{\tbox{erg}}$. For the `piston'  
example, that will be discussed in Sec.\ref{s_piston}, 
the correlation time $\tau_{\tbox{cl}}$ is equal to the 
duration of a collision with the wall, 
while $t_{\tbox{erg}}$ is determined by the 
ballistic time $\tau_{\tbox{bl}}$.

If $x$ is time dependent rather than a constant,  
for example $x(t)=Vt$, then for any finite $V$ and $0<t$ 
the actual distribution of $(Q(t),P(t))$ 
is no longer microcanonical. 
The statistical properties of the 
fluctuating force ${\cal F}(t)$ are 
expected to be different from the $V{=}0$ case. 
The average $\langle {\cal F}(t) \rangle$ 
is no longer expected to be zero. 
Rather, we shall argue (See (\ref{e_23}) 
that $\langle {\cal F}(t) \rangle = -\mu V$, 
where $\mu$ is the dissipation coefficient. 
This implies that the correlator 
$C(t,\tau)$ acquires an offset $(\mu V)^2$. 
The offset term can be neglected for 
a limited time $t<t_{\tbox{frc}}$ 
\begin{eqnarray} \label{e_9}
t_{\tbox{frc}} \ = \ \nu/(\mu V)^2
\hspace*{2cm} \mbox{[classical breaktime],}
\end{eqnarray} 
provided $(\mu V)^2 \ll C_{\tbox{E}}(0)$. 
The latter condition, which is equivalent to having 
\begin{eqnarray} \label{e_10}  
\tau_{\tbox{cl}} \ \ll \ t_{\tbox{frc}}
\hspace*{2cm} \mbox{[non-trivial slowness condition].}
\end{eqnarray}
implies that the velocity $V$ should be small enough. 
There is another possible reason for the correlator 
$C(t,\tau)$ to be different from  $C_{\tbox{E}}(\tau)$. 
Loss of correlation may be either due to the 
dynamics of $(Q(t),P(t))$ or else due to the 
parametric change of $x(t)$. The correlation time 
which is associated with the dynamics is 
$\tau_{\tbox{cl}}$. The correlation time 
which is associated with the parametric time dependence 
is $\tau_c^{\tbox{cl}}=\delta x_c^{\tbox{cl}}/V$. 
We always assume that   
\begin{eqnarray} \label{e_11}   
\tau_{\tbox{cl}} \ \ll \ \tau_c^{\tbox{cl}}
\hspace*{2cm} \mbox{[trivial slowness condition].}
\end{eqnarray}
meaning that loss of correlations is predominantly 
determined by the chaotic nature of the 
dynamics rather than by the (slow) parametric 
change of the Hamiltonian. 
Since we always assume that that 
the classical slowness conditions ((\ref{e_10})
and (\ref{e_11})) are being satisfied,  
it follows that we can make the approximation 
\begin{eqnarray} \label{e_12}  
C(t,\tau) \ = \ C(\tau) \ \approx \ C_{\tbox{E}}(\tau) 
\hspace*{2cm} \mbox{for $t<t_{\tbox{frc}}$}
\end{eqnarray}

\section{Energy spreading and dissipation}
\label{s_frc}

\begin{figure}
\begin{center}
\leavevmode
\epsfysize=1.7in 
\epsffile{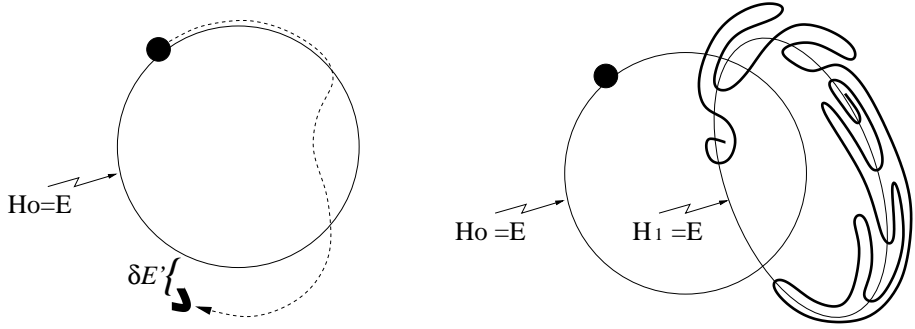}
\end{center}
\caption{\protect\footnotesize 
Schematic illustration of the dynamics.
An initially localized distribution is launched in phase space.  
For a limited time (left plot) it travels upon the 
initial energy surface. But then it departs from it. 
After much longer times (right plot) the evolving 
distribution is concentrated across an  
instantaneous energy surface. } 
\label{f_evolving}
\end{figure}

For $x(t)=\mbox{\it const}$ the energy ${\cal E}(t)$ is a constant 
of the motion. For time dependent $x(t)$ and 
any particular time-history we can write 
\begin{eqnarray} \nonumber 
\frac{d{\cal E}}{dt} \ = 
\ \frac{\partial {\cal H}}{\partial t} \ = \ 
- (F(x(t)) + {\cal F}(t)) \ \dot{x}
\end{eqnarray}
The first term implies reversible change of energy 
due to a conservative force that equals $F(x)$. 
In what follows we shall see that the second term 
is responsible for an irreversible dissipation process. 
Integrating over time and using (\ref{e_5}) we get
\begin{eqnarray} \label{e_13}  
{\cal E}(t) \ = \ E(x(t)) 
\ - \ V \int_0^t {\cal F}(t)dt 
\end{eqnarray}
where $E(x)\equiv E_{m}(x)$ is the 
energy that correspond to the initial 
phase-space volume $m$. The difference 
$E(x(t)){-}E(x(0))$ is due to the 
reversible work done by the generalized force $F(x)$.  
If we disregard the fluctuating term, then 
we come to the conclusion that the trajectory is 
approximately bounded to the evolving 
energy surface $|n(x(t))\rangle$. 
Thus the phase-space volume $n=\Omega({\cal E}(t),x(t))$ 
is an approximate constant of the motion, 
the so-called `adiabatic invariant'.  
Using (\ref{e_13}) we can estimate the energy dispersion 
which is associated with the fluctuating force:
\begin{eqnarray} \label{e_14} 
\langle ({\cal E}(t)- E(x(t)))^2 \rangle \ = \ 
V^2 \int_0^t dt' \int_{-t'}^{t'} C(t',\tau) d\tau 
\end{eqnarray}
Hence we get a crossover from ballistic 
to diffusive behavior:
\begin{eqnarray} \label{e_15} 
\langle ({\cal E}(t)-E(x(t)))^2 \rangle 
\ \approx \ C_{\tbox{E}}(0) \cdot (Vt)^2
\ \ \ \ \ \ \ \ \ \ 
& \mbox{for $t \ll \tau_{\tbox{cl}}$} \\
\label{e_16} 
\langle ({\cal E}(t)-E(x(t)))^2 \rangle 
\ \approx \ 2D_{\tbox{E}} \ t
\ \ \ \ \ \ \ \ \ \ \ \ \ \ \ \ \ \  
& \mbox{for $\tau_{\tbox{cl}} \ll t \ll t_{\tbox{frc}}$} 
\end{eqnarray} 
The ballistic spreading on short time scales just 
reflects the parametric change of the energy surfaces. 
This is the essence of the {\em sudden approximation} 
which is illustrated in Fig.\ref{f_evolving} and further 
explained in App.\ref{a_evolving}. 
The diffusive spreading on longer times reflects 
the deviation from the {\em adiabatic approximation}. 
See illustration in Fig.\ref{f_evolving} and 
further details in App.\ref{a_evolving}.   
The diffusion coefficient is 
\begin{eqnarray} \label{e_17} 
D_{\tbox{E}} \ \ = \ \ 
\frac{1}{2}V^2 \int_{-t}^t C_{\tbox{E}}(\tau)d\tau
\ \ \rightarrow \ \
\frac{1}{2} \ \nu_{\tbox{E}} \ V^2
\end{eqnarray} 
If $C_{\tbox{E}}(\tau)$ is short range in 
nature, then $D_{\tbox{E}}$ will tend eventually 
to the well defined constant value which is 
indicated in the right hand side of (\ref{e_17}).   
Note that the adiabatic approximation becomes exact in the formal 
limit $V\rightarrow 0$, keeping $Vt$ constant.

For {\em intermediate} times $(Q(t),P(t))$ are 
distributed ergodically across the evolving energy 
surface, within a shell of thickness $\sqrt{2D_{\tbox{E}} t}$.  
We shall argue later (Sec.\ref{s8}) that more generally, 
for any $t\gg t_{\tbox{erg}}$,  the spreading profile $\rho(E)$ 
obeys the following diffusion equation 
\begin{eqnarray} \label{e_18} 
\frac{\partial \rho}{\partial t} \ = \ 
\frac{\partial}{\partial E}
\left(g(E)D_{\tbox{E}} \frac{\partial}{\partial E}
\left(\frac{1}{g(E)}\rho\right)\right)
\end{eqnarray}
For simplicity we assume here that there is 
no conservative work ($F(x){=}0$).
The energy dependence of the diffusion process 
implies a systematic growth of the mean energy 
$\langle {\cal E}(t) \rangle = \int E\rho(E)dE $.
Namely, 
\begin{eqnarray} \label{e_19} 
\dot{{\cal Q}} \ \equiv \ \frac{d}{dt}\langle {\cal E} \rangle 
\ = \ - \int_0^{\infty} dE \ g(E) \ D_{\tbox{E}}  
\ \frac{\partial}{\partial E}
\left(\frac{\rho(E)}{g(E)}\right)
\end{eqnarray}
Substituting $D_{\tbox{E}}=\half\nu_{\tbox{E}}V^2$ and 
integrating by parts one obtains $\dot{{\cal Q}}=\mu V^2$, 
along with ${\cal F\!D}$ relation that can be written 
schematically as $\mu={\cal F\!D}[\nu]$. The result for 
$\mu$ depends on $\rho(E)$. If $\rho(E)$ is well concentrated 
around some  energy $E$, one obtains 
\begin{eqnarray} \label{e_20}  
\mu_{\tbox{E}} \ = \ \frac{1}{2} \frac{1}{g(E)} 
\frac{\partial}{\partial E}
(g(E) \nu_{\tbox{E}}) 
\ \ \ \ \ \mbox{[microcanonical version]} 
\end{eqnarray}
Another, more familiar variation of the ${\cal F\!D}$ relation 
is obtained if one assumes a canonical distribution   
$\rho(E) \propto g(E)\exp(-E/(k_{\tbox{B}}T))$.  
By substitution into (\ref{e_19}), or simply by 
canonical averaging over (\ref{e_20}), one obtains:  
\begin{eqnarray} \label{e_22}  
\mu_{\tbox{T}} \ = \ \frac{1}{2k_{\tbox{B}} T} \nu_{\tbox{T}}
\ \ \ \ \ \mbox{[canonical version]}
\end{eqnarray}
where $\nu_{\tbox{T}}$ is related to $C_{\tbox{T}}(\tau)$, and  
the latter is defined the same way as $C_{\tbox{E}}(\tau)$, 
but with canonical rather than microcanonical averaging.

Having energy dissipation implies that for $V\ne0$ the 
fluctuating quantity ${\cal F}(t)$ has a non-zero average:  
From (\ref{e_13}), and recalling that we assume $F(x)=0$, we get 
$\dot{{\cal Q}}=\langle {\cal F}(t) \rangle V$. 
Therefore $\dot{{\cal Q}}=\mu V^2$ implies 
\begin{eqnarray} \label{e_23} 
\langle {\cal F}(t) \rangle \ = \ -\mu V 
\end{eqnarray}
In case of the `piston' example $V$ is the velocity 
of the `piston' and  $\langle {\cal F}(t) \rangle$ is the 
associated friction force. In case of conductivity 
calculation (See Sec.1.9), the parameter $x$ represents 
(time-dependent) magnetic flux via a ring, 
$V$ is the electro-motive-force, and $\langle {\cal F}(t)\rangle$   
is the drift velocity.

\section{Quantal energy-spreading: Linear response theory}
\label{s_elementary}

At first sight it seems that the classical derivation 
in Sec.\ref{s_frc} applies also to the QM case 
provided ${\cal F}(t)$ is treated as an operator. 
This is essentially the so-called `linear response theory'.
The only approximation involved is 
$C(t,\tau) \approx C_{\tbox{E}}(\tau)$. This approximation 
should be valid as long as the evolving state $\rho_t(Q,P)$ 
is similar to the initial {\em microcanonical preparation}. 
It is more difficult to satisfy this condition in the QM case. 
The similarity $\rho_t(E)\approx\rho_0(E)$  
is not a sufficient condition: It is also required  
that off-diagonal elements of the probability-matrix could  
be ignored, meaning that a superposition could be treated as 
if it were an incoherent mixture of the corresponding 
energy-eigenstates.  The classical considerations 
(Sec.\ref{s_flc}) lead to the time restriction $t\ll t_{\tbox{frc}}$. 
The QM considerations will lead to 
a stronger time restriction $t\ll t_{\tbox{prt}}$. 
Accordingly, the classical slowness condition (\ref{e_10}) 
is replaced by:
\begin{eqnarray} \label{e_24} 
\tau_{\tbox{cl}} \ \ll \ t_{\tbox{prt}}
\ \ \ \ \ \mbox{[quantal slowness condition].}
\end{eqnarray} 
The determination of $t_{\tbox{prt}}$, which is the 
breaktime for QM perturbation theory,  
will be discussed in later sections. 
For non-slow velocities, (meaning throughout this paper
that (\ref{e_10}) and (\ref{e_11}) are satisfied 
but (\ref{e_24}) is violated), 
the following elementary considerations does not apply. 
The quantal slowness condition (\ref{e_24}) should not be 
confused with the QM-adiabaticity condition 
(to be discussed later). QM-adiabaticity  
requires extremely slow velocities. 

The QM version of the derivation in 
Sec.\ref{s_frc} gives a classical look-alike result 
for the energy spreading. The only implicit modification 
is that the classical $C_{\tbox{E}}(\tau)$ should be replaced 
by the corresponding QM object.
For the purpose of concise presentation, the formula 
for the energy spreading can be written as follows:
\begin{eqnarray} \label{e_25} 
\hspace*{-2cm} 
\delta E^2 \ = \  
V^2\int_0^t\int_0^t C_{\tbox{E}}(t_2{-}t_1)dt_1dt_2
\ = \ 
V^2 t \int_{-\infty}^{+\infty}\frac{d\omega}{2\pi}
\tilde{C}_{\tbox{E}}(\omega) \ \tilde{F}_t(\omega)
\end{eqnarray} 
where 
\begin{eqnarray}  \label{e_26} 
\tilde{F}_t(\omega) \ = \ t{\cdot}(\mbox{sinc}(\omega t /2))^2
\end{eqnarray} 
Now $C_{\tbox{E}}(\tau)$ is a QM object, and its 
Fourier transform $\tilde{C}_{\tbox{E}}(\omega)$ can 
be expressed as
\begin{eqnarray} \label{e_27}
\tilde{C}_{\tbox{E}}(\omega) \ = \ 
\sum_n'
\left|\left(\frac{\partial {\cal H}}{\partial x}\right)_{nm}\right|^2
\ 2\pi\delta\left(\omega-\frac{E_n{-}E_m}{\hbar}\right)
\end{eqnarray} 
One observes that the power-spectrum of the QM fluctuations 
has a discrete nature, and consequently the correlation 
function $C_{\tbox{E}}(\tau)$ is characterized by the 
the additional time scale $t_{\tbox{H}}$.  We assume that 
$\hbar$ is reasonably small such that $\tau_{\tbox{cl}}\ll t_{\tbox{H}}$. 
Correspondence considerations imply that the 
quantal $C_{\tbox{E}}(\tau)$ is similar to the 
classical $C_{\tbox{E}}(\tau)$ as long as $t \ll t_{\tbox{H}}$. 
Equivalently, as long as $t \ll t_{\tbox{H}}$ the discrete nature of 
the quantal $\tilde{C}_{\tbox{E}}(\omega)$ can be ignored, 
and we can effectively use the classical $\tilde{C}_{\tbox{E}}(\omega)$. 
Recall that the power-spectrum of the classical fluctuations 
looks like that of white noise: 
It satisfies $\tilde{C}_{\tbox{E}}(\omega) \approx \nu_{\tbox{E}}$ 
for $|\omega|\ll 1/\tau_{\tbox{cl}}$ and decays rapidly to zero 
outside of this regime. Thus, for $t \ll \tau_{\tbox{cl}}$ 
we can make the replacement $\tilde{F}_t(\omega)\rightarrow t$, 
and we obtain the ballistic result 
$\delta E^2 = C_{\tbox{E}}(0){\cdot}(Vt)^2$,  
while for  $t \gg \tau_{\tbox{cl}}$ we can make the replacement 
$\tilde{F}_t(\omega)\rightarrow 2\pi\delta(\omega)$, and 
we get then the diffusive behavior 
$\delta E^2 = \nu_{\tbox{E}}V^2 t$.

We can get a semiclassical estimate for the matrix elements 
in (\ref{e_27}) by exploiting the correspondence that has 
been mentioned above \cite{mario}. 
The function $\tilde{C}_{\tbox{E}}(\omega)$ 
is assumed to be vanishingly small for $\omega\gg 1/\tau_{\tbox{cl}}$ 
which implies that $({\partial {\cal H}}/{\partial x})_{nm}$  is a 
banded matrix. Energy levels are coupled by 
matrix elements provided $|E_n-E_m|<\Delta_b$ where
\begin{eqnarray} \label{e_28}
\Delta_b \ = \ 
\frac{2\pi\hbar}{\tau_{\tbox{cl}}} \ = \ 
\mbox{band width}
\end{eqnarray} 
For $\omega \ll 1/\tau_{\tbox{cl}}$ the smoothed 
$\tilde{C}_{\tbox{E}}(\omega)$ should be equal to 
the classical noise intensity $\nu_{\tbox{E}}$.   
Consequently one obtains the following estimate for 
individual matrix elements within the band:
\begin{eqnarray} \label{e_29}  
\sigma^2 \ = \ 
\left|\left(\frac{\partial {\cal H}}{\partial x}\right)_{nm}\right|^2
\ \approx \ \frac{\Delta}{2\pi\hbar} \ \nu_{\tbox{E}} 
\ \ \ \ \ \mbox{for} \ \ |E_n{-}E_m| < \Delta_b
\end{eqnarray}
It is important to specify the minimal number of 
(generic) parameters that are involved in the 
above analysis.  In the classical problem there 
are just two generic parameters: Namely, 
$\tau_{\tbox{cl}}$ and $\nu_{\tbox{E}}$.  
Quantum-mechanics requires the specification of 
{\em two} additional parameters: Namely,  
the band width $\Delta_b$  and the 
mean level spacing $\Delta$. 
The associated dimensionless parameters are: 
\begin{eqnarray}  
v_{\tbox{PR}} \ & = \ \mbox{\small scaled velocity} \ & = \ 
\sqrt{2D_{\tbox{E}} \ \tau_{\tbox{cl}}} \ / \ \Delta_b \\
b \ & = \ \mbox{\small scaled band width} \ & = \ {\Delta_b} \ / \ {\Delta}
\end{eqnarray} 
The specification of $\Delta$ is not dynamically significant as long 
as $t\ll t_{\tbox{H}}$.  Longer times are required in order 
to resolve individual energy levels. Thus we come to the 
conclusion that in the time regime $t\ll t_{\tbox{H}}$
there is a {\em single} generic dimensionless 
parameter, namely $v_{\tbox{PR}}$, that controls QCC.
We shall see that the QM definition 
of slowness (\ref{e_24}) can be cast into 
the form $v_{\tbox{PR}} \ll 1$.

\section{Quantal energy spreading: The conventional FGR picture}
\label{s_further}

Equation (\ref{e_25}) for the QM energy spreading 
can be derived using linear response theory, i.e. by 
following the same steps as in Sec.\ref{s_frc}. 
However, the simplicity of linear response theory is lost once 
we try to formulate a controlled version of it. 
It is difficult to derive and to get a good understanding 
for the breaktime scale $t_{\tbox{prt}}$. 
It is better to use the conventional version of 
time-dependent first-order perturbation theory (FOPT), 
and to view the energy spreading as arising 
from {\em transitions between energy levels}. 

The choice of basis for the representation of the 
dynamics is a crucial step in the analysis. The 
proper basis for the understanding of energy spreading 
is the $x$-dependent set of eigenstates $|n(x)\rangle$ 
of the Hamiltonian ${\cal H}(Q,P;x(t))$.  This is the 
basis that we are going to use later in this paper. 
In a sense we are going to introduce an {\em improved} version 
of FGR picture. However, for sake of completeness, we 
would like to discuss in this section the capabilities 
and the limitations of the {\em conventional} FGR picture. 
The conventional FGR picture is using a {\em fixed basis} 
that is determined by the unperturbed Hamiltonian 
${\cal H}(Q,P;x(0))$. It should be clear that transitions 
between unperturbed energy levels reflect reduced-energy-changes 
rather than actual-energy-changes (see corresponding 
classical definitions in App.\ref{a_evolving}). 
Therefore the description of the crossover 
from ballistic to diffusive energy spreading is 
out-of-reach for this version of perturbation theory.  

The {\em conventional} FGR picture can be used in order 
to determine the diffusion coefficient $D_{\tbox{E}}$, 
as well as the perturbative breaktime $\tau_{\tbox{prt}}$. 
A detailed derivation can be found in App.\ref{a_FGR}).
We use the notation $\tau_{\tbox{prt}}$ rather than 
$t_{\tbox{prt}}$, because a different version of perturbation 
theory is involved here. The final result is 
$D_{\tbox{E}} = \half\nu_{\tbox{E}}^{\tbox{eff}}V^2$, 
with the effective noise intensity:  
\begin{eqnarray} \label{e_32} 
\nu_{\tbox{E}}^{\tbox{eff}} \ = \  
\int_{-\infty}^{+\infty} 
C_{\tbox{E}}(\tau) \ F(\tau) \ S(\tau) \ d\tau 
\end{eqnarray} 
where $F(\tau)$ is the correlation function of the 
driving source, $S(\tau)=\exp(-(\Gamma/2)t)$ 
is the survival amplitude, and $\tau_{\tbox{prt}}=1/\Gamma$.   
The introduction of $S(\tau)$ is a common ad-hoc improvement 
of equation (\ref{e_FGRD}). Such type of improvement 
is used in other contexts to get the Wigner-Weisskopf 
Lorentzian line shape. It approximates the effect of higher 
orders of time-dependent perturbation theory.   
It is also important to specify the validity condition of 
the FGR picture. It is not difficult to be convinced 
that the requirement (\ref{e_FGRrq}) can be relaxed slightly, 
and the actual condition is:
\begin{eqnarray} \label{FGRc}
\mbox{\bf FGR-condition:} \ \ \ \  
\mbox{Either} \ \tau_{\tbox{cl}} \ \mbox{or} \ \tau_c  
\ \ \ \ll \ \ \ \tau_{\tbox{prt}}   
\end{eqnarray} 
Here $\tau_c$ characterizes the correlation 
function $F(\tau)$, namely, it is the correlation time 
of the driving source. It should be realized that having 
$\tau_c\gg\tau_{\tbox{cl}}$ constitutes an obvious variation 
of the trivial slowness condition (\ref{e_11}). 
Furthermore, using (\ref{ea_c4}) one can easily conclude that 
a necessary condition for the applicability of  
FGR picture is $v_{\tbox{PR}}\ll 1$, which coincides 
with the quantal slowness condition (\ref{e_24}).  
Thus, it is {\em not possible in principle} to 
apply the FGR picture in the limit $\hbar\rightarrow 0$.

The consistency of the FGR result (\ref{e_32})
with the linear response result (\ref{e_25})
is not obvious. It is true that (\ref{e_25}) 
gives a crossover from ballistic to diffusive behavior,  
where indeed $D_{\tbox{E}} = \half\nu_{\tbox{E}}^{\tbox{eff}}V^2$.  
But if one takes (\ref{e_25}) seriously 
for $t\gg t_{\tbox{H}}$, one will come 
to the conclusion that this diffusion will 
stop due to recurrences \cite{berry}.  
The FGR result (\ref{e_32}), due to the presence 
of $S(\tau)$, does not imply such a conclusion,   
provided $\tau_{\tbox{prt}} \ll t_{\tbox{H}}$. 
A vanishingly small result for $D_{\tbox{E}}$ 
is obtained only in the QM-adiabatic regime 
where we may have $\tau_{\tbox{prt}}\gg t_{\tbox{H}}$.
In the QM-adiabatic regime Landau-Zener transitions 
between neighboring levels become the predominant 
mechanism for energy spreading \cite{wilk1}, and 
the FGR picture becomes of minor importance.

\section{The `piston' example and the wall formula}
\label{s_piston}

The above picture and considerations become much more 
transparent once applied to cavities with moving walls. 
The Hamiltonian ${\cal H}=E(\mbf{p})+{\cal V}(\mbf{x})$ describes 
the free motion of a `gas particle', whose canonical coordinates 
are $Q=\mbf{x}$ and $P=\mbf{p}$, inside a $d$-dimensional space 
which is confined by some boundary. Unless otherwise specified 
$E(\mbf{p})=\mbf{p}^2/2m$ where $m$ is the mass of the gas particle.  
The corresponding velocity will be denoted by $v_{\tbox{E}}$. 
The boundary is composed of wall-elements, and may have 
few components. For example it may consist of some 
`static' component that defines an interior space, 
and an additional `moving' component that defines an 
excluded space of an impenetrable `piston' as in Fig.\ref{f_pistons}. 
The displacement of the moving wall-elements is parameterized   
by $x$. The gas particle undergoes elastic collisions with 
the boundary. The ballistic time will be denoted 
by $\tau_{\tbox{bl}}$. It is determined by the collision rate 
with the walls. The derivation in App.\ref{a_Ft} 
gives the result:
\begin{eqnarray} \label{e_39} 
\frac{1}{\tau_{\tbox{col}}} \ = \ 
\left\langle\sum_{\tbox{col}}\delta(t-t_{\tbox{col}})\right\rangle
\ = \ 
\frac{1}{2}\frac{\mboxs{Area}}{\mboxs{Volume}} 
\langle|\cos\theta|\rangle \ v_{\tbox{E}}
\end{eqnarray}
The $d$-dependent geometrical factor $\langle|\cos\theta|\rangle$ is
defined in App.\ref{a_spherical}.
For the purpose of `ballistic-time' definition we should 
take the total \mboxs{Area} of all the wall elements. 
For the purpose of calculating an effective 
collision rate the {\em effective} \mboxs{Area} should be defined as 
in (\ref{e_41}).  For the `piston' example the total 
effective \mboxs{Area} of the moving-faces of the `piston' 
may be much smaller compared with the total \mboxs{Area} 
of the walls, and consequently the effective time between 
collisions will be much larger than the ballistic time.

For the later QM considerations 
it is essential to consider `soft walls'. For concreteness we 
may assume that the wall is realized by a constant force field 
$f$. If $z$ is a coordinate perpendicular to a wall elements, 
than the potential barrier is ${\cal V}(z)=0$ for $z<0$ and 
${\cal V}(z)=f{\cdot}z$ for $z>0$ up to some maximal value 
${\cal V}_{\tbox{wall}}$ well inside the barrier.  We assume 
that the energy $E$ of the particle is much lower than
${\cal V}_{\tbox{wall}}$, and therefore the latter energy scale 
should be of no significance. For strongly chaotic billiards 
successive collisions with the `piston' are uncorrelated and 
therefore the classical correlation time $\tau_{\tbox{cl}}$ is 
equal to the collision time with the wall. Namely 
\begin{eqnarray} \label{e_40} 
\tau_{\tbox{cl}} \ = \ (2mv_{\tbox{E}})/f 
\end{eqnarray}
Obviously, in the hard wall limit 
$f\rightarrow\infty$ we have $\tau_{\tbox{cl}}\rightarrow 0$. 
The displacement of the walls is parameterized by 
some parameter $x$. With each surface-element $ds$ 
we can associate a normal unit vector $\mbf{n}$, 
and a `propagation velocity' which will be denoted  
by $\hat{\mbf{V}}(\mbf{s})$. 
The latter is simply the derivative 
of the wall-element displacement by the controlling 
parameter $x$. The effective moving-wall area 
is defined as follows:
\begin{eqnarray} \label{e_41} 
\mboxs{Area} \ = \ \oint (\mbf{n}{\cdot}\hat{\mbf{V}})^2 ds
\end{eqnarray}
Due to ergodicity there are two different strategies 
that can be applied in order to calculate $\nu_{\tbox{E}}$. 
One possibility is to average ${\cal F}(t){\cal F}(t{+}\tau)$
over $(\mbf{x},\mbf{p})$, as implied by the definition (\ref{e_7}),  
and then to integrate over $\tau$. 
Schematically the calculation goes as follows: 
$\langle {\cal F}^2 \rangle$
simply equal to $f^2$ multiplied by the 
ratio between the collision-volume 
($\mboxs{Area}\times (v_{\tbox{E}} \tau_{\tbox{cl}}))$ and the 
total $\mboxs{Volume}$. Note that the latter ratio 
simply equals $\tau_{\tbox{cl}}/\tau_{\tbox{col}}$. 
The noise intensity 
is obtained by further multiplication 
with $\tau_{\tbox{cl}}$. 
The proper calculation  
should be done as in App.\ref{a_CE}.
The other possibility  
to calculate $\nu_{\tbox{E}}$
is to write ${\cal F}(t)$ as a sum over short 
impulses and to perform the averaging 
in time domain. The magnitude of 
the impulses is $2mv_{\tbox{E}}$. The 
noise intensity is simply equal to the 
square of the impulses 
multiplied by the collision rate.  
The exact calculation is done 
in App.\ref{a_Ft}.  Both approaches give 
obviously the same result:
\begin{eqnarray} \label{e_42} 
\nu_{\tbox{E}} \ = \ 
2 \langle|\cos\theta|^3\rangle \  
\frac{\mboxs{Area}}{\mboxs{Volume}} 
\ m^2 v_{\tbox{E}}^3  
\end{eqnarray}
This result is quite general, but it assumes that successive 
collisions with the `piston' are uncorrelated. 
More generally, correlations between 
successive collisions should be taken into account.
The derivation can be done by following 
an essentially identical computation by 
Koonin \cite{koonin} and the result is cast into the form   
\begin{eqnarray} \label{e_43} 
\hspace*{-1cm}
\nu_{\tbox{E}} \ = \ \oint\oint ds_2 ds_1 \ 
(\mbf{n}{\cdot}\hat{\mbf{V}}(\mbf{s}_2)) \ 
\nu(\mbf{s}_2,\mbf{s}_1) \ 
(\mbf{n}{\cdot}\hat{\mbf{V}}(\mbf{s}_1))   
\end{eqnarray}
If one ignores correlations between successive collisions, 
then one obtains 
$\nu(\mbf{s}_2,\mbf{s}_1) \ \propto 
\delta(\mbf{s}_2-\mbf{s}_1)$ 
and (\ref{e_43}) reduces to (\ref{e_42}).
Using the ${\cal F\!D}$ relation one obtains the following 
generalized `wall formula':
\begin{eqnarray} \label{e_44}  
\mu_{\tbox{E}} \ = \ 2 \langle|\cos\theta|\rangle \  
\frac{\mboxs{Area}}{\mboxs{Volume}} \ mv_{\tbox{E}} 
\end{eqnarray}
The familiar $d{=}3$ version of the 
wall formula is obtained by substituting 
$\langle|\cos(\theta)|\rangle=1/2$. 
As in the the case of $\nu_{\tbox{E}}$ we 
can try to derive this result using 
a simple-minded time-domain approach. 
See App.\ref{a_Ft}. It turns out that only 
{\em half} of the correct result is obtained.  
Alternatively, the correct result (\ref{e_44}) 
can be obtained by extending the standard 
`kinetic' derivation. See App.\ref{a_kinetic}. 
The kinetic derivation demonstrates that (\ref{e_44}) 
is more general than it seems at first sight. It  
applies to any velocity-momentum dispersion relation.
The mass $m=(dv/dp)^{-1}$ may be energy dependent.

The QM calculation of $\nu_{\tbox{E}}^{\tbox{eff}}$
requires the knowledge of the quantal $C_{\tbox{E}}(\tau)$, 
and we should also have a proper understanding of $F(\tau)$. 
For the time being let us assume that the effective 
$\tau_c$ is much larger than $\tau_{\tbox{cl}}$. It means 
that the transitions are resonance-limited, and 
detailed knowledge of $F(\tau)$ becomes irrelevant. 
If the De-Broglie wavelength 
$\lambda_{\tbox{E}}=2\pi\hbar/(mv_{\tbox{E}})$ 
is much smaller compared with 
other (classical) scales, then it is expected to have 
QCC, as discussed previously with respect to (\ref{e_27}).  
However, it is also possible 
to  make a direct estimate of the matrix elements that 
appear in (\ref{e_27}). See App.\ref{a_matrix}.  The results 
are in complete agreement with our semiclassical 
expectations. Indeed, the bandwidth is determined by the 
collision time with the (soft) walls (see App.\ref{a_softwall}), 
and the power-law decay of matrix elements outside the band 
can be associated with the discontinuity 
(see App.\ref{a_CE})
in the derivative of the classical $C_{\tbox{E}}(\tau)$ at $\tau=0$.  
The expression for the effective noise intensity can be cast 
into the form of (\ref{e_43}) with 
\begin{eqnarray} \label{e_45}  
\hspace*{-2cm} 
\nu(\mbf{s}_2-\mbf{s}_1) \ = \ 
\Omega_d \ \frac{\mboxs{Area}}{\mboxs{Volume}} 
\ m^2 v_{\tbox{E}}^3 \ \frac{1}{\lambda_{\tbox{E}}^{d{-}1}} \ 
\left(\mbox{Sinc}\left( 
\frac{2\pi}{\lambda_{\tbox{E}}} |\mbf{s}_2-\mbf{s}_1|\right)\right)^2 
\end{eqnarray}
The Sinc function, as well as other notations, are defined 
in App.\ref{a_spherical}.  The QM result coincides 
with the classical result if $\lambda_{\tbox{E}}$ 
is small compared with the classical length scales  
that describe the $\mbf{s}$-dependence of $\mbf{n}{\cdot}\mbf{V}(\mbf{s})$. 
It is also assumed that $\lambda_{\tbox{E}}$ is small compared 
with the surface radius-of-curvature, else further 
corrections are required \cite{koonin}.

\newpage
\section{The route to stochastic behavior}

\label{s8}

In this section we shall introduce a general phase-space 
formulation for the theory of energy-spreading.    
The main mathematical object of the study, 
namely the kernel $P_t(n|m)$, will be defined. 
In order to go smoothly from the classical theory 
to the QM theory it is essential 
to use proper notations. From now on we 
use the variable $n = \Omega(E)$ instead of $E$. 
See definitions in Sec.\ref{s_surfaces}. 
The transition probability kernel $P_t(n|m)$ is 
defined as the projection of an evolving state 
on the instantaneous set of energy-states. 
It is also possible to define a parametric kernel
$P(n|m)$. The latter depends on the displacement 
$\delta x$ but not on the actual time that it takes 
to realize this displacement. The definitions are:
\begin{eqnarray} \label{e_47}
P_t(n|m) \ & = & \ \mbox{trace}
( \ \rho_{n,x(t)} \ {\cal U}(t) \ \rho_{m,x(0)} \ ) \\
P(n|m) \ \  & = & \ \mbox{trace}
( \ \rho_{n,x(t)} \ \rho_{m,x(0)} \ )
\end{eqnarray}
In the above definitions the initial energy-surface
is $|m(x(0))\rangle$, and the associated  
phase-space density is $\rho_{m,x(0)}(Q,P)$. 
In the QM-case $|m(x(0))\rangle$ is an energy-eigenstate 
and $\rho_{m,x(0)}(Q,P)$ is the associated 
Wigner function. The evolving surface/state is 
represented by  ${\cal U}(t) \ \rho_{m,x(0)}$, where  
${\cal U}(t)$ is either the classical Liouville propagator 
or its QM version. In the classical case 
it simply re-positions points in phase-space. 
In the QM case it propagates a Wigner function 
and it may have a more complicated structure.
The trace operation is just a $dQdP$ integral over 
phase-space. In the QM-case 
the definitions of $P(n|m)$ and  $P_t(n|m)$ can be cast  
into a much simpler form using Dirac's notations: 
See (\ref{e_68}) and (\ref{e_69}).

\begin{figure}
\begin{center}
\leavevmode 
\epsfysize=6.0in 
\epsffile{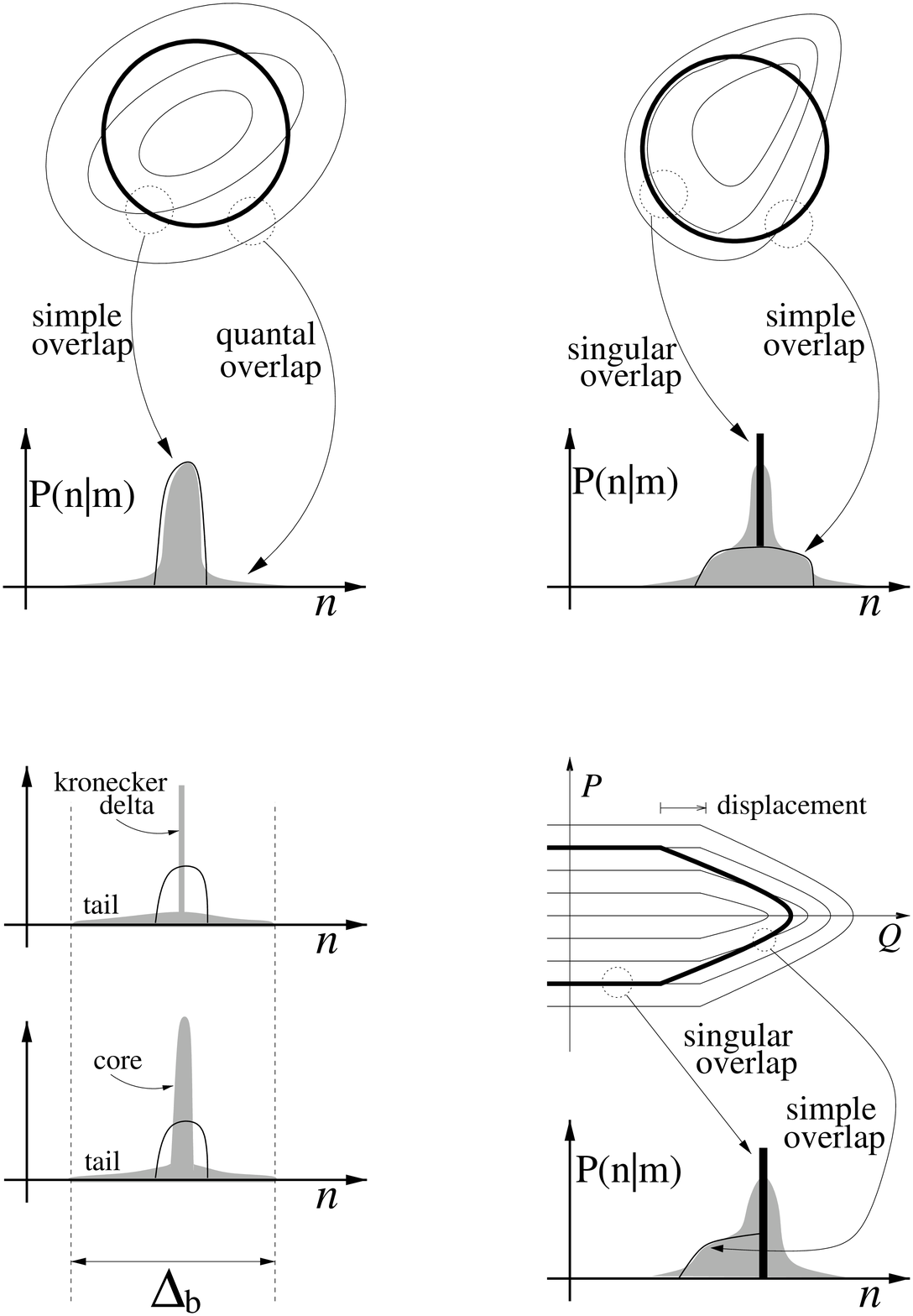}
\end{center}
\caption{\protect\footnotesize 
{\em Upper Left:} Phase space illustration of the 
initial and of the instantaneous set of parametric 
energy surfaces; Plot of the associated $P(n|m)$, 
where the classical behavior is indicate 
by the black lines, and the QM  
behavior is represented by the grey filling.   
Detailed QCC is assumed.  In the QM case  
classical sharp-cutoffs are being smeared.  
{\em Upper Right:} Illustration of a typical 
non-generic feature. In the QM case 
the classical delta-singularity is being smeared. 
{\em Lower Right:} The same non-generic feature 
manifests itself in the `piston' example. 
{\em Lower Left:} In the perturbative case there 
is no detailed QCC. 
The kernel is characterized by a core-tail 
structure. The tail is limited by the bandwidth 
of the coupling matrix-elements. If $\delta x$ is 
sufficiently small the core is just a kronecker's delta.} 
\label{f_surfaces}
\end{figure}

In the classical case 
the kernel $P(n|m)$ reflects the parametric correlations 
between two sets of energy surfaces 
(Fig.\ref{f_surfaces} upper left). 
Consequently non-Gaussian features may manifest themselves.  
An important special non-Gaussian feature is encountered 
in many specific examples where $x$ affects only 
a tiny portion of the energy surface. 
(Fig.\ref{f_surfaces} upper right). 
In the `piston' example this is the case 
because $({\partial {\cal H}}/{\partial x}) = 0$ unless 
$Q$ is near the face of the piston. Consequently 
$P(n|m)$ will have a $\delta$-singularity for $n=m$.

The classical scenario for $P_t(n|m)$ consists 
of three time regimes. For short times 
we have the classical {\em sudden approximation}:  
\begin{eqnarray} \label{e_49}
P_t(n|m) \ \approx \ P(n|m)
\ \ \ \ \ \ \mbox{for $t \ll \tau_{\tbox{cl}}$}
\end{eqnarray}
See Fig.\ref{f_evolving} and App.\ref{a_evolving} 
for more details. For longer times we have the classical 
{\em adiabatic approximation}, or more precisely we have 
diffusive spreading:  
\begin{eqnarray} \label{e_50}
P_t(n|m) \ \approx \ \mbox{Gaussian}(n{-}m)
\ \ \ \ \ \ \mbox{for $t_{\tbox{erg}} \ll t \ll t_{\tbox{frc}}$}
\end{eqnarray}
For $t \gg t_{\tbox{frc}}$ the kernel $P_t(n|m)$ is no-longer 
a narrow Gaussian that is centered around $n=m$. 
Using (\ref{e_9}) it is easily observed that 
$t>t_{\tbox{frc}}$ is equivalent to 
$\dot{{\cal Q}}t > \sqrt{2D_{\tbox{E}}t}$, meaning that the 
systematic energy change becomes larger than the 
width of the spreading. Thus $t_{\tbox{frc}}$  should be 
regarded as the {\em breaktime} for the classical 
adiabatic approximation.

On time scales larger than $t_{\tbox{erg}}$
one may argue that the energy spreading is 
like a {\em stochastic process}, and consequently 
the diffusive behavior should persist beyond $t_{\tbox{frc}}$. 
A precise formulation of this point will be 
presented now. 
For $t\gg t_{\tbox{erg}}$ the evolving surface 
${\cal U}(t) |m(x(0))\rangle$  
becomes very convoluted due to `mixing'. 
As long as one does not insist on looking 
for fine structures, $\rho_t(Q,P)$ can be replaced 
by its smeared version. 
Any `tangential' non-homogeneity in phase space 
will be washed away due to the ergodic behavior, 
and therefore the smeared $\rho_t(Q,P)$ is
fully characterized by the projected 
distribution $\rho_t(n)$. Obviously this statement 
is true only if $t_{\tbox{erg}}$ is much smaller than 
the time scale $t_{\tbox{frc}}$ that characterizes the 
`transverse' spreading.    
Thus the dynamics acquires the following 
stochastic property:  
\begin{eqnarray} \label{e_51}
\hspace*{-2cm}   
P_{t_1{+}t_2}(n|m) \approx 
\int P_{t_2}(n|n') \ dn' \ P_{t_1}(n'|m)
\ \ \ \ \ 
\mbox{provided} \ \ \ 
t_1,t_2 \gg t_{\tbox{erg}} 
\end{eqnarray}
Assuming that $t_{\tbox{erg}} \ll t_{\tbox{frc}}$, 
it is possible to define an intermediate time 
$t_1=t/N$, where $N$ is some large integer, such that 
$t_{\tbox{erg}} \ll t_1 \ll t_{\tbox{frc}}$. 
Using the stochastic property (\ref{e_51}) one can write 
$P_t(n|m)$ as a convolution of the $N$ kernels 
$P_{t_1}(n|m)$. Applying the same considerations 
as in the derivation of the central limit theorem, 
we come to the conclusion that $P_t(n|m)$ will become 
a spreading Gaussian that obeys the diffusion 
equation 
\begin{eqnarray} \label{e_18a} 
\frac{\partial \rho}{\partial t} \ = \ 
\frac{\partial}{\partial n}
\left(D_n \frac{\partial}{\partial n} \rho \right)
\end{eqnarray}
which is equivalent to (\ref{e_18}). Note that   
$D_n=g(E)^2 D_{\tbox{E}}$. This description 
holds on time scales larger than $t_{\tbox{erg}}$, 
irrespective of the detailed structure of $P_{t_1}(n|m)$. 
Only the second moment of the latter is important for the 
determination of the diffusion coefficient.

The argumentation in favor of long-time stochastic 
behavior is more subtle in the QM case. 
Using obvious notations the stochastic 
assumption (\ref{e_51}) is:  
\begin{eqnarray} \label{e_53}
\hspace*{-2cm}   
\Big\langle \Big|  
\langle n | \mbf{U}_{t_2} \mbf{U}_{t_1} | m \rangle
\Big|^2 \Big\rangle  
\ \approx \  \sum_{n'}
\Big\langle \Big| \langle n | \mbf{U}_{t_2} | n' \rangle \Big|^2 \Big\rangle
\Big\langle \Big| \langle n' | \mbf{U}_{t_1} | m \rangle \Big|^2 \Big\rangle
\end{eqnarray}
Later we shall argue that $\mbf{U}$ is a banded matrix. 
It is true in general that in the absence of correlations 
between successive unitary operations we will always have  
a stochastic diffusive behavior \cite{qkr}. Thus, in order 
to establish a stochastic behavior in the QM 
case we should look for a time scale 
$\tau_c$ that marks the loss of phase-correlation.
For $t_1,t_2 \gg \tau_c$ we can argue that the off-diagonal 
'interference' terms in the matrix multiplication 
$\mbf{U}_{t_2}\mbf{U}_{t_1}$ will be averaged to zero.

One way to establish the existence of a 
time scale $\tau_c$ is just to assume {\em irregular} driving. 
As an example let us assume that we have a `piston' 
that is pushed back and forth in arbitrary a-periodic 
fashion, meaning that $\dot{x}(t)$ becomes uncorrelated on 
a time scale that will be denoted by $\tau_c^{\tbox{drv}}$. 
The irregular driving is like noise and consequently 
the interference contribution is averaged to zero \cite{qkr}. 
For {\em periodic} driving, there may be limitation of 
diffusion due to `localization' effect as in the 
quantum-kicked-rotator model \cite{qkr}. The study of this 
latter issue is beyond the scope of this paper. 
As in the classical case we will be able to 
establish a diffusive behavior on an {\em intermediate} 
time scale. QM considerations will be limited either by 
a semiclassical breaktime $t_{\tbox{scl}}$ or by a 
perturbative breaktime $t_{\tbox{prt}}$, which are 
analogous to the classical breaktime $t_{\tbox{frc}}$. 
If we do not have the separation of time scales 
($\tau_c^{\tbox{drv}} \ll t_{\tbox{prt}}$ 
or $\tau_c^{\tbox{drv}} \ll t_{\tbox{scl}}$) 
then we should wonder whether there is an {\em intrinsic}  
$\tau_c$. An intrinsic $\tau_c$ is expected to be 
either equal or larger than the classical time $t_{\tbox{erg}}$.  
Indeed, in the perturbative regime, 
where $t_{\tbox{erg}}\ll t_{\tbox{prt}}$, 
it will be argued that effectively $\tau_c \gg t_{\tbox{erg}}$. 
In the perturbative regime we are not able to give 
a general mathematical proof for having 
an effective $\tau_c$ such that the separation of 
time scales requirement ($\tau_c \ll t_{\tbox{prt}}$) is 
being satisfied. However, we are going to demonstrate that 
a crossover to a diffusive-growth of the second moment 
happens {\em before} the breaktime  $t_{\tbox{prt}}$. 
The {\em assumption} that the diffusive behavior 
{\em persists} beyond $t_{\tbox{prt}}$ with the {\em same} 
diffusion coefficient is the cornerstone of the 
common FGR picture.  We are not going to study in this paper 
the general conditions for having stochastic-like behavior.

\begin{figure}
\begin{center}
\leavevmode 
\epsfysize=2.5in 
\epsffile{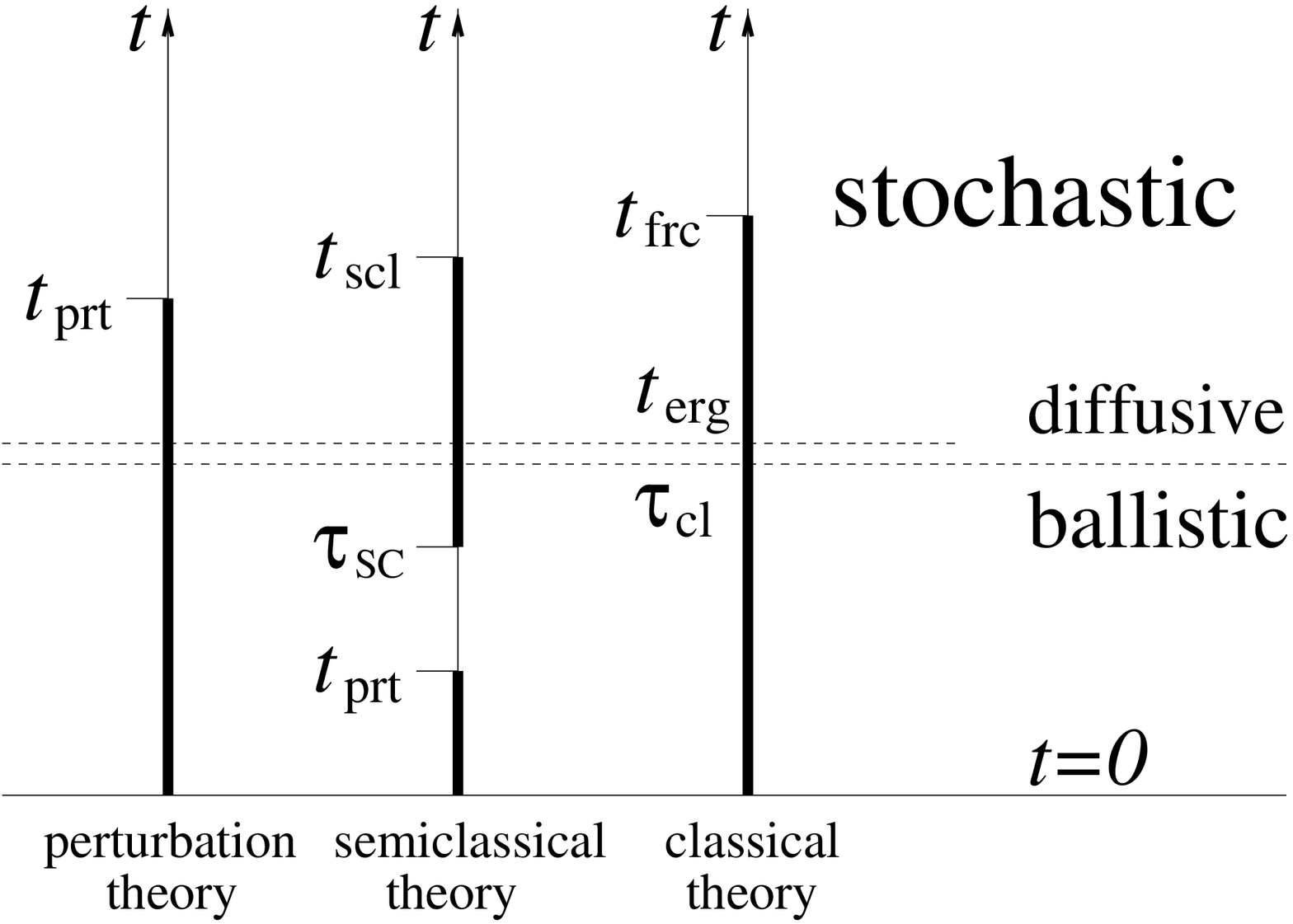} 
\vspace*{5mm}
\frame{\ \ \ \ \ \ \mpg{9}{ \ \\
$t_{\tbox{frc}} \ = \ \mbox{breakdown of classical perturbation theory}$ \\
$t_{\tbox{prt}} \ = \ \mbox{breakdown of quantal perturbation theory}$ \\ 
$t_{\tbox{scl}} \ = \ \mbox{breakdown of semiclassical approximation}$ \\ 
$\tau_{\tbox{cl}} < t_{\tbox{frc}}$
$ \ \ \ \ \ \ \leadsto \ \ \ \ \ \ $
\mbox{classical definition of slowness} \\
$\tau_{\tbox{cl}} < t_{\tbox{prt}}$
$ \ \ \ \ \ \ \leadsto \ \ \ \ \ \ $
\mbox{quantal definition of slowness} \\
$\tau_{\tbox{cl}} < t_{\tbox{scl}}$
$ \ \ \ \ \ \ \leadsto \ \ \ \ \ \ $
\mbox{not restrictive condition} \\
$\tau_{\tbox{SC}} < \tau_{\tbox{cl}}$
$ \ \ \ \ \ \ \leadsto \ \ \ \ \ \ $
\mbox{quantal definition of fastness}
\ \\ }}
\end{center}
\caption{\protect\footnotesize 
Illustration of the various time scales involved in 
constructing either classical, semiclassical or 
perturbation theory of dissipation. 
We use the notation $\tau_{\tbox{SC}}=\delta x_{\tbox{SC}}/V$.  
The accompanying table summarizes the associated requirements for the 
applicability of each of those theories. Note that 
the quantum mechanical definitions of slowness and 
of fastness are not complementary. Note also that 
slowness in the classical sense is always assumed. 
The rest of this paper (Sections 9-20) is devoted to 
the study of the ballistic-diffusive crossover. 
This crossover is `captured' either by perturbation 
theory or by the semiclassical theory, provided the 
respective slowness or fastness conditions are 
satisfied.  For simplicity we assume a generic system where 
$\tau_{\tbox{cl}}$ can be identified with $t_{\tbox{erg}}$. 
The persistence of the stochastic behavior for 
arbitrarily long times is assumed. We are not going to study 
in this paper the general conditions for having 
such long-time stochastic behavior.} 
\label{f_route}
\end{figure}

The derivation of the classical ${\cal F\!D}$ relation consists of two 
steps: The {\em first step} establishes the local 
diffusive behavior for short ($t\ll t_{\tbox{frc}}$) 
time scales, and $D_{\tbox{E}}$ is determined;
The {\em second step} establishes the global stochastic 
behavior on large ($t\gg t_{\tbox{erg}}$) time scales. 
The various time scales involved are illustrated 
in Fig.\ref{f_route}. The validity of the classical 
derivation depends on the slowness condition 
(\ref{e_10}). The validity of the analogous QM 
theory is further restricted by the quantal slowness condition 
(\ref{e_24}).  However, an optional derivation 
of the ${\cal F\!D}$ relation in the QM case 
can be based on semiclassical considerations. 
The limitations of the latter strategy are 
illustrated in Fig.\ref{f_route}, and further discussed 
in the next section.

\section{The semiclassical picture and detailed QCC}

The main objects of our discussion are the 
transition probability kernel $P_t(n|m)$ and 
the parametric kernel $P(n|m)$ which have been 
introduced in the previous section.
Recall that we are measuring phase-space volume 
($n{=}\Omega(E)$) in units of $(2\pi\hbar)^d$.  
This way we can obtain a `classical approximation' for 
the QM kernel, simply by making $n$ and $m$ integer 
variables. If the `classical approximation' is  
similar to the QM kernel, then we say that there is 
{\em detailed} QCC. If only the second-moment is 
similar, then we say that there is {\em restricted} QCC. 
In the present section we are going to discuss the 
conditions for having detailed QCC, using simple  
{\em semiclassical} considerations.  
In the next paragraph we discuss the conditions for 
having detailed QCC in the computation 
of the parametric kernel $P(n|m)$. 
Then we discuss the further restrictions on detailed QCC, 
that are associated with the computation of the 
actual kernel $P_t(n|m)$.

Wigner function $\rho_{n,x}(Q,P)$, unlike its classical 
microcanonical analog, has a non-trivial 
transverse structure. For a curved energy surface 
the transverse profile looks like Airy function and 
it is characterized by a width \cite{airy}
\begin{eqnarray} \label{e_56}
\Delta_{\tbox{SC}} \ = \ 
\left( \varepsilon_{\tbox{cl}} 
\left( \frac{\hbar}{\tau_{\tbox{cl}}} 
\right)^2 \right)^{1/3} 
\end{eqnarray}
where $\varepsilon_{\tbox{cl}}$ is a classical 
energy scale. For the `piston' example  
$\varepsilon_{\tbox{cl}}=E$ is the kinetic 
energy of the gas particle. 
The classical $P(n|m)$ has a dispersion 
\begin{eqnarray} \label{e_57_0}
\delta E_{\tbox{cl}} \ =  \  
\sqrt{\left\langle
\left(\frac{\partial {\cal H}}{\partial x}\right)^2
\right\rangle_{\tbox{E}}}
\ \delta x
\end{eqnarray}
which characterizes the transverse 
distance between the intersecting energy-surfaces 
$|m(x)\rangle$ and $|n(x{+}\delta x)\rangle$. 
In the generic case, it should be legitimate to 
neglect the transverse profile of Wigner function 
provided $\delta E_{\tbox{cl}} \gg \Delta_{\tbox{SC}}$. 
This condition can be cast into the form 
$\delta x \gg \delta x_{\tbox{SC}}$ where 
\begin{eqnarray} \label{e_57_1}
\delta x_{\tbox{SC}} \  =  \  
\frac{\Delta_{\tbox{SC}}}
{\sqrt{\nu_{\tbox{E}}/\tau_{\tbox{cl}}}}
\ \propto \ \hbar^{2/3}
\end{eqnarray}
For the `piston' example see \cite{wls}.  
Another important parametric scale 
is defined in a similar fashion: 
We shall see that it is {\em not} 
legitimate to ignore the transverse profile 
of Wigner function if 
$\delta E_{\tbox{cl}} < \Delta_b$.
This latter condition can be cast into the form 
$\delta x \ll \delta x_{\tbox{prt}}$ where  
\begin{eqnarray} \label{e_57_2}
\delta x_{\tbox{prt}} \  =  \  
\frac{\Delta_b}
{\sqrt{\nu_{\tbox{E}}/\tau_{\tbox{cl}}}}
\ = \ 
\frac{2\pi\hbar}
{\sqrt{\nu_{\tbox{E}} \tau_{\tbox{cl}}}} 
\end{eqnarray}
Typically the two parametric scales are well separated 
($\delta x_{\tbox{prt}} \ll \delta x_{\tbox{SC}}$). 
If we have $\delta x \ll \delta x_{\tbox{prt}}$ then the 
parametric kernel $P(n|m)$ is characterized by a 
perturbative core-tail structure 
which is illustrated in Fig.\ref{f_surfaces} and 
further discussed in the next section. 
If we have $\delta x \gg \delta x_{\tbox{SC}}$ 
then the transverse profile of Wigner function 
can be ignored, and we get detailed QCC. 
Obviously, `detailed QCC' does not mean 
complete similarity.  The classical kernel is 
typically characterized by various non-Gaussian 
features, such as sharp cutoffs, delta-singularities 
and cusps. These features are expected to be 
smeared in the QM case. 
The discussion of the latter issue is beyond 
the scope of the present paper \cite{wls}.

We turn now to discuss the actual transition 
probability kernel $P_t(n|m)$. Here we encounter 
a new restriction on QCC:  
The evolving surface ${\cal U}(t)|m\rangle$ becomes more 
and more convoluted as a function of time. 
This is because of the mixing behavior that  
characterizes chaotic dynamics. For $t\gg t_{\tbox{scl}}$ 
the intersections with a given instantaneous energy 
surface $|n\rangle$ become very dense, and associated 
QM features can no longer be ignored. 
The time scale $t_{\tbox{scl}}$ can be related to the failure 
of the stationary phase approximation \cite{heller}.

The breaktime scale $t_{\tbox{scl}}$ of the 
semiclassical theory is analogous to 
the breaktime scale $t_{\tbox{prt}}$ of perturbation 
theory, as well as to the breaktime scale 
$t_{\tbox{frc}}$ of the classical theory.
In order to establish the crossover from 
ballistic to diffusive energy spreading 
using a semiclassical theory we should satisfy 
the condition $\tau _{\tbox{cl}} < t_{\tbox{scl}}$. 
This velocity-independent condition is not very restrictive. 
On the other hand we should also satisfy 
the condition $\delta x \gg \delta x_{\tbox{SC}}$, 
with $\delta x = V\tau _{\tbox{cl}}$. 
The latter condition implies that the applicability 
of the semiclassical theory is restricted to relatively 
high velocities. We can define: 
\begin{eqnarray} \label{e_58}
v_{\tbox{SC}} \  =  \ 
\sqrt{ D_{\tbox{E}} \ \tau_{\tbox{cl}} } 
\ / \ \Delta_{\tbox{SC}}
\end{eqnarray}
If $v_{\tbox{SC}}\gg 1$ then the above semiclassical analysis 
is applicable in order to analyze the crossover from 
ballistic to diffusive energy spreading.

\section{The perturbative picture and restricted QCC}

Detailed QCC between the quantal $P(n|m)$ 
and the classical $P(n|m)$ is not guaranteed if 
$\delta x < \delta x_{\tbox{SC}}$. A-fortiori, this 
statement holds also for $P_t(n|m)$.  
For sufficiently small parametric changes $\delta x$, 
or for sufficiently short times $t$,  
perturbation theory becomes a useful tool for the 
analysis of these kernels. A detailed formulation of perturbation 
theory is postponed to later sections. 
In the present section we are going to sketch the main observations. 
We are going to argue that for small enough
$\delta x$ there is no {\em detailed} QCC between the 
quantal and the classical kernels, but there is still   
{\em restricted} QCC that pertains to the second moment 
of the distributions.  Large enough  $\delta x$ 
is a necessary condition for getting detailed QCC.
The following paragraph discuss the parametric evolution  
of $P(n|m)$, and the rest of this section discuss the 
actual evolution of $P_t(n|m)$.

For extremely small $\delta x$  the parametric kernel $P(n|m)$ 
has a standard `first-order' perturbative structure, namely: 
\begin{eqnarray} \label{e_59}
P(n|m) \ \approx \ \delta_{nm} 
\ + \ \mbox{Tail}(n{-}m)
\ \ \ \ \ \mbox{for $\delta x \ll \delta x_c^{\tbox{qm}}$} 
\end{eqnarray}
where $\delta x_c^{\tbox{qm}}$ is defined as parametric 
change that is needed in order to mix neighboring levels. 
For larger values of $\delta x$ neighbor levels 
are mixed non-perturbatively and consequently we have 
a more complicated spreading profile: 
\begin{eqnarray} \label{e_60}
P(n|m) \ \approx \ \mbox{Core}(n{-}m)
\ + \ \mbox{Tail}(n{-}m)
\ \ \ \ \ \mbox{for $\delta x \ll \delta x_{\tbox{prt}}$} 
\end{eqnarray}
In the perturbative case ($\delta x \ll \delta x_{\tbox{prt}}$) 
the second moment of $P(n|m)$ is generically determined by the `tail'.
It turns out that the QM expression for the 
second-moment is classical look-alike, and consequently 
{\em restricted} QCC is satisfied. 
The {\em core} of the quantal $P(n|m)$ is of non-perturbative 
nature. The {\em core} is the component that is expected to become 
similar (eventually) to the classical $P(n|m)$.  
A large perturbation $\delta x \gg \delta x_{\tbox{prt}}$ 
makes the {\em core} spill over the perturbative {\em tail}.  
If we have also $\delta x \gg \delta x_{\tbox{SC}}$, then we 
can rely on detailed QCC in order to estimate $P(n|m)$. 
The parametric scales $\delta x_c^{\tbox{qm}}$ and $\delta x_{\tbox{prt}}$ 
are easily estimated in case of the `piston' example. 
The displacement which is needed in order to mix levels is  
much smaller than De-Broglie wavelength, namely 
$\delta x_c^{\tbox{qm}}\approx
(\lambda_{\tbox{E}}^{d{+}1}/\mbox{\small Area})^{\tbox{1/2}}$.  
The displacement which is needed in order to mix core and tail is 
much larger than De-Broglie wavelength, namely  
$\delta x_{\tbox{prt}}\approx
(\tau_{\tbox{col}}/{\tau_{\tbox{cl}}})^{\tbox{1/2}}
\lambda_{\tbox{E}}$. For a more careful discussion of 
these parametric scales see \cite{wls,prm} and the 
concluding section.

\begin{figure}
\begin{center}
\leavevmode 
\epsfysize=3.0in 
\epsffile{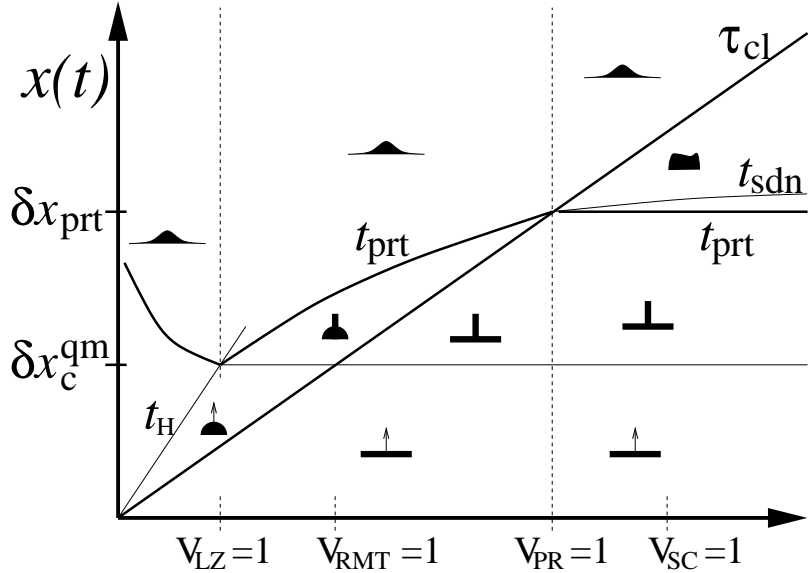}
\end{center}
\caption{\protect\footnotesize 
The various crossovers in the time evolution 
of $P_t(n|m)$. The vertical axis is $x(t)=Vt$.
The parametric scales $\delta x_c^{\tbox{qm}}$ 
and $\delta x_{\tbox{prt}}$ are indicted by 
horizontal lines. The horizontal  
axis is the velocity $V$. 
It is divided by vertical dashed lines 
to various velocity regimes. In each 
velocity regime there is a different dynamical 
route. The various crossovers are explained 
in the text and the various symbols are easily 
associated with having either Gaussian or 
some non-Gaussian spreading profile. In particular 
the perturbative spreading profile is either 
with or without non-trivial core, and its tail 
is either band-limited or resonance-limited. 
} 
\label{f_regimes}
\end{figure}

The dynamical evolution of $P_t(n|m)$ is related 
to the associated parametric evolution of $P(n|m)$.  
We can define a perturbative time scale $t_{\tbox{prt}}$ 
which is analogous to $\delta x_{\tbox{prt}}$.  
For $t\ll t_{\tbox{prt}}$ the kernel $P_t(n|m)$ is characterized 
by a core-tail structure that can be analyzed using perturbation theory.
In particular we can determine the second moment of 
the energy distribution, and we can establish {\em restricted} QCC. 
If the second moment for the core-tail structure is proportional 
to $t^2$, we shall say that there is a ballistic-like behavior. 
If it is proportional to $t$, we shall say that there is a 
diffusive-like behavior. {\em In both cases the actual 
energy distribution is not classical-like}, and therefore 
the term `ballistic' and `diffusive' should be used with care. 
We are going now to give a brief overview of the various 
scenarios in the time evolution of $P_t(n|m)$. These are 
illustrated in Fig.\ref{f_regimes}. 
In later sections we give a detailed account of the theory. 

For {\em slow velocities} such that 
$\tau_{\tbox{cl}}\ll t_{\tbox{prt}}$, 
there is a crossover from ballistic-like spreading 
to diffusive-like spreading at $t\sim \tau_{\tbox{cl}}$. 
In spite of the lack of detailed QCC there is still 
restricted QCC as far as this ballistic-diffusive 
crossover is concerned. 
If the breakdown of perturbation theory happens before 
the Heisenberg time ($t_{\tbox{prt}} \ll t_{\tbox{H}}$)  
it is implied that there is a second crossover 
at $t \sim t_{\tbox{prt}}$ from a diffusive-like spreading 
to a genuine diffusive behavior. Once a stochastic 
behavior is established, the time scale $t_{\tbox{H}}$ 
for recurrences becomes non-effective, and we expect 
a long-time classical-like behavior.

{\em extremely slow velocities} are defined by the 
the inequality $t_{\tbox{H}} \ll t_{\tbox{prt}}$. This 
inequality implies that there are QM recurrences {\em before} 
the expected crossover from diffusive-like spreading  
to genuine-diffusion. This is the QM adiabatic regime. In the 
$t\rightarrow\infty$ limit Landau-Zener transitions 
will dominate the energy spreading, and consequently 
neither detailed nor restricted QCC is a-priori expected \cite{wilk1}.

For {\em fast velocities} we have  
$t_{\tbox{prt}} \ll \tau_{\tbox{cl}}$. 
There is a crossover at $t\sim t_{\tbox{prt}}$ from 
ballistic-like spreading to a genuine ballistic behavior, 
and at $t \sim \tau_{\tbox{cl}}$ there is a second  
crossover from genuine-ballistic to 
genuine-diffusive spreading. The description of  
this classical-type crossover is out-of-reach for 
perturbation theory, but we can use the semiclassical 
picture instead.  Note that 
the semiclassical definition of `fastness' and the 
perturbative definition of `slowness' imply 
that there is a `gap' between the corresponding regimes. 
However, the interpolation is smooth, and therefore 
for simple systems surprises are not expected.

\newpage
\section{Actual versus Parametric Evolution}

The QM time evolution is governed by the
time dependent Schroedinger equation with the time dependent 
Hamiltonian ${\cal H}(x(t))$.  In practice it is quite 
unnatural to use a fixed basis. For example, in case of the 
`piston' example one may propose to use the fixed-basis 
that consists of the eigenfunctions of the empty cavity. 
However, the matrix elements of the `piston' may be very large and 
even infinite if we assume impenetrable walls. Thus, it is 
much more natural to use the so called adiabatic basis,  
though the time evolution is not necessarily of adiabatic 
nature. The evolving state-vector is expanded as follows:  
\begin{eqnarray} \label{e_64} 
|\psi(t)\rangle \ = \ \sum_{n} a_n(t) \ |n(x(t))\rangle 
\end{eqnarray}
Using standard manipulation we obtain the
Schroedinger-like equation  
\begin{eqnarray} \label{e_65} 
\frac{da_n}{dt} \ = \ 
-\frac{i}{\hbar}E_n \ a_n 
-\frac{i}{\hbar}\sum_m \mbf{W}_{nm}(x(t)) \ a_m
\end{eqnarray}
Were the off diagonal elements of ${\mathbf W}_{nm}$ are
\begin{eqnarray} \label{e_66} 
\mbf{W}_{nm} \ = \ 
\frac{\hbar}{i} 
\Big\langle n \Big| \frac{d}{dt} m \Big\rangle \ = \ 
i\frac{\hbar\dot{x}}{E_n{-}E_m} 
\Big\langle n \Big| \frac{\partial{\cal H}}{\partial x} \Big| m \Big\rangle
\end{eqnarray}
and we use the `gauge' convention 
${\mathbf W}_{nm}{=}0$ for $n{=}m$. 
(Only one parameter is being changed and 
therefore Berry's phase is not an issue).

Equation (\ref{e_65}) will be now our starting point. 
It is defined in terms of $\mbf{W}_{nm}(x)$ and in terms 
of a set of numbers $\{E_n\}$. Note that as long 
as $t \ll t_{\tbox{H}}$ we can {\em ignore} the dependence of 
$E_n$ on the changing parameter $x$. 
The formal solution of (\ref{e_65}) 
will be written as follows:
\begin{eqnarray} \label{e_67} 
a_n(t) \ = \ \sum_{m} \mbf{U}_{nm}(t) \ a_m(0)   
\end{eqnarray}
If all the $E_n$ are set equal to the same constant, 
(or without loss of generality to zero), then (\ref{e_65}) 
describes the time evolution of a frozen wavefunction.
In other words, (\ref{e_65}) without the $\{E_n\}$ is equivalent 
to the trivial equation $d\psi/dt=0$.  In this special 
case the formal solution (\ref{e_67}) will be 
written with $\mbf{T}_{nm}(x(t))$ instead of $\mbf{U}_{nm}(t)$. 
Note that if the $\{E_n\}$ are taken away from (\ref{e_65}), 
then $\dot{x}$ can be scaled out and therefore the 
$x$ dependence rather than the $t$ dependence become significant. 
The transition probability kernel and the 
parametric kernel can be written as:
\begin{eqnarray} \label{e_68} 
P_t(n|m) \ = \ |{\mathbf U}_{nm}(t)|^2
\ = \ \left|\langle n(x(t))| {\mathbf U}(t) | m(x(0)) \rangle \right|^2 
\\ \label{e_69} 
P(n|m) \ = \ |{\mathbf T}_{nm}(x)|^2
\ = \ \left|\langle n(x)| m(x(0)) \rangle \right|^2
\end{eqnarray} 
From now on we shall refer to the $t$-dependent evolution 
which is represented by $\mbf{U}_{nm}(t)$ as 
the {\em actual evolution} (AE). To the $t$-dependent 
evolution which is represented by $\mbf{T}_{nm}(x(t))$
will shall refer as {\em parametric evolution} (PE). 
For PE the velocity $\dot{x}=V$ plays no role, 
and it can be scaled out from the above 
equation. Consequently, for PE, parametric scales and 
temporal scales are trivially related via the scaling 
transformation  $\delta x = V \tau$.

It is important to realize that in a certain sense, 
defined below, the AE coincided with the PE for short times 
$t \ll t_{\tbox{sdn}}$. This is the QM-{\em sudden} approximation. 
The detailed picture is as follows: We start with some 
initial state $|m\rangle$.  After time $t$ there will be 
some non-vanishing probability to find the system 
in a certain energy range $\delta E(t)$ around $E_m$. 
As long as $\delta E(t) \ll \hbar/t$ the corresponding 
energy levels $E_n$ within $\delta E$ are not 
resolved. The latter condition defines a time interval 
$t \ll t_{\tbox{sdn}}$.  By definition, for $t \ll t_{\tbox{sdn}}$
it is as if the energy-levels were degenerated.    
Therefore we can say that the AE coincides with the PE, 
implying that the evolving state (\ref{e_64}) remains 
approximately unchanged.  The QM sudden approximation will 
be further discussed in Sec.\ref{s_sdn}.

\section{Application of perturbation theory}

We can use Equation (\ref{e_65}) as a starting point for 
a standard first-order perturbation theory (FOPT). 
For short times, such that $P_t(m|m) \sim 1$, 
the transition probability from level $m$ to level $n$ 
is determined by the coupling strength $|\mbf{W}_{nm}|^2$, 
by the energy difference $(E_n{-}E_m)$ and 
by the correlation function $F(\tau)$. 
The latter describes loss of correlation between 
$\mbf{W}_{nm}(x(0))$ and $\mbf{W}_{nm}(x(t))$. 
It is defined via 
\begin{eqnarray} \label{e_xcorr}
\left\langle
\ \mbf{W}_{nm}^{\star}(t{+}\tau)  
\ \mbf{W}_{nm}(t) \ 
\right\rangle \ = \ 
\ |W_{nm}|^2 \ F(\tau) 
\end{eqnarray} 
with the convention $F(0)=1$. Using FOPT one obtains the following result: 
\begin{eqnarray} \nonumber
P_t(n|m) \ &=& \ \left|\int_{0}^{t} 
\frac{\mbf{W}_{nm}(t')}{\hbar} 
\ \mbox{e}^{i\frac{E_n{-}E_m}{\hbar}t'} 
\ dt' \right|^2 
\\ \nonumber
\ &=& \ 
\left( \frac{W_{nm}}{\hbar} \right)^2
\int_0^t\int_0^t dt_2 dt_1 \ F(t_2{-}t_1) 
\ \mbox{e}^{i\frac{E_n{-}E_m}{\hbar}(t_2{-}t_1)} 
\\ 
\ &=& \ \label{e_Pnm}
t\tilde{F}_t\left(\frac{E_n{-}E_m}{\hbar}\right) \times 
\left( \frac{W_{nm}}{\hbar} \right)^2  
\ \ \ \ \ \ \mbox{for $n\ne m$} 
\end{eqnarray} 
The function $\tilde{F}_t(\omega)$ describes the 
spectral content of the perturbation. For a {\em constant}  
perturbation ($F(\tau)=1$) it is just given 
by equation (\ref{e_26}). For a {\em noisy} perturbation 
$F(\tau)$ is characterized by some finite correlation-time 
$\tau_c$, and therefore the definition of $\tilde{F}_t(\omega)$
is modified as follows:
\begin{eqnarray} 
\tilde{F}_t(\omega) \ = \ \left\{ \matrix{
t{\cdot}(\mbox{sinc}(\omega t /2))^2 &
\mbox{for} & t<\tau_c  \cr
\tilde{F}(\omega) &
\mbox{for} & t>\tau_c} \right.
\end{eqnarray} 
where $\tilde{F}(\omega)$ is the Fourier transform of 
the correlation function $F(\tau)$.  Now it is a simple 
matter to calculate the {\em second-moment} of the spreading: 
\begin{eqnarray} \label{e_43a}
\hspace*{-1cm}
\delta E^2 \ = \  
\sum_n (E_n{-}E_m)^2 \ P_t(n|m) \ = \ 
V^2 t \int_{-\infty}^{+\infty}\frac{d\omega}{2\pi}
\tilde{C}_{\tbox{E}}(\omega) \ \tilde{F}_t(\omega)
\end{eqnarray} 
This result coincides with the linear-response result (\ref{e_25})
only if the coupling matrix-elements could have been treated 
as {\em constant} in time, meaning $F(\tau){=}1$ 
and accordingly $\tau_c=\infty$.  For $t>\tau_c$ it becomes 
{\em formally equivalent} to the FGR result (\ref{e_32}).  
However, a {\em practical equivalence} seems unlikely 
because $F(\tau)$ of (\ref{e_xcorr}) is not necessarily 
determined by the correlations of the external driving source.  
The critical discussion of this point is going to be the main 
issue of the subsequent sections.

In order to use (\ref{e_43a}) we should determine how 
$F(\tau)$ look like, and in particular we should determine 
what is the correlation-time $\tau_c$. We postpone 
this discussion, and assume that $F(\tau)$ and hence 
$\tau_c$ are known from some calculation. The total 
transition probability is $p(t)=\sum_n'P(n|m)$, where
the prime indicates omission of the term $n=m$. 
FOPT is valid as long as $p(t)\ll 1$. 
This defines a breaktime $t_{\tbox{prt}}'$  
for the {\em standard} FOPT treatment.  
The above derivation imply that we can trust (\ref{e_43a}) 
only during the short time $t\ll t_{\tbox{prt}}'$. 
However, later we shall argue that with a proper 
(modified) definition of $F(\tau)$ we can trust (\ref{e_43a})
during a longer time $t\ll t_{\tbox{prt}}$. 
The breaktime $t_{\tbox{prt}}$ will be determined 
by using an {\em improved} perturbation theory (IMPT).

It is now possible to formulate the conditions for having 
{\em restricted} QCC. By `restricted' QCC we mean that only the 
{\em second-moment} of the spreading (\ref{e_43a}) 
is being considered.  It is essential to distinguish between 
two different possible scenarios:
\begin{eqnarray}
\mbox{Resonance-limited transitions:} \ \ \ \ \ &
\tau_c \gg \tau_{\tbox{cl}} \\
\mbox{Band-limited transitions:} \ \ \ \ \ &
\tau_c \ll \tau_{\tbox{cl}} 
\end{eqnarray}
For resonance-limited transitions,  
finite $\tau_c$ has no consequence as far as 
$\delta E^2$ is concerned. The crossover 
to diffusive behavior $\delta E^2 \propto t$ 
will happen at $t\sim\tau_{\tbox{cl}}$. This 
diffusive behavior will persist for $t>\tau_c$ with 
the same diffusion coefficient. On the other hand, for 
band-limited transitions we will have at 
$t\sim\tau_c$  a pre-mature crossover from 
ballistic to diffusive behavior. 
Consequently the classical result will be  
suppressed by a factor $(\tau_c/\tau_{\tbox{cl}}) \ll 1$. 
This is due to the fact that the transitions 
between levels are limited not by 
the resonance width (embodied~by~$\tilde{F}(\omega)$), 
but rather by the band-width of the coupling matrix elements 
(embodied~by~$\tilde{C}_{\tbox{E}}(\omega)$).

We have realized that the perturbative result (\ref{e_43a}) 
can be used in order to establish a {\em diffusive} growth of 
the {\em second moment}. Obviously the applicability of 
this picture requires a separation of time scales:   
\begin{eqnarray} \label{e_FGRc}
\mbox{\bf FGR-condition:} \ \ \ \  
\mbox{Either} \ \tau_{\tbox{cl}} \ \mbox{or} \ \tau_c  
\ \ \ \ll \ \ \ t_{\tbox{prt}}   
\end{eqnarray} 
The long time {\em stochastic} behavior of the spreading is 
determined by the short-time dynamics, as explained in 
Section \ref{s8}.  The FGR condition guarantees that 
the {\em diffusive} growth of the {\em second-moment} 
is established {\em before} the breakdown of the short-time analysis. 
Therefore, the correct determination of the breaktime $t_{\tbox{prt}}$ 
is extremely important, and it is going to be the main issue of the 
subsequent sections.

\section{The applicability regime of the standard FOPT treatment}

In order to have practical estimates for applicability regime 
of the standard FOPT treatment, we should look on the 
matrix ${\mathbf W}_{nm}$.  
This matrix is banded, and its elements satisfy:
\begin{eqnarray}
\left\langle 
\left|\frac{{\mathbf W}_{nm}}{\hbar}\right|^2
\right\rangle 
\ \approx \ 
\left(\frac{V}{\delta x_c^{\tbox{qm}}}\right)^2 
\ \frac{1}{(n{-}m)^2} 
\ \ \ \ \ \mbox{for} \ \ |n{-}m|<b/2 
\end{eqnarray}
where $\delta x_c^{\tbox{qm}}=\Delta/\sigma$.  
From the above expression, once used in (\ref{e_Pnm})
for the calculation of the kernel $P(n|m)$, 
it follows that  $\delta x_c^{\tbox{qm}}$ 
is the parametric change which is required in order to mix neighboring 
levels. Similarly, in the calculation of $P_t(n|m)$ the related 
$\tau_c^{\tbox{qm}}=\delta x_c/V$ is the time which is  
required in order to mix neighboring levels. Given two 
distant levels $n$ and $m$, and taking the mixing on ``small'' 
scale into account, one realizes that $\delta x_c^{\tbox{qm}}$ 
also determines the correlation time $\tau_c=\tau_c^{\tbox{qm}}$ 
of the matrix-element ${\mathbf W}_{nm}(x(t))$, as defined 
in (\ref{e_xcorr}). These observations can be summarized as follows: 
\begin{eqnarray} \label{e_simple}
t_{\tbox{prt}}' \ = \ \tau_c \ = \ \tau_c^{\tbox{qm}} 
\ \ \ \ \ \mbox{for the standard FOPT.}  
\end{eqnarray}
The standard perturbative structure (\ref{e_59}) of either $P(n|m)$ 
or $P_t(n|m)$ is maintained as long as neighboring levels are not 
being mixed. This structure obviously does not correspond to 
the classical structure since it is characterized by 
the non-classical energy scale $\Delta_b$. Still, there is 
{\em restricted} QCC which is implied by (\ref{e_43a}).

A sufficient condition for the applicability of 
the {\em standard} FOPT treatment is $v_{\tbox{RMT}} \ll 1$.
The argument goes as follows: By definition $v_{\tbox{RMT}} \ll 1$ 
implies $\tau_{\tbox{cl}} \ll \tau_c^{\tbox{qm}}$. 
Using (\ref{e_simple}) we observe that it is equivalent  
to $\tau_{\tbox{cl}} \ll t_{\tbox{prt}}'$. By definition  
$t_{\tbox{prt}}$ is either equal or larger than 
$t_{\tbox{prt}}'$. Therefore the FGR condition (\ref{e_FGRc}) 
is satisfied. The converse however is not true.  
Having $v_{\tbox{RMT}} \gg 1$ does not imply that the 
FGR condition cannot be satisfied. Therefore, for 
$v_{\tbox{RMT}} \gg 1$,  we cannot tell on the basis 
of the {\em standard} FOPT treatment whether or when 
there is a crossover to a diffusive behavior. We shall 
try to overcome this difficulty in the next sections.

\section{The over-simplified RMT (ORMT) picture}
\label{s_RMT}

Recall that the matrix $\mbf{W}_{nm}$  is a banded.   
The bandwidth $b{=}1,3...$ corresponds to diagonal, 
tridiagonal matrix and so on.  
In the spirit of RMT we can think of $\mbf{W}_{nm}$ as 
a particular realization which is taken out from some 
large ensemble of (banded) random matrices.  
In order to go beyond standard FOPT it is essential 
to further specify the {\em cross-correlation between matrix elements}. 
Following \cite{wilk2} the simplest statistical assumption is 
{\em absence} of cross-correlations, namely, 
\begin{eqnarray} \label{e_crosscorr} 
\left\langle
\ \mbf{W}_{n'm'}^{\star}(t{+}\tau)  
\ \mbf{W}_{nm}(t) \ 
\right\rangle \ = \ 0 
\ \ \ \ \ \ \mbox{if} \ \ \ 
\{n',m'\} \ne \{n,m\}
\end{eqnarray} 
Equation (\ref{e_65}) with the statistical assumptions 
(\ref{e_crosscorr}) and (\ref{e_xcorr}), where $\tau_c=\tau_c^{\tbox{qm}}$, 
is a well defined RMT model. We shall refer to it as 
the over-simplified RMT (ORMT) picture. The main observation 
of \cite{wilk2} can be summarized as follows: 
The FOPT result (\ref{e_43a}), assuming an ORMT picture, 
can be trusted for classically long times. Namely, 
\begin{eqnarray} \label{e_rmtpic}
t_{\tbox{prt}}' = t_{\tbox{frc}}  
\ \ \ \ \ \ \mbox{and} \ \ \ \ \ \ 
\tau_c \ = \ \tau_c^{\tbox{qm}} 
\ \ \ \ \ \mbox{for the ORMT picture.}  
\end{eqnarray}
The ORMT picture reduces to FOPT and implies restricted QCC 
provided $v_{\tbox{RMT}}\ll 1$. However, this condition 
is {\em not} satisfied in the classical limit 
($\hbar\rightarrow 0$).  In the classical limit  
$v_{\tbox{RMT}}\gg 1$, and consequently transitions 
are band-limited.  In particular it follows from (\ref{e_43a}), 
that the classical diffusion  ($D_{\tbox{E}}^{\tbox{cl}}$) 
is suppressed by a factor $\tau_c/\tau_{\tbox{cl}}$, leading to  
\begin{eqnarray} \label{e_79} 
D_{\tbox{E}}^{\tbox{ORMT}} \ \approx \ 
\frac{1}{v_{\tbox{RMT}}} D_{\tbox{E}}^{\tbox{cl}} 
\hspace*{2cm} \mbox{for \ $v_{\tbox{RMT}} \gg 1$}
\end{eqnarray}
This result, which is the main result of \cite{wilk2}, 
is obviously inconceivable, because 
it is implied that the classical limit does not 
coincide with the classical result, and that 
the correspondence principle is actually violated.

The ORMT picture predicts (for $v_{\tbox{RMT}}\gg 1$) 
a premature crossover from ballistic to diffusive 
behavior once  $\delta x$ becomes larger than $\delta x_c^{\tbox{qm}}$. 
It is important to realize that (\ref{e_79}), if it were true,  
would reflects a property of PE.  This statement becomes 
more transparent if we write: 
\begin{eqnarray} \label{e_79a} 
\delta E^2\Big|_{\tbox{ORMT}} 
\ \ = \ \  2D_{\tbox{E}}^{\tbox{ORMT}} \times t
\ \ \approx \ \ 
\left\langle
\left(\frac{\partial {\cal H}}{\partial x}\right)^2
\right\rangle_{\tbox{E}}
\delta x_c^{\tbox{qm}} \times \delta x
\end{eqnarray}
Exactly the same result would be obtained if we 
start with (\ref{e_65}) without the the first term in the right hand side.  
The value of $V$ has no significance in the above analysis.

\section{The core-tail structure}

There is no detailed QCC between the classical 
and the quantal $P_t(n|m)$ for short times.
We are going to explain that for a limited time 
$P_t(n|m)$ consists of a {\em core} whose width 
will be denoted by $b(t)$, and a {\em tail} 
whose main component is contained within 
the bandwidth $b$. In order to analyze 
this core-tail structure we are going to use   
an improved perturbation theory (IMPT).  
The IMPT treatment assumes that out-of-band transitions 
($b/2<|n{-}m|$) can be neglected. 
The IMPT is useful as long as the {\em second-moment} 
of the energy distribution is predominated by the 
tail component. This determines the breaktime $t_{\tbox{prt}}$.  
The breakdown of IMPT at $t \sim t_{\tbox{prt}}$ happens 
once the core spills over the tail region, and the 
FOPT-like structure of $P_t(n|m)$ is completely washed away.

It is important to have a proper intuitive understanding 
of how the {\em core} is being formed. 
For \mbox{$t \ll \tau_c^{\tbox{qm}}$} we have 
$P_t(m|m)\sim 1$ and $P_t(n|m) \ll 1$ for $n \ne m$. 
It means that the core width is $b(t)=1$.  
For \mbox{$t \sim \tau_c^{\tbox{qm}}$} few levels 
are expected to be mixed by the perturbation, 
meaning that $b(t)>1$. We may have the tendency to 
associate this mixing with an avoided crossing. However, 
this aspect should not be over-emphasized. The 
mixing of neighboring levels is {\em not} conditioned by having 
exceptionally small energy difference 
(having $(E_{m{+}1}{-}E_m)\ll\Delta$ is not required). 
Moreover, one should not over-emphasize the importance 
of near-neighbor transitions, unless $v_{\tbox{LZ}}\ll 1$. 
(See further discussion of the QM-adiabatic regime 
in Sec.\ref{s_LZ}). If near-neighbor transitions were 
the dominant mechanism for energy spreading, it would 
imply that $b(t) \approx (t/\tau_c^ {\tbox{qm}})$. Rather, 
we shall argue that the core develops much more 
rapidly, namely  $b(t) \approx (t/\tau_c^ {\tbox{qm}})^2$.

It is also important to have a proper intuitive understanding 
of how the {\em tail} is being formed. 
Let us assume for simplicity that only three levels 
$m=1$, $m=2$ and $n=100$ are actually coupled by matrix elements. 
Let us start at $t=0$ with all the probability concentrated in $m=1$.  
As we go at $t=t_1$ via an avoided-crossing of $m=1$ and $m=2$, 
the matrix element $\mbf{W}_{\! n,\tbox{1}}$ may change sign. 
This change of sign may be viewed as a loss of correlations. 
However, at the same time, assuming a diabatic transition, almost 
all the probability is transfered to $m=2$ in such a way that 
$\mbf{W}_{\! n,\tbox{2}}a_{\tbox{2}}$ at $t>t_1$ 
is strongly correlated with 
$\mbf{W}_{\! n,\tbox{1}}a_{\tbox{1}}$ at $t<t_1$.  
It is implied that the effective 
correlation time $\tau_c$ for $m\rightarrow n$ transitions 
is larger than the time between avoided crossings. 
This is not captured by the statistical assumption (\ref{e_crosscorr}) 
of the previous section.  A proper transformation, that effectively 
removes the in-core transitions between $m$-states, 
can be used for the purpose of tail-formation analysis.  
The associated perturbative treatment is characterized by 
an effective $\tau_c \gg \tau_c^{\tbox{qm}}$ 
as well as by $t_{\tbox{prt}}' \gg \tau_c^{\tbox{qm}}$.

\begin{figure}
\begin{center}
\leavevmode  
\epsfysize=1.5in 
\epsffile{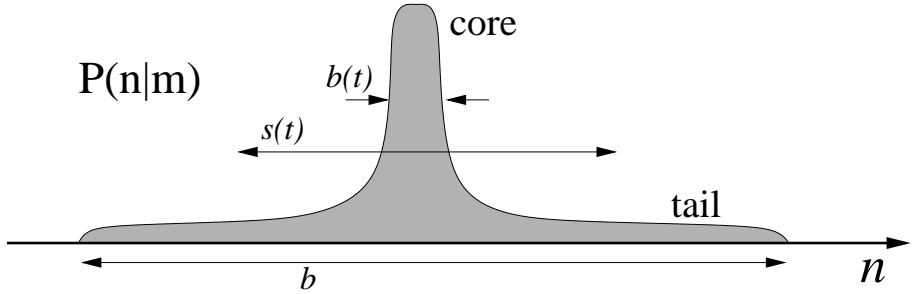}
\end{center}
\caption{\protect\footnotesize 
Schematic illustration of a generic core-tail spreading profile.
The core-width $b(t)$ is defined by the participation-ratio. 
The second-moment should satisfy $b(t) \ll s(t) \ll b$, 
where $b$ is that bandwidth. In case of $P_t(n|m)$ the 
tail becomes  (for $t>\tau_{\tbox{cl}}$) resonance-limited 
rather than band-limited.  In the resonance-limited case 
the bandwidth $b$ in the above figure should be replaced 
by $(\hbar/t)/\Delta$, and accordingly the requirement is 
$b(t) \ll s(t) \ll (\hbar/t)/\Delta$. }
\label{f_coretail}
\end{figure}

For $t>\tau_c^{\tbox{qm}}$ neighboring levels are 
being mixed and consequently the transition kernel 
acquires a non-trivial core-tail structure which is illustrated 
in Fig.\ref{f_coretail}. The expression for $P_t(n|m)$ can 
be written schematically as follows:
\begin{eqnarray} \label{e_80} 
P_t(n|m) \ \approx \ \mbox{Core}(n{-}m) 
\ + \ \mbox{Tail}(n{-}m) 
\ \ \ \ \ \mbox{for $t \ll  t_{\tbox{prt}}$}
\end{eqnarray}
The kernel is characterized now by two scales:
\begin{eqnarray} \label{e_81} 
b(t) \ = \ \mbox{\small core width} \ &= \ 
\left( \sum_n (P_t(n|m))^2 \right)^{-1}  
\\  \label{e_82} 
s(t) \ = \ \mbox{\small spreading} \ &= \
\left( \sum_n (n{-}m)^2 \ P_t(n|m) \right)^{-1/2}
\end{eqnarray}
such that $b(t) \ll s(t) \ll b$. 
For $t<\tau_c^{\tbox{qm}}$ we have a trivial 
core with $b(t) \approx 1$, whereas for $t \gg \tau_c^{\tbox{qm}}$
we have a non-trivial core with $b(t) \gg 1$.  
The matrix elements satisfy 
$\langle |{\mathbf W}_{nm}|^2 \rangle 
\ \propto \ 1/(n{-}m)^2$. 
We shall see that in the `band-limited tail' case 
we have $P_t(n|m) \sim \mbox{const}/(n{-}m)^2$  
up to the cutoff $b$, while for the `resonance-limited tail'
case we have $P_t(n|m) \sim \mbox{const}/(n{-}m)^2$ 
up to the cutoff $(\hbar/t)/\Delta$.  One should realize  
that the power-law behavior of the tail is `fast' enough
in order to guarantee that $b(t)$ is independent of the  
tail's cutoff.  The cutoff does not have any effect on 
the evolving core. On the other hand, the second moment $s(t)$, 
unlike $b(t)$, is predominantly determined by the tail's cutoff, 
and it is independent of the core structure.  

\section{An improved perturbation theory (IMPT)}

\label{s16}
 
In order to extend perturbation theory beyond 
$\tau_c^{\tbox{qm}}$ it is essential to eliminate 
the non-perturbative transitions within the core. 
This can be done by making a transformation to 
an appropriate basis as follows:
\begin{eqnarray} \label{e_83} 
a_n(t) \ = \ \sum_m \tilde{\mathbf{T}}_{nm} \ c_m(t) 
\\ \label{e_84} 
\tilde{\mathbf{T}}_{nm}  =  {\mathbf{T}}_{nm}
\ \ \ \mbox{if $|n{-}m|<b'/2$, else zero.} 
\end{eqnarray}
The amplitudes $c_n(t)$ satisfy the same Schroedinger 
equation as the $a_n(t)$, with a transformed matrix
$\tilde{\mathbf{W}}$. The general expression 
for $\tilde{\mathbf{W}}$ is 
\begin{eqnarray} \label{e_85} 
\tilde{\mbf{W}}=
\tilde{\mbf{T}}^{\dagger}\mbf{W}\tilde{\mbf{T}}
-i\hbar\tilde{\mbf{T}}^{\dagger}(d\mbf{T}/dt)+
\tilde{\mbf{T}}^{\dagger}\mbf{E}\tilde{\mbf{T}}
\end{eqnarray}
where $\mbf{E}$ is a diagonal matrix of the energies. 
This is a quite complicated expression. However, we are 
interested only in the core-to-tail transitions for which 
\begin{eqnarray} \label{e_86} 
(\tilde{\mathbf{W}})_{nm}=
(\tilde{\mathbf{T}}^{\dagger}\mathbf{W}\tilde{\mathbf{T}})_{nm}
\ \ \ \mbox{for $|n{-}m|>b'$}
\end{eqnarray}
(no approximation is involved). 
Once this transformation is performed the `new' Schroedinger 
equation will be characterized by a new correlation 
time $\tau_c$ and by a new perturbative time $t_{\tbox{prt}}'$. 
Both $\tau_c$ and $t_{\tbox{prt}}'$ will depend on the 
free parameter $b'$. Our choice of the 
course-graining parameter $b'$ is not completely arbitrary.  
The restrictions are: \\ 
\ \\
$\hspace*{2mm} \bullet \ \ $ 
Unitarity is approximately preserved: 
\ \ \ \ \ \ \ \ \ $b(t) \ \ll \ b'$. \\
$\hspace*{2mm} \bullet \ \ $ 
Core-to-Tail transitions are preserved:
\ \ \ \ \ \ \ \ \ $b' \ \ll \ b$. \\
$\hspace*{2mm} \bullet \ \ $ 
Long effective correlation time is attained:
\ \ \ \ $\tau_{\tbox{cl}} \ \ll \ \tau_c$ \\
\ \\
The feasibility of the last requirement deserves further 
discussion. Only matrix elements with $|n{-}m|\gg b'$ 
are of interest, and therefore in the multiplication 
$(\tilde{\mathbf{T}}^{\dagger})_{n'n}
(\mathbf{W})_{nm}(\tilde{\mathbf{T}})_{mm'}$ 
we can substitute (\ref{e_66}) with $(E_{n}{-}E_{m})$ 
replaced by $(E_{n'}{-}E_{m'})$. As $b'$ becomes 
closer to $b$, the matrix elements  
$(\tilde{\mathbf{W}})_{n'm'}$ become correlated 
on a time scale of the order $\tau_c^{\tbox{cl}}$.
This is because $b'= b$ implies  
transformation to an $x$-independent basis. 
The time scale $\tau_c^{\tbox{cl}}$ has been 
defined in Sec.\ref{s_flc}. 
As we change $b'$ from $b$ back to smaller values, 
we expect $\tau_c$ to become smaller. By continuity, we 
expect no difficulty in satisftying the conditions 
$b'\ll b$ and $\tau_c\gg\tau_{\tbox{cl}}$ simultaneously. 
The validity of the improved perturbative treatment   
is further discussed in the next section.

The usefullness of the above transformation stems from 
the fact that due to the elimination of non-perturbative 
transitions within the core, $t_{\tbox{prt}}'$  becomes much 
longer than $\tau_c^{\tbox{qm}}$. At the same time the 
information which is required in order to determine 
the second moment $s(t)$ is not lost. 
We have $|\tilde{\mathbf{W}}_{nm}| \approx |{\mathbf{W}}_{nm}|$
for core-to-tail transitions, and a practical approximation 
for the `renormalized' spreading profile is 
\begin{eqnarray} \label{e_87}
\hspace*{-2cm} 
P_t(n|m) \ \sim \ \delta_{nm} \ + \ 
t \tilde{F}_t\left( \frac{E_n{-}E_m}{\hbar}\right)
\times\left(\frac{1}{\tau_c^{\tbox{qm}}}\right)^2 
\ \frac{1}{(b')^2+(n{-}m)^2}   
\end{eqnarray}
Breakdown of the improved perturbative treatment 
happens once the total transition probability becomes 
non-negligible (of order 1). Thus   
\begin{eqnarray} \label{e_88}
t_{\tbox{prt}}' \ = \ (b')^{\tbox{1/2}} \times \tau_c^{\tbox{qm}} 
\end{eqnarray}
The behavior for $|n{-}m| \le b'$ is an artifact of 
the transformation and contains false information. 
However, for the calculation of the second moment 
only the tail is significant. The tail is not affected
by our transformation and therefore we will 
obtain the {\em same} result (\ref{e_43a}) for $\delta E^2$ with 
one important modification: a different effective 
value for $\tau_c$. Moreover, since $b'$ is 
chosen such that $\tau_c\gg\tau_{\tbox{cl}}$, it follows 
that the transitions are resonant-limited and consequently 
QCC is established also 
in the domain $v_{\tbox{RMT}}>1$. Obviously, at the same time we 
should satisfy the condition $\tau_{\tbox{cl}}\ll t_{\tbox{prt}}'$. 
It is easily verified that the latter condition cannot be 
satisfied if $v_{\tbox{PR}}>1$.  This is not just a technical 
limitation of the IMPT strategy, but reflects a real 
difference between two distinct routes towards QCC. 
This point is further illuminated in the next section.

\section{Consequences of the IMPT treatment}

\label{s17}

The IMPT is capable of giving information about the tail, 
and hence about the second moment. 
Given $t$, one wonders how much $b'$  can 
be `pushed down' without violating 
the validity conditions of our procedure. 
It is quite clear that $b'\gg b(t)$ is 
a necessary condition for {\em not} having a breakdown   
of perturbation theory. If we {\em assume} that the 
energy-spreading-profile is characterized just by 
the single parameter $b(t)$, then the condition 
$b'\gg b(t)$ should be equivalent to $t\ll t_{\tbox{prt}}'$.  
Hence the following estimate is suggested:  
\begin{eqnarray} \label{e_89}
b(t) \ = \ (t/\tau_c^{\tbox{qm}})^2
\end{eqnarray}

We turn now to determine the $\delta x_{\tbox{prt}}$ of the 
parametric evolution (PE), and then the $t_{\tbox{prt}}$ of 
the actual evolution (AE).
Recall that PE is obtained formally by 
ignoring the differences $(E_n{-}E_m)$, 
which implies that we can 
make in (\ref{e_87}) the replacement 
$\tilde{F}_t \mapsto t$.  Thus the 
tail of $P(n|m)$ is band-limited and 
consequently the second moment is 
\begin{eqnarray} \label{e_90}
s(t)^2 \ = \ b \times (1/\tau_c^{\tbox{qm}})^2 \ t^2
\hspace*{2cm} \mbox{[band-limited tail]}
\end{eqnarray}
in agreement with the classical ballistic result (\ref{e_15}). 
Our procedure for analyzing the core-tail structure 
of $P(n|m)$ is meaningful as long as 
we have $b(t) \ll s(t) \ll b$. This defines 
an upper time limitation $t_{\tbox{prt}}$ 
which is related via $\delta x = Vt$
to the following parametric scale: 
\begin{eqnarray} \label{e_91}
\delta x_{\tbox{prt}} \ = \
b^{\tbox{1/2}} \ \delta x_c^{\tbox{qm}} \ = \ 
\frac{\hbar}{\sqrt{\nu_{\tbox{E}}
\tau_{\tbox{cl}}}}
\end{eqnarray}
At $t=t_{\tbox{prt}}$ we have $b(t) \sim s(t) \sim b$, 
and we expect a crossover from a ballistic-like 
spreading to a genuine ballistic spreading.

In the perturbative regime the AE departs 
from the PE once the energy scale $\Delta_b$ is resolved. 
This happens when $t\sim\tau_{\tbox{cl}}$. 
The perturbative approach is applicable for the 
analysis of the crossover at $t\sim\tau_{\tbox{cl}}$ 
provided  $V\tau_{\tbox{cl}}\ll\delta x_{\tbox{prt}}$. 
This is precisely the condition $v_{\tbox{PR}}\ll 1$. 
For $t\gg\tau_{\tbox{cl}}$ the tail becomes resonance 
limited ($|n{-}m|<(\hbar/t)/\Delta$) 
rather than band limited  ($|n{-}m|<b$) and  we obtain:
\begin{eqnarray} \label{e_92}
s(t)^2 \ = \ 
(1/\tau_c^{\tbox{qm}})^2 \ t_{\tbox{H}} \ t 
\hspace*{2cm} \mbox{[resonance-limited tail]} 
\end{eqnarray}
in agreement with the classical diffusive result (\ref{e_16}). 
Our procedure for analyzing the core-tail structure 
of $P_t(n|m)$ is meaningful as long as 
we have $b(t) \ll s(t) \ll b$. This defines 
a {\em modified} upper time limitation 
\begin{eqnarray} \label{e_93}
t_{\tbox{prt}} \ = \ 
(\tau_c^{\tbox{qm}})^{\tbox{2/3}} \ t_{\tbox{H}}^{\tbox{1/3}} 
\ = \ 
\left( \frac{\hbar^2}
{\nu_{\tbox{E}}V^2}\right)^{1/3} 
\ \ \ \ \ \mbox{[applies to $v_{\tbox{PR}}\ll 1$]}
\end{eqnarray}
At $t=t_{\tbox{prt}}$ we have $b(t) \sim s(t) \ll b$, 
and we expect a crossover from a diffusive-like 
spreading to a genuine diffusive spreading.

\section{Validity of the IMPT picture}

It is important to have a clear understanding of the 
difference between the IMPT picture and the 
ORMT picture. In both cases we can argue that 
$P(n|m)$ has, for short times, a core-tail structure.  
The fundamental difference 
is the assumption concerning the effective $\tau_c$ 
for core-to-tail transitions. The ORMT picture 
assumes that effectively  $\tau_c=\tau_c^{\tbox{qm}}$, 
and equivalently the tails grow like $\delta x$. 
This is the reason for having the pre-mature crossover 
to diffusive growth (\ref{e_79}) of the second moment. 
The IMPT picture assumes that the effective $\tau_c$ is 
scale-dependent, and that the tails grow predominantly 
like $\delta x^2$. 
In other words, the tails grow as if we are still in 
the regime of FOPT. Therefore, from practical point of view, 
all we have to do in order to establish the validity of 
the IMPT picture is to verify that indeed the tails grow 
in a ballistic-like fashion ($\propto \delta x^2$), and not 
in diffusive-like fashion ($\propto \delta x$).  
A related observation is that also the core-width $b(t)$ 
grows as ($\propto \delta x^2$).   

The argumentation (Sec. \ref{s16}) in favor of the IMPT picture 
is not mathematically rigorous. It is therefore important to 
study specific examples. The obvious example to begin with 
has been defined by Wigner forty years ago \cite{wigner,casati,flamb}. 
Let ${\cal H} = \mbf{E} + x\mbf{B}$ where $\mbf{E}$ is a diagonal 
matrix and $\mbf{B}$ is a banded random matrix. 
The IMPT picture should apply to the 
analysis of the PE of this Wigner model. Indeed, it is well known 
that $P(n|m)$ for Wigner's model is a Lorentzian, and we may 
view this Lorentzian as a special case of core-tail structure. 
The width of the Wigner's Lorentzian, in energy units, is  
$\Gamma=2\pi ((x{\cdot}\sigma)/\Delta)^2\times\Delta$. Thus we 
have indeed $\propto \delta x^2$ for both the core and the 
tails. If the ORMT picture were true we would expect 
to get $\propto \delta x$ dependence.  All the results of 
the previous sections are consistent with the established 
results of Wigner. Having established the IMPT picture 
for PE, and observing that going from PE to AE involves no 
additional assumptions, it follows that we can safely  
proceed with the analysis as in Sec.\ref{s17}.

The validity of the the IMPT also has been verified 
numerically for the PE of the 'piston' example \cite{prm}, 
and for the PE of a 2D nonlinear oscillator \cite{lds}.   
It has been verified that indeed the tails grow in 
a ballistic-like fashion ($\propto \delta x^2$), and not 
in diffusive-like fashion ($\propto \delta x$).   
Obviously, the assumption of having a structure-less {\em core} 
does not universally apply, and also having 
$b(t)\propto\delta x^2$ is a quite fragile result.  
If we want to have a better idea about 
the core structure we should apply, in any special  
example, specific (non-perturbative) considerations. 
In case of the `piston' example 
we can use semiclassical considerations \cite{wls} 
in order to argue that the core has a Lorentzian shape 
whose width is $\hbar/\tau_{\tbox{col}}$.  
This semiclassical Lorentzian has nothing to do with Wigner's 
Lorentzian. The semiclassical Lorentzian is a purely 
non-perturbative structure. This structure is exposed provided  
$(\hbar/\tau_{\tbox{col}})/\Delta \ll b(t)$,  
leading to the condition $\delta x \gg \lambda_{\tbox{E}}$. 
Else we have a structure-less core whose 
width is characterized by the single 
parameter $b(t)$ of (\ref{e_89}).

\section{The quantum mechanical sudden approximation}
\label{s_sdn}

\begin{table}[h]
\begin{center}
\leavevmode  
\fbox{\mpg{12}{
\begin{eqnarray} 
\mbox{perturbative route ($v_{\tbox{PR}} \ll 1$):} 
\hspace*{-2cm} & \ \nonumber\\
t_{\tbox{sdn}} = \tau_{\tbox{cl}} \ll t_{\tbox{prt}} & \ \nonumber\\ 
\mbox{At} \ \ t = \tau_{\tbox{cl}} &
b(t) \ll s(t) \ll b \sim (\hbar/t)/\Delta  \nonumber\\   
\mbox{At} \ \ t = t_{\tbox{prt}}  &  
b(t) \sim s(t) \sim  (\hbar/t)/\Delta  \ll  b \nonumber\\
\ & \ & \ \nonumber\\
\mbox{Non-perturbative route ($v_{\tbox{PR}} \gg 1$):} 
\hspace*{-2cm} & \ \nonumber\\ 
t_{\tbox{prt}} \ll t_{\tbox{sdn}} \ll \tau_{\tbox{cl}} & \ \nonumber\\
\mbox{At} \ \ t = t_{\tbox{prt}} &  
b(t) \sim s(t) \sim b \ll (\hbar/t)/\Delta \nonumber\\   
\mbox{At} \ \ t = t_{\tbox{sdn}} & 
b \ll s(t) \sim  (\hbar/t)/\Delta  \nonumber\\
\mbox{At} \ \ t = \tau_{\tbox{cl}} & 
b \sim  (\hbar/t)/\Delta \ll s(t)  \nonumber
\end{eqnarray}
\vspace*{0.1mm} }} 
\end{center}
\caption{\protect\footnotesize 
Various time scales in the route to stochastic behavior.}
\end{table}

It is now appropriate to discuss the QM  
sudden approximation. For the perturbative scenario 
($v_{\tbox{PR}}\ll 1$) we have already mentioned 
that the AE departs from the PE at $t_{\tbox{sdn}}=\tau_{\tbox{cl}}$, 
which is the time to resolve the energy scale $\Delta_b$. 
In case of the non-perturbative scenario ($v_{\tbox{PR}}\gg 1$)
there will be an {\em earlier breakdown} of the 
QM sudden approximation. This is 
because we have $\tau_{\tbox{cl}} \gg t_{\tbox{prt}}$ 
and consequently at $t=\tau_{\tbox{cl}}$ we 
already have $s(t)\gg b$. Therefore $t_{\tbox{sdn}}$ should 
be defined as the time to resolve the energy scale 
which is associated with $s(t)$. It leads to
\begin{eqnarray} \label{e_94}
t_{\tbox{sdn}} \ = \ b^{\tbox{1/4}} 
(\tau_c^{\tbox{qm}} \tau_{\tbox{cl}})^{\tbox{1/2}}
\ = \ 
\left( \frac{\hbar^2 \tau_{\tbox{cl}}}
{\nu_{\tbox{E}}V^2}\right)^{1/4}
\hspace*{1cm} \mbox{for $v_{\tbox{PR}}\gg 1$}
\end{eqnarray}
The various time scales are summarized in Table 4.   
The non-perturbative crossover from genuine-ballistic 
to genuine-diffusive behavior in not trivial. 
If  $v_{\tbox{SC}} \gg 1$ we can relay on semiclassical 
considerations in order to establish the existence 
of this crossover. More generally, for $v_{\tbox{PR}} \gg 1$, 
we would like to have an appropriate effective RMT model. 
This effective RMT model should support genuine-ballistic 
motion with an elastic scattering time~$\tau_{\tbox{cl}}$.

\section{The quantum mechanical adiabatic approximation}
\label{s_LZ}

The previous analysis has emphasized the 
role of core-to-tail transitions in energy spreading. 
An implicit assumption was that these transitions are not 
suppressed by recurrences. This is not true  
in the QM adiabatic regime 
($v_{\tbox{LZ}}\ll 1$). Following \cite{wilk1} 
it is argued that energy spreading in the 
latter regime is dominated (eventually) by 
Landau-Zener transitions between near-neighbor 
levels. For completeness, the present section is 
devoted to the clarification of this observation. 

As a preliminary exercise it interesting to estimate 
the contribution of transitions between near-neighbor 
levels. The time scale that characterize these 
transitions is $\tau_c^{\tbox{qm}}$, and the 
`step' size is $\Delta$. Disregarding all other 
transitions, we have a random-walk process with 
diffusion coefficient $(\Delta)^2/\tau_c^{\tbox{qm}}$, 
leading to 
\begin{eqnarray} \label{e_95} 
D_{\tbox{E}}^{\tbox{NN}} \ \sim \ 
\frac{1}{v_{\tbox{LZ}}} D_{\tbox{E}}^{\tbox{cl}} 
\hspace*{2cm} \mbox{[not applicable]}
\end{eqnarray}
Thus for $v_{\tbox{LZ}} \gg 1$ the contribution of 
near-neighbor (NN) transitions is indeed negligible as 
expected. In the QM adiabatic regime 
($v_{\tbox{LZ}}\ll 1$) the above result should be 
modified as follows \cite{wilk1}:
\begin{eqnarray} \label{e_96} 
D_{\tbox{E}}^{\tbox{LZ}} \ \approx \ 
\left(\frac{1}{v_{\tbox{LZ}}}\right)^{1{-}(\beta/2)}
D_{\tbox{E}}^{\tbox{cl}} 
\hspace*{2cm} \mbox{for \ $v_{\tbox{LZ}}\ll 1$}       
\end{eqnarray}
This result takes into account the no-trivial 
nature of Landau-Zener transitions and the 
statistics of the avoided-crossings. One should use 
$\beta=1$ for the Gaussian unitary ensemble (GUE) 
and $\beta=2$ for the Gaussian orthogonal ensemble (GOE). 
Recalling the stochastic considerations 
that lead to (\ref{e_95}) one deduces that 
the perturbative breaktime is
\begin{eqnarray}  
\hspace*{-1cm}
t_{\tbox{prt}} \ = \ 
\left(\frac{1}{v_{\tbox{LZ}}}\right)^{\beta/2}
\tau_c^{\tbox{qm}} \ \propto \ V^{-(1+(\beta/2))} 
\hspace*{1cm} \mbox{[applies to \ $v_{\tbox{LZ}}\ll 1$]}       
\end{eqnarray}

In the QM adiabatic regime 
energy spreading is dominated by near-neighbor 
level transitions for two distinct reasons. 
The first reason applies to the $\beta=1$ case, namely 
$D_{\tbox{E}}^{\tbox{LZ}} \gg D_{\tbox{E}}^{\tbox{cl}}$. 
The other reason is that 
$D_{\tbox{E}}^{\tbox{cl}} \gg D_{\tbox{E}}^{\tbox{FGR}}$. 
In the latter inequality, $D_{\tbox{E}}^{\tbox{FGR}}$ 
is based on the FGR result (\ref{e_32}).  
The FGR result becomes very small, compared 
with the classical result, once $\tilde{F}(\omega)$ 
becomes much narrower than the average level-spacing. 
The QM-adiabaticity condition $t_{\tbox{H}} \ll \tau_c^{\tbox{qm}}$ 
means that individual energy levels are being 
resolved before the breakdown of first-order perturbation theory. 
Having no 'systematic' transitions to 'other' levels 
implies that the energy-distribution remains localized 
in the initial level for a very long time. 
The above argumentation implies that 
$D_{\tbox{E}}^{\tbox{LZ}} \gg D_{\tbox{E}}^{\tbox{FGR}}$, 
meaning that for extremely slow velocities energy spreading, 
and the eventual breakdown of the QM adiabatic approximation,  
is predominantly due to Landau-Zener mechanism.

\section{Open questions and future directions}

The purpose of this paper was to make the first 
steps towards a theory for energy spreading 
and quantum dissipation.  In particular we wanted 
to demonstrate that perturbation theory, 
and semiclassical theory have different regimes 
of validity. There are still a lot of open 
questions that have to be answered.  

An important issue is the specification of the 
general conditions for having a genuine stochastic behavior 
in the QM case. For fast velocities it is 
suggested (but not proved) that the 
stochastic behavior persists beyond the 
semiclassical breaktime. For slow velocities, 
it is suggested (but again not proved) 
that the stochastic behavior persists beyond the 
breaktime of perturbation theory.  The latter suggestion  
is indirectly supported by common-wisdom and 
by various numerical experiments 
with banded matrices \cite{kottos,wilk2,wbr}.  
For the generic RMT picture, 
which is still lacking, it is implied that both,  
breakdown of perturbation theory and resolving 
the bandwidth of first-order transitions, are 
necessary conditions for having genuine stochastic behavior. 
In any case, stochasticity {\em can be established}  
if we assume irregular a-periodic driving with 
an appropriate correlation scale. For periodic driving, 
further considerations are required in order to analyze  
the possible manifestation of localization effect.

A better understanding of the core-tail 
structure is required. Only in the case 
of Wigner's model \cite{wigner,flamb,casati} 
we have an established result: Namely, 
the core-tail structure is simply a Lorentzian. 
For real systems the core-tail structure 
is not necessarily a Lorentzian \cite{wls}.
The determination of the border between 
the core and the tail may be problematic. 
One cannot exclude the existence of a distinct 
tail component, in the vicinity of the core, 
that does not grow like $\delta x^2$.   
A strongly related issue is to get an analytical 
understanding of the $b'$ dependence of the 
effective (`renormalized') correlation time $\tau_c$.

The `piston' example is non-generic in many respects.
There are three classical length scales: 
The penetration distance upon collision with the piston; 
The mean path-length between collisions with the piston; 
And the ballistic length scale that characterizes the 
volume of the cavity. 
Quantum-mechanics adds two additional length scales: 
one is related to the Airy structure in the vicinity 
of the turning points, and the other is the De-Broglie wavelength.   
Having all these scales has some non-universal 
consequences \cite{wls} that we have  not considered   
in this paper. 
The application of the general theory  of sections 8-20 
to the 'piston' example is quite straightforward, but 
these non-universal features should be taken into account. 
In the generic theory there are only two parametric scales: 
The displacement $\delta x_c^{\tbox{qm}}$  that is needed 
in order to mix neighboring levels; 
And the displacement $\delta x_{\tbox{prt}}$ that is needed 
in order to mix the core with the tail. 
The former is much smaller than De-Broglie wavelength, 
and the latter is much larger than De-Broglie wavelength. 
It turns out that in the `piston' example there is a third, 
non-universal parametric scale $\delta x_{\tbox{NU}}$ 
that roughly equals to De-Broglie wavelength \cite{wls}. 
Consequently, the perturbative (slow-velocity) regime is 
further divided into a universal slow-velocity regime, and 
a non-universal slow-velocity regime.

Of particular importance is the understanding of 
the {\em hard-wall limit}. In the generic theory 
$\tau_{\tbox{cl}}$ determines the bandwidth $\Delta_b$ 
of the matrix $\mbf{W}_{nm}$. Having {\em finite bandwidth}  
is essential in order to understand that there is 
a crossover to a {\em non-perturbative} regime in the 
$\hbar\rightarrow 0$ limit.  We cannot treat 
$\mbf{W}_{nm}$ as a banded matrix if we take first 
the limit $\tau_{\tbox{cl}}\rightarrow 0$.  
The walls of the `piston' should be regarded 
as `hard' once $\Delta_b$ becomes equal or 
larger than $E$. This is equivalent to having 
(classical) penetration distance smaller than 
De-Broglie wavelength.  The consequence of taking 
the hard wall limit is that the non-perturbative 
regime ($v_{\tbox{PR}}>1$) disappears. This 
state-of-affairs is possibly responsible to the  
{\em illusion} that a {\em general} theory for 
quantum dissipation can be base on a perturbative 
approach.  At first sight it looks strange that 
hard-walls are `better' for perturbation theory. 
It looks even more strange that for hard walls 
we cannot apply the semiclassical theory. In order 
to make the latter observation less strange 
recall that solving the one-dimensional 
Schroedinger equation near a sharp step, 
and then taking $\hbar\rightarrow 0$, never  
corresponds to the WKB approximation.

The problem of quantum dissipation, in the 
sense of this paper, is a preliminary stage 
in the construction of a theory \cite{vrn} of 
quantal Brownian motion (QBM).  In the 
classical case it is known \cite{jar} that 
the motion of a 'heavy' particle that is coupled to  
{\em chaotic} degrees-of-freedom is quite generally described 
by the classical Langevin equation. The effect of 
the environment is represented by 
a friction force plus a noise term. The friction 
leads to dissipation of energy and the noise 
term is essential for having diffusion.  
Furthermore, the friction coefficient is related 
to the noise intensity via the universal 
${\cal F\!D}$ relation. 
The fact that there is no general theory for 
quantum dissipation, and a-fortiori there is no 
general theory of QBM, has not been universally 
recognized in the literature.  It is true that 
there is a vast literature that comes under those  
headings, but actually this literature is commonly 
based on an {\em effective-bath approach}. 
In previous studies \cite{dld,qbm,dph} the 
common effective-bath strategy has been applied in order 
to develop a universal description of QBM and dephasing.  
Another possibility is to use an effective RMT bath \cite{rmt}. 
The results of the latter study agree with \cite{dld,qbm}. 
A future theory of QBM should clarify whether 
effective-bath methods universally apply.


\ack{
I thank {\em Eric Heller} for useful and stimulating discussions. 
I also thank {\em Shmuel Fishman} for fruitful interaction in early 
stages, for his comments on an earlier version of this paper, 
and for his generous hospitality while visiting the Technion. 
The ITAMP in the Harvard-Smithsonian Center is acknowledged 
for support, and the MPI f\"ur Komplexer Systeme in Dresden is 
acknowledged for support and for the kind hospitality during 
the workshop and conference {\it Dynamics of Complex Systems}.
I thank {\em Felix Izrailev} for making me aware of the fact that 
Wigner's Lorentzian is the obvious and the simplest example 
for a core-tail structure, and for drawing my attention to the 
strong relation between the QCC considerations in this paper 
and the classical approximation of \cite{felix2}.}

\newpage
\appendix

\section{The sudden and the adiabatic approximations}
\label{a_evolving}

For the time dependent Hamiltonian ${\cal H}(Q,P;x(t))$ 
energy is not a constant of the motion. 
Changes in the actual energy ${\cal E}(t)$ reflect 
`real' dynamical changes as well as parametric changes. 
Therefore it is useful to introduce the following definitions: 
\begin{eqnarray}
{\cal E}(t) \ \ = \ {\cal H}(Q(t),P(t);x(t)) \\
{\cal E}'(t) \ = \ {\cal H}(Q(t),P(t);x(0)) 
\end{eqnarray}
For simplicity we assume that the phase space volume 
$\Omega(E;x)$ is independent of $x$, thus $F(x)=0$. 
The actual energy change can be calculated as follows:
\begin{eqnarray}
\delta{\cal E} \ = \ {\cal E}(t)-{\cal E}(0) \ = \ 
- V \int_0^t {\cal F}(t') dt' 
\end{eqnarray}
The actual energy change $\delta{\cal E}$ can be 
viewed as a sum of parametric-energy-change~$\delta{\cal E}_o$, 
and reduced-energy-change~$\delta{\cal E}'$.
\begin{eqnarray}
\delta{\cal E} \ \ = \
{\cal H}(Q(t),P(t);x(t)) - {\cal H}(Q(0),P(0);x(0)) 
\nonumber\\
\delta{\cal E}_o  \ = \
{\cal H}(Q(t),P(t);x(t)) - {\cal H}(Q(t),P(t);x(0)) 
\nonumber\\
\delta{\cal E}' \ = \ 
{\cal H}(Q(t),P(t);x(0)) - {\cal H}(Q(0),P(0);x(0))
\nonumber
\end{eqnarray}
The reduced energy change $\delta{\cal E}'$ reflects 
the deviation of $(Q,P)$ from the original energy surface. 
It can be calculated as follows:
\begin{eqnarray}
\delta{\cal E}' \ = \  
V \cdot \left[{\cal F}(t)\times t 
\ - \ \int_0^t {\cal F}(t')dt' \right]
\end{eqnarray}
On the other hand, the actual energy change 
$\delta{\cal E}$ reflects the deviation of 
$(Q,P)$ from the instantaneous energy surface.

By inspection of the expressions for the reduced 
energy change we arrive at the conclusion that for 
short times we have the so called `sudden approximation':
\begin{eqnarray}
\delta{\cal E}' \ \approx \ 0 
\ \ \ \ \ \mbox{for $t\ll \tau_{\tbox{cl}}$}
\end{eqnarray}
By inspection of the expression for the actual energy 
change we arrive at the conclusion that for 
long times we have the so called `adiabatic approximation':
\begin{eqnarray}
\delta{\cal E} \ \sim \ 0 
\ \ \ \ \ \mbox{for $t\ll t_{\tbox{frc}}$}
\end{eqnarray}
The time evolution of an initially localized 
phase-space distribution $\rho_0(Q,P)$ is illustrated 
in Fig.\ref{f_evolving}.  
For short times we have in general non-stationary time evolution:
\begin{eqnarray} \nonumber
{\cal U}(t) \ \rho_0(Q,P) \ \ne \ \rho_0(Q,P)
\ \ \ \ \ \ \ \ \ \ \ \ \ \ \ \  
\mbox{for $t\ll\tau_{\tbox{cl}}$}
\end{eqnarray}
Here ${\cal U}(t)$ is the classical propagator of 
phase space points. However, if we operate  
with the same ${\cal U}(t)$ on a microcanonical 
distribution, then
\begin{eqnarray} \nonumber
{\cal U}(t) \ \rho_{\tbox{E,x(0)}}(Q,P) 
\ \approx \ \rho_{\tbox{E,x(0)}}(Q,P)
\ \ \ \ \ \ \ \mbox{for $t\ll\tau_{\tbox{cl}}$}
\end{eqnarray}
Thus, the sudden approximation implies that for 
short times ${\cal U}(t)$ can be replaced by 
{\em unity} if it operates on an initial 
microcanonical distribution. For long times we have 
\begin{eqnarray} \nonumber
{\cal U}(t) \ \rho_0(Q,P) 
\ \sim \ \rho_{\tbox{E,x(t)}}(Q,P) \ \ \ \ \ \ \ 
\mbox{for $t_{\tbox{erg}} \ll t \ll t_{\tbox{frc}}$}
\end{eqnarray} 
It is not required to start with 
a microcanonical distribution, unless 
$\tau_{\tbox{cl}} \ll t \ll t_{\tbox{erg}}$. 
The adiabatic approximation becomes 
worse and worse as time elapse due to the 
transverse spreading across the energy surface.
$t_{\tbox{frc}}$ is the breaktime of  
the adiabatic approximation. See discussion after (\ref{e_50}).

\newpage
\section{The conventional FGR Picture} 
\label{a_FGR}

The simplest version of time-dependent perturbation 
theory is base on the approximated Hamiltonian 
\begin{eqnarray} \label{ea_c1} 
{\cal H}(Q,P;x(t)) \ \approx \
{\cal H}(Q,P;x(0))+
\frac{\partial {\cal H}}{\partial x} x(t)
\end{eqnarray}
and using a {\em fixed basis} that is determined 
by the unperturbed Hamiltonian ${\cal H}(Q,P;x(0))$. 
For simplicity we set $x(0){=}0$. A limitation that follows 
from using fixed basis is that the crossover from ballistic 
to diffusive spreading is out-of-reach for this version 
of perturbation theory.  

Another limitation of the present FGR picture stems 
from the fact that some additional assumptions 
should be imposed on $x(t)$, else the treatment 
may be not valid. We shall assume that at any time 
$x(t) \ll \delta x_c^{\tbox{cl}}$, so that the 
expansion in (\ref{ea_c1}) is valid. Moreover we shall 
assume that $x$ is being changed in a an 
arbitrary a-periodic fashion, such that $\dot{x}(t)$
becomes uncorrelated on a time scale $\tau_c^{\tbox{drv}}$. 
The assumption $\tau_{\tbox{cl}} \ll \tau_c^{\tbox{drv}}$ 
is implied by the trivial definition of slowness 
(\ref{e_11}). The loss of velocity-velocity correlation will be 
described by a function $F(\tau)$, with the convention 
$F(0)=1$.  This function should satisfy the 
normalization $\int F(\tau)d\tau = 0$, else we will have  
an unbounded growth of $(x(t)-x(0))^2$.  
We still assume that the typical `velocity' is $V$, 
meaning that 
$\langle \dot{x}(t) \dot{x}(t{+}\tau) \rangle = V^2 F(\tau)$. 
The above assumptions implies that the 
correlator  $\langle x(t) x(t{+}\tau) \rangle = {\rm F}(\tau)$
is well defined. Its Fourier transform satisfies 
$\tilde{{\rm F}}(\omega)=V^2\tilde{F}(\omega)/\omega^2$. 
Note that the normalization of the frequency 
distribution $\tilde{F}(\omega)$ is $1$, 
while the normalization of $\tilde{{\rm F}}(\omega)$ 
is $(V\tau_c^{\tbox{drv}})^2$.

The FGR expression for the transition rate from  
the energy levels $m$ to some other energy level $n$ is:
\begin{eqnarray} \label{e_FGR} 
\Gamma_{nm} \ = \ \frac{1}{\hbar^2} 
\ \left|\left(
\frac{\partial {\cal H}}{\partial x}
\right)_{nm}\right|^2
\ \tilde{{\rm F}}\left( 
\frac{E_n{-}E_m}{\hbar} \right)
\end{eqnarray}
From here follows an expression for the 
diffusion constant. This expression is 
easily cast into a classical look-alike formula: 
\begin{eqnarray} \label{e_FGRD}    
\hspace*{-2cm}
D_{\tbox{E}} \ = \ \frac{1}{2}\frac{1}{{\cal N}} 
\sum_{nm} \Gamma_{nm} (E_n{-}E_m)^2 
\ = \ \frac{1}{2}V^2  
\int_{-\infty}^{\infty} C_{\tbox{E}}(\tau) F(\tau) d\tau
\end{eqnarray}
It is argued that (\ref{e_FGR}) is valid 
for any $t>\tau_c^{\tbox{drv}}$ provided 
there is a separation of time scales 
$\tau_c^{\tbox{drv}} \ll \tau_{\tbox{prt}}$. 
This is the `golden-rule' condition, namely,  
breaktime of first-order perturbation theory 
should be after $\tau_c^{\tbox{drv}}$. 
We use the notation $\tau_{\tbox{prt}}$ 
rather than $t_{\tbox{prt}}$ in order to 
emphasize that fixed-basis perturbation theory 
is being used.
The persistence of transitions with the 
{\em same} rate for  $t>\tau_{\tbox{prt}}$ 
is guaranteed due the stochastic nature 
of the dynamics: The irregular driving is 
like noise, and interference contribution 
is averaged to zero once time intervals 
larger than $\tau_c^{\tbox{drv}}$ are being composed. 
Let us assume further that 
$\tau_{\tbox{prt}}\ll t_{\tbox{H}}$, 
in order to guarantee that there are no 
recurrences irrespective of $\tau_c^{\tbox{drv}}$. 
Gathering all our assumption together 
(classical slowness condition, FGR condition, 
and the non-recurrence condition) we get:  
\begin{eqnarray} \label{e_FGRrq} 
\tau_{\tbox{cl}} \ll \tau_c^{\tbox{drv}} \ll   
\tau_{\tbox{prt}} \ll t_{\tbox{H}}
\end{eqnarray}
It should be noted that the classical slowness condition 
implies resonance-limited transitions and hence 
restricted QCC is guaranteed. 

We turn now to get an actual estimate for the 
perturbative breaktime  $\tau_{\tbox{prt}}$, 
and for the associated slowness condition.  
The total transition rate from a level $m$ is 
$\Gamma=\sum_n'\Gamma_{nm}$, and we have 
\begin{eqnarray}  \label{ea_c4}  
\hspace*{-2cm}
\tau_{\tbox{prt}} \ = \ \frac{1}{\Gamma} \ = \
\frac{\hbar^2}{\nu_{\tbox{E}}V^2  
\ (\tau_c^{\tbox{drv}})^2} 
\hspace*{2cm} \mbox{fixed-basis, $v_{\tbox{PR}} \ll 1$}  
\end{eqnarray}
The maximal value of $\tau_{\tbox{prt}}$ 
is attained if $\tau_c^{\tbox{drv}}=\tau_{\tbox{cl}}$. 
The minimal value 
$\tau_{\tbox{prt}}=t_{\tbox{prt}}$  
is attained for $\tau_c^{\tbox{drv}}=t_{\tbox{prt}}$.
One easily concludes that a necessary condition for 
the applicability of the FGR picture is 
\begin{eqnarray} \label{ea_c5} 
\hspace*{-1.3cm}
V \ \ll \ \frac{\hbar/\tau_{\tbox{cl}}}
{\sqrt{\nu_{\tbox{E}}\tau_{\tbox{cl}}}}
\hspace*{1.3cm} 
\mbox{QM definition of slowness}
\end{eqnarray}
The latter condition is always violated 
in the classical limit. Thus, it is 
{\em not possible in principle} to establish QCC 
in the limit $\hbar\rightarrow 0$
by using FGR picture. Note that (\ref{ea_c5}) is equivalent 
to (\ref{e_24}) leading to $v_{\tbox{PR}}\ll 1$.

\section{Alternative derivation of the `sudden time'}  
\label{a_sdnt}

Simple considerations based on fixed-basis 
perturbation theory can be used in order to 
derive a result for $t_{\tbox{sdn}}$. 
By `sudden approximation' we mean that 
for a limited time we can `ignore' the 
dynamical changes that are generated 
by the Hamiltonian. A necessary condition 
is that the transition probability 
between levels is much less than 
unity, or equivalently $t \ll \tau_{\tbox{prt}}$. 
It is essential to use the breaktime $\tau_{\tbox{prt}}$
of {\em fixed-basis} perturbation theory,  
since we want to avoid the fake transitions due to 
non-trivial parametric-evolution.
However, $t \ll \tau_{\tbox{prt}}$  is not 
a sufficient condition. If $|\psi\rangle$ 
is a superposition of few levels, then 
we should also require that the 
corresponding energies will not be resolved. 
Namely we should have $|E_n{-}E_m|t \ll \hbar$ for any 
$n$ and $m$ in the superposition. We are interested 
in the way in which energy is re-distributed. 
Therefore, as long as first order perturbation 
theory applies, it is not important whether 
energy levels are resolved unless 
they are directly coupled ($|E_n{-}E_m|<\Delta_b$). 
It follows that 
\begin{eqnarray}   
t_{\tbox{sdn}} \ \ = \ \ 
\mbox{minimum}( \ \tau_{\tbox{prt}} \ , \ \tau_{\tbox{cl}} \ )  
\end{eqnarray}
where $\tau_{\tbox{prt}}$ is the breaktime of 
fixed-basis perturbation theory. It is a trivial 
matter to estimate the transition probabilities. 
Since we are interested in short times 
($t<\tau_{\tbox{cl}}<\tau_c^{\tbox{drv}}$) 
we can substitute $x(t)=Vt$ in (\ref{ea_c1}), and we get: 
\begin{eqnarray}   
\hspace*{-2cm}
\mbox{transition probability} \ = \ 
\left| \ \frac{1}{2} \ \frac{1}{\hbar} \left(
\frac{\partial {\cal H}}{\partial x}
\right)_{nm} Vt^2 \ \right|^2 
\ \ \ \ \ \ \ \mbox{for $n \ne m$} 
\end{eqnarray}
Note that the FGR expression for the 
transition probability, namely $\Gamma_{nm}t$, 
is valid in a different time regime   
($t\gg\tau_c^{\tbox{drv}}\gg\tau_{\tbox{cl}}$). 
The transition-probability to each of the levels 
within the band is approximately the same, and 
upon multiplication by $b$, an expression for the   
total transition-probability is easily obtained. 
It can be written as $(t/\tau_{\tbox{prt}})^4$ where 
\begin{eqnarray}   \label{ea_d3}
\hspace*{-2cm}
\tau_{\tbox{prt}} \ = \ 
\left( \frac{\hbar^2 \tau_{\tbox{cl}}}
{\nu_{\tbox{E}}V^2}\right)^{1/4}
\hspace*{2cm} \mbox{fixed-basis, $v_{\tbox{PR}} \gg 1$}
\end{eqnarray}
If the velocity $V$ is not slow (in the sense of (\ref{ea_c5})), 
then we have $t_{\tbox{sdn}}=\tau_{\tbox{prt}} < \tau_{\tbox{cl}}$.
For slow velocities we have $\tau_{\tbox{prt}}\gg\tau_{\tbox{cl}}$
and therefore $t_{\tbox{sdn}}=\tau_{\tbox{cl}}$.
In the latter case expression (\ref{ea_d3}) underestimates 
the perturbative breaktime: Once $t\gg\tau_{\tbox{cl}}$ 
the transitions become resonance-limited, and consequently 
a diffusive-like spreading develops.  Thus, in the 
slow velocity regime the FGR expression (\ref{ea_c4}) for 
the perturbative breaktime $\tau_{\tbox{prt}}$ should be used. 
Taking into account the restrictions on 
$\tau_c^{\tbox{drv}}$ one observes that going     
from (\ref{ea_d3}) to (\ref{ea_c4}) does not involve 
any discontinuity.

\newpage
\section{Spherical coordinates and related results}
\label{a_spherical}

The solid angle in $d$ dimension and the 
volume element in spherical coordinates are
\begin{eqnarray}   
\Omega_d \ = \ \frac{2\pi^{d/2}}{\Gamma(d/2)} 
\ = \ 2, \ 2\pi, \ 4\pi, \ ... 
\ \\
d\Omega_d \ r^{d{-}1}dr \ = \ 
d\Omega_{d{-}1} \ (\sin(\theta))^{d{-}2} 
d\theta \ r^{d{-}1}dr
\end{eqnarray}
The following results are easily derived:
\begin{eqnarray}  \label{ea_e3} 
\langle|\cos(\theta)|\rangle \ = \ 
\frac{1}{\sqrt{\pi}}
\frac{\Gamma(d/2)}{\Gamma((d{+}1)/2)} \ = \ 
1, \ \frac{2}{\pi}, \ \frac{1}{2}, \ ... \\
\langle|\cos(\theta)|^3\rangle \ = \ 
\frac{1}{\sqrt{\pi}}
\frac{\Gamma(d/2)}{\Gamma((d{+}3)/2)} \ = \ 
1, \ \frac{4}{3\pi}, \ \frac{1}{4}, \ ... 
\end{eqnarray}
Note the relation 
$2\langle|\cos(\theta)|\rangle = 
(d{+}1)\langle|\cos(\theta)|^3\rangle$.

It is useful to define a generalized cosine function 
by averaging $\exp(i\mbf{n}{\cdot}\mbf{r})$ over the orientation of 
the unit vector $\mbf{n}$. The averaging is easily 
performed by using spherical coordinates:  
\begin{eqnarray}   
\mbox{Cos}(r) \ = \ 
\langle \mbox{e}^{i r \cos(\theta)} \rangle \ = \ 
2^{\frac{d}{2}{-}1}\Gamma(\mboxs{$\frac{d}{2}$})
\ \frac{J_{\frac{d}{2}{-}1}(r)}{r^{\frac{d}{2}{-}1}}
\end{eqnarray}
It is also useful to define a generalized sinc function 
as follows:
\begin{eqnarray}   
\mbox{Sinc}(r) \ = \ -\frac{\mbox{Cos}'(r)}{r} \ = \ 
\frac{1}{d} \  
2^{\frac{d}{2}}\Gamma(\mboxs{$\frac{d}{2}{+}1$})
\ \frac{J_{\frac{d}{2}}(r)}{r^{\frac{d}{2}}}
\end{eqnarray}
The function $(\mbox{Sinc})^2$ will appear 
in an integral over a $(d{-}1)$ dimensional surface. 
It will be possible to replace  
it by an effective delta-function:
\begin{eqnarray}   \label{ea_e7}
(\mbox{Sinc}(k|\mbf{s}_2-\mbf{s}_1|))^2 
\ \longrightarrow \ 
\left(\frac{2\pi}{k}\right)^{d{-}1} 
\frac{2 \langle|\cos(\theta)|^3\rangle}{\Omega_d} 
\ \delta(\mbf{s}_2-\mbf{s}_1) 
\end{eqnarray}
In order to derive this result one should use 
the following:
\begin{eqnarray}   
\int_0^{\infty} \frac{(J_{\nu}(x))^2}{x^2}dx \ = \
\frac{4}{\pi (4\nu^2{-}1)}  \\
\int_0^{\infty} (\mbox{Sinc}(r))^2 
\ \Omega_{d{-}1} \ r^{d{-}2}dr \ = \ 
\frac{ \Omega_{d{-}1} }{ \pi } \
\frac{ 2^d }{ d^2{-}1 } \ 
(\Gamma( \mboxs{$\frac{d}{2}$} ))^2 
\end{eqnarray}

\newpage
\section{Collisions with a wall: phase space approach}
\label{a_CE}

The position of a particle inside a cavity can 
be described by $Q=(z,\mbf{x}_{\perp})$ where $z$ is  
perpendicular to the surface.  
The potential which is experience by the particle 
is assumed to be ${\cal V}(Q)=0$ inside 
the cavity where $z<0$, and ${\cal V}(Q)=f{\cdot}z$ for $z>0$.  
Let the variable $x$ parameterize the perpendicular 
displacement of the surface (hence $\mbf{n}{\cdot}\mbf{V}=1$). 
With this parameterization we we have simply 
${\cal F}(Q,P;x)=f$ provided $z>0$, else ${\cal F}(Q,P;x)=0$.  
For an isolated collision we have 
\begin{eqnarray} 
z(t \pm \half\tau) \ = \ 
z \pm \frac{p_z}{m}\cdot \left(\half \tau \right)
- \frac{1}{2} \frac{f}{m} \left(\half \tau \right)^2
\end{eqnarray}
The correlation function is  
\begin{eqnarray} \label{ea_f2} 
\hspace*{-2cm} 
C_{\tbox{E}}(\tau) \ = \ 
\frac{f^2 \ \int d\mbf{p}_{\perp} \int dp_z 
\int d\mbf{x}_{\perp} \int dz 
\ \delta\left(E-\left(\frac{\mbf{p}^2}{2m}+f\cdot z\right)\right)}
{\int d\mbf{p} \int d\mbf{x} 
\ \delta\left(E-\frac{\mbf{p}^2}{2m}\right)}
\end{eqnarray}
The integration in the numerator is restricted by the 
conditions $z(t+\frac{1}{2}\tau)>0$ and 
$z(t-\frac{1}{2}\tau)>0$. The $dx_{\perp}$ integration gives 
an {\it Area} factor. After the $dz$ integration 
one is left with a $d\mbf{p}$ integration that represents 
the volume $\mbf{p}^2+2(\half f\tau)p_z+(\half f\tau)^2 < 2mE$. 
Using spherical coordinates one obtains 
\begin{eqnarray} \nonumber
\hspace*{-2cm} 
C_{\tbox{E}}(\tau) \ = \ 
\frac{ \mboxs{Area} \cdot f \cdot 2 \int \Omega_{d{-}1} 
(\sin\theta)^{d{-}2} d\theta \int p^{d{-}1}dp}  
{\mboxs{Volume} \cdot \half \Omega_d \ 
((2mE)^{\tbox{1/2}})^d \ / \ E} \ \equiv \  
\nu_{\tbox{E}}\hat{C}_{\tbox{E}}
\left(\frac{\tau}{\tau_{\tbox{cl}}}\right)
\end{eqnarray}
The $d\theta$  and the $dp$ integrations in the numerator are 
restricted by the conditions \ 
$(\half f\tau)/(2mE)^{\tbox{1/2}} < \cos\theta < 1$ \ and \ 
$(\half f\tau)/\cos\theta < p < (2mE)^{\tbox{1/2}}$ \ 
respectively. The noise intensity is 
\begin{eqnarray}  
\nu_{\tbox{E}} \ = \ 2\langle|\cos\theta|^3\rangle   
\ \frac{\mboxs{Area}}{\mboxs{Volume}} 
\ m^2 v_{\tbox{E}}^3
\end{eqnarray}
The properly normalized $\hat{C}_{\tbox{E}}(\tau)$ 
equals zero for $|\tau|>1$ and otherwise can be 
expressed using an hypergeometrical function as follows: 
\begin{eqnarray} 
\hspace*{-2cm}
\hat{C}_{\tbox{E}}(\tau) \ &=& \ 
\frac{d^2{-}1}{2d}\int_{\tau{<}\cos\theta}
\hspace*{-0.5cm} (\sin\theta)^{d{-}2}d\theta
\left[1-\left(\frac{|\tau|}{\cos\theta}\right)^d\right] \\
\nonumber \hspace*{-2cm} &=& \
\frac{d+1}{2d}
\left(\frac{1}{\tau^2}-1\right)
^{\frac{d{-}1}{2}}
\left[ _2F_1 \left(\frac{d{-}1}{2},\frac{d}{2};
\frac{d{+}1}{2};{-}\left(\frac{1}{\tau^2}-1\right)\right)
-|\tau|^d\right] \\
\nonumber \hspace*{-2cm} &=& \ 
\left\{ \matrix{ 
\frac{3}{4}
\left(\arctan\left(\sqrt{\frac{1}{\tau^2}-1}\right)
-\tau^2\sqrt{\frac{1}{\tau^2}-1} \right) 
\ & \ \ \ \mbox{for $d=2$} \cr
\frac{2}{3}(|\tau|^3-3|\tau|+2) \ & \ \ \ \mbox{for $d=3$}  
}\right.
\end{eqnarray}
It is much easier to obtain an explicit expression 
for $C_{\tbox{T}}(\tau)$.  The $\delta(\mboxs{energy})$ function 
in (\ref{ea_f2}) should be replaced by 
$\exp(-\mboxs{energy}/(k_{\tbox{B}}T))$. The 
integration is factorized, and one obtains
\begin{eqnarray} 
\hspace*{-2cm}
C_{\tbox{T}}(\tau) \ = \ \frac{\mbox{Area}}{\mbox{Volume}}
\ k_{\tbox{B}}T f \cdot  \mbox{erfc}\left(\frac{1}{2}
\frac{f}{\sqrt{2mk_{\tbox{B}}T}} \ |\tau|\right)    
\end{eqnarray}
 The noise intensity $\nu_{\tbox{T}}$ can be found 
either by integration over this function, or else by 
performing thermal average over $\nu_{\tbox{E}}$. 
In both cases the result is:
\begin{eqnarray} 
\nu_{\tbox{T}} \ = \ \frac{4}{\sqrt{\pi}}
\ \frac{\mboxs{Area}}{\mboxs{Volume}} 
\ k_{\tbox{B}}T \ (2mk_{\tbox{B}}T)^{1/2}
\end{eqnarray}

\newpage
\section{Collisions with a wall: time domain approach}  
\label{a_Ft}

Consider the motion of a particle inside a chaotic cavity. 
we can define and estimate the collision rate with 
a wall element as follows:
\begin{eqnarray}  \label{ea_g1} 
\frac{1}{\tau_{\tbox{col}}} \ = \ 
\left\langle\sum_{\tbox{col}}\delta(t-t_{\tbox{col}})\right\rangle
\ = \ 
\frac{1}{2}\frac{ds}{\mboxs{Volume}} 
\langle|\cos\theta|\rangle \ v_{\tbox{E}}
\end{eqnarray}
Above $t_{\tbox{col}}$ is the time of of a collision, 
and $ds$ is the area of the wall element. 
The derivation of last equality is based on   
ergodicity considerations which we are going to 
explain now. Let us define the coordinate 
$z$ as in the previous appendix. We have
\begin{eqnarray} \nonumber
\hspace*{-1.0cm} 
\delta(z(t)) \ = \ 
\sum_{\tbox{col}}\sum_{\pm} 
\frac{1}{|(v_z)_{\tbox{col}}|} 
\delta(t-t_{\tbox{col}}^{\pm}) \ = \ 
\sum_{\tbox{col}} 
\frac{2}{v_{\tbox{E}} |\cos\theta_{\tbox{col}}|} 
\delta(t-t_{\tbox{col}})
\end{eqnarray}
where $t_{\tbox{col}}^{\pm}$ correspond to the crossing times 
of the $z=0$ surface, and 
$t_{\tbox{col}}=(t_{\tbox{col}}^{+}+t_{\tbox{col}}^{-})/2$ 
is the time of the collision. 
The duration of the collision
$\Delta t_{\tbox{col}} = (t_{\tbox{col}}^{+}-t_{\tbox{col}}^{-})$ 
is assumed to be extremely short.   
Equivalently we can write 
$\sum\delta(t{-}t_{\tbox{col}}) = \half v_{\tbox{E}} 
\ |\cos\theta| \ \delta(z(t))$. 
In order to get (\ref{ea_g1}), the latter expression should be 
averaged over the (implicit) initial conditions $(Q(0),P(0))$. 
As always $\langle...\rangle$ indicates this 
type of averaging. 
Due to the chaos, the coordinate $z$ and $\cos\theta$ 
can be treated as independent variables. 
Due to ergodicity we have   
$\langle \delta(z(t)) \rangle  = 
\langle \delta(z) \rangle_{\tbox{E}} = 
{ds}/{\mboxs{Volume}}$. 
As always $\langle...\rangle_{\tbox{E}}$ 
stands for a microcanonical average. 
For $\langle|\cos\theta|\rangle$ one 
may substitute (\ref{ea_e3}).

Exactly the same procedure can be used in order to 
estimate $\nu_{\tbox{E}}$. The first step is to realize that 
${\partial{\cal H}}/{\partial x}=(\mbf{n}{\cdot}\hat{\mbf{V}})\times f$ 
for $z>0$, and {\em zero} otherwise. The fluctuating force  
${\cal F}(t)=-{\partial{\cal H}}/{\partial x}$ looks 
like a train of short impulses. The duration of each 
impulse is $\Delta t_{\tbox{col}}=2m|(v_z)_{\tbox{col}}|/f$. 
Therefore we can write  
\begin{eqnarray} 
\hspace*{-1.0cm} 
{\cal F}(t) \ = \ 
-\frac{\partial {\cal H}}{\partial x} \ = \ 
\sum_{\tbox{col}} \ 
[\ (\mbf{n}{\cdot}\hat{\mbf{V}}) \ 
2mv_{\tbox{E}} \ \cos(\theta_{\tbox{col}}) \ ] \ \ 
\delta(t-t_{\tbox{col}}) 
\end{eqnarray}  
The intensity of the fluctuations due to this 
random-like sequence of delta-impulses is 
\begin{eqnarray} \nonumber 
\hspace*{-1.0cm} 
\nu_{\tbox{E}} \ = \ 
\left\langle \sum_{\tbox{col}} \ 
[ \ (\mbf{n}{\cdot}\hat{\mbf{V}}) \ 
2mv_{\tbox{E}} \ \cos(\theta_{\tbox{col}}) \ ]^2 \ \
\delta(t-t_{\tbox{col}}) \right\rangle 
\end{eqnarray}
The last expression is manipulated as in  
(\ref{ea_g1}) and leads to the correct result.

It is more illuminating to repeat the last derivation 
using a kinetic point of view. 
The velocity of the particle prior to the 
collision with the wall element is $\mbf{v}_{\tbox{col}}$. 
The orientation of the wall 
element is represented by a normal unit 
vector $\mbf{n}$. The velocity of the wall 
element is denoted by $\mbf{V}$. 
Only the $\mbf{n}{\cdot}\mbf{V}$ component of the 
wall velocity is significant. 
After the collision the $z$ component of the particle's 
velocity is $\mbf{n}{\cdot}(-\mbf{v}_{\tbox{col}}+2\mbf{V})$.  
The corresponding energy gain due to the collision is 
\begin{eqnarray}  \label{ea_g3} 
\Delta E_{\tbox{col}} \ = \ -2m \ (\mbf{n}{\cdot}\mbf{V})
\ ( \mbf{n}{\cdot}(\mbf{v}_{\tbox{col}}-\mbf{V}) ) \ \ \ \ . 
\end{eqnarray}
The stochastic-like force ${\cal F}(t)$ is a sum over impulses 
that are created by collisions with the surface of the cavity.  
The energy gain due to this collisions is 
obtained by integrating $-V{\cal F}(t)$ over time (see Sec.4). 
It follows that each collision involves an impulse 
$-\Delta E_{\tbox{col}}/V$ leading to 
\begin{eqnarray}  \label{ea_Ft}   
{\cal F}(t) \ = \ \sum_{\tbox{col}}
[ \ 2m \ (\mbf{n}{\cdot}\hat{\mbf{V}}) 
\ (\mbf{n}{\cdot}(\mbf{v}_{\tbox{col}}-\mbf{V})) \ ] 
\ \ \delta(t-t_{\tbox{col}})  
\end{eqnarray}
For $V{=}0$ we can assume that the 
velocities $v_{\tbox{col}}$ are uncorrelated. 
For a moving slab-shaped `piston' half of the collisions 
will be from the `left side' with 
$\mbf{n}{\cdot}\mbf{V}=-V$, and half of the 
collisions will be from the `right' side 
with $\mbf{n}{\cdot}\mbf{V}=+V$. 
Consequently we have 
$\langle {\cal F}(t) \rangle = 0$. 
More generally, we have 
$\langle {\cal F}(t) \rangle = 0$ 
for any parametric deformation that keeps 
constant the total volume of the cavity.

It is now tempting to make an attempt for a direct 
estimate of $\langle {\cal F}(t) \rangle$ for $V\ne 0$. 
Let us assume for a moment that we have a slab-shaped 
`piston' which is moving from `left' to `right'.  
At first sight it seems that friction is due to the fact that 
energy-gain due to collisions from the `right' is larger than 
energy-loss due to collisions from the `left'. 
Our ``hypothesis'' implies that the ${\cal O}(V)$ term
in (\ref{ea_Ft}) should be responsible for 
the friction. However, it turns out that only 
{\em half} of the correct result is obtained: 
\begin{eqnarray} \nonumber 
\left\langle \sum_{\tbox{col}} 
2m(\mbf{n}{\cdot}\hat{\mbf{V}})^2 
\ \delta(t{-}t_{\tbox{col}}) \right\rangle \ = \ 
\langle|\cos\theta|\rangle \ 
\frac{\mboxs{Area}}{\mboxs{Volume}} \ mv_{\tbox{E}}
\ = \ \frac{1}{2}\mu_{\tbox{E}}
\end{eqnarray}
We conclude that the first term in (\ref{ea_Ft}), 
the term that includes $v_{\tbox{col}}$,  
has a non-zero average. Once $V \ne 0$ we can no-longer treat 
the collisions as uncorrelated. The statistical effect (that has 
been  emphasized in the derivation of the ${\cal F\!D}$ relation) is important 
also in time-domain analysis.

Thus we see that the ${\cal F\!D}$ indirect approach is a quite 
powerful tool.  This becomes manifestly evident once we consider 
a variation of the above example: 
If successive collisions with the `piston' are correlated, for example 
due to bouncing behavior, then it is still a relatively 
easy task to estimate $\nu_{\tbox{E}}$ for the $V{=}0$ case, and 
then to obtain $\mu_{\tbox{E}}$ via the ${\cal F\!D}$ relation. 
On the other hand, a direct evaluation of $\mu_{\tbox{E}}$ 
using kinetic considerations is extremely difficult, because  
in calculating  $\langle{\cal F}(t)\rangle$ it is essential to 
take into account subtle correlations between successive collisions.

\section{The wall formula: the standard kinetic approach}
\label{a_kinetic}

Here we generalize the derivation of the wall formula to 
any dimension, and to any dispersion relation $E=E(\mbf{p})$. 
We use the standard kinetic approach. The momentum 
probability distribution of the particles inside the cavity 
is uniform in space and $\rho(\mbf{p})$ in momentum. 
(Later we are using many-particles jargon, 
but the actual meaning is always probabilistic). 
We assume isotropic distribution, and therefore the energy 
distribution is just $\rho(E)=g(E)\rho(p)$. 
The velocity which is associated with a momentum $\mbf{p}$ is 
$\mbf{v}{=}dE/d\mbf{p}$. The mass is defined in the 
differential sense $m(E)=(dv/dp)^{-1}$. 

With a surface area $ds$ we associate a normal unit 
vector $\mbf{n}$, and a displacement velocity $\mbf{V}$. 
The particles that actually collide with the wall-element 
during a time interval $dt$ must satisfy the condition      
$(\mbf{v}{-}\mbf{V}){\cdot}\mbf{n}>0$. Those with 
a velocity $\mbf{v}$ are contained in a volume element 
\mbox{$d\mboxs{Volume}=
((\mbf{v}{-}\mbf{V}){\cdot}\mbf{n}) \ dtds$}.
The energy gain due to a collision, assuming $V \ll v$, 
and expanding up to ${\cal O}(V^2)$, is still given 
by (\ref{ea_g3}). The total energy change is 
\begin{eqnarray}   
dE \ = \ \int\rho(\mbf{p})d\mbf{p} 
\times d\mboxs{Volume} \times \Delta E_{\tbox{col}} 
\end{eqnarray}
The above integral can be manipulated as follows: We make 
a transformation to an integration variable 
$\mbf{p}'=\mbf{p}{-}m\mbf{V}$ and 
correspondingly $\mbf{v}'=\mbf{v}{-}\mbf{V}$. Then we expand 
$\rho(\mbf{p}'{+}m\mbf{V}) = \rho(\mbf{p}') + 
(\partial\rho(p')/\partial E) m\mbf{v}'{\cdot}\mbf{V}$, 
and throw away the first term since it 
is associated with reversible work. 
From here on we omit the primes. Due to symmetry 
consideration we may replace $\mbf{v}{\cdot}\mbf{V}$ 
by $(\mbf{v}{\cdot}\mbf{n})(\mbf{V}{\cdot}\mbf{n})$. 
Now we obtain 
\begin{eqnarray}   
dE \ = \ -\int \frac{\partial\rho(p)}{\partial E} d\mbf{p} 
\ m^2 (\mbf{V}{\cdot}\mbf{n})^2 
|\mbf{v}{\cdot}\mbf{n}|^3 \ dsdt 
\end{eqnarray}
The latter integral is easily cast into the 
form of (\ref{e_19}) with the identification 
\begin{eqnarray}   
D_{\tbox{E}} \ = \  
\frac{\langle|\cos(\theta)|^3\rangle}{\mboxs{Volume}} 
m^2 v_{\tbox{E}}^3 \oint (\mbf{V}{\cdot}\mbf{n})^2 ds 
\end{eqnarray}
The latter expression incorporates integration 
over all the wall-elements.

\newpage
\section{Chaotic eigenstates}       
\label{a_RW}

The average density of energy eigenstates for a particle 
in $d$ dimensional cavity is 
\begin{eqnarray}   
\frac{1}{\Delta} \ = \ \frac{1}{\hbar v_{\tbox{E}}} \ 
\frac{\mboxs{Volume}}{(2\pi)^d} \ \Omega_d \ k^{d{-}1}
\end{eqnarray}
where $k$ is the wavenumber and $v_{\tbox{E}}$ is 
the corresponding velocity of the particle.
An eigenstate that corresponds to a chaotic cavity 
looks like a `random wave'. More precisely, it has 
the same statistical properties as those of 
an un-constrained random superposition of 
plane waves whose wavenumber is $|\mbf{k}|=k$.  
It is characterized by the correlation function 
\begin{eqnarray}   
\langle \psi_{\tbox{R}}(\mbf{x}_1)\psi_{\tbox{R}}(\mbf{x}_2)
\rangle \ = \ \frac{1}{\mboxs{Volume}} 
\ \mbox{Cos}(k|\mbf{x}_2-\mbf{x}_1|)
\end{eqnarray}
The function Cos is defined in App.\ref{a_spherical}.
We can constrain a random wave to be zero along 
a boundary by making an antisymmetric superposition 
of an un-constrained random wave with its mirror image. 
The normal derivative along the boundary 
$\varphi(\mbf{s})=\mbf{n}{\cdot}\nabla\psi$ satisfies  
\begin{eqnarray}   \label{ea_i3}
\langle \varphi_{\tbox{R}}(\mbf{s}_1)
\varphi_{\tbox{R}}(\mbf{s}_2)
\rangle \ = \ \frac{2k^2}{\mboxs{Volume}} 
\ \mbox{Sinc}(k|\mbf{s}_2-\mbf{s}_1|) 
\end{eqnarray}
This expression ignores curvature effect. We assume 
that the boundary's radius-of-curvature is very large 
compared with De-Broglie wavelength, as well as   
compared with the distance of interest $r=|\mbf{s}_2-\mbf{s}_1|$.

\section{Matrix elements for hard walls}       
\label{a_matrix}

The position of a particle in the vicinity of 
a wall element can be described by $Q=(z,\mbf{s})$ 
where $\mbf{s}$ is a surface coordinate 
and $z$ is a perpendicular `radial' coordinate. 
The potential which is experience 
by the particle is assumed to be ${\cal V}(Q)=0$ inside 
the cavity where $z<0$, and ${\cal V}(Q)=f{\cdot}z$ for 
$z>0$ up to some maximal value ${\cal V}_{\tbox{wall}}$ 
well inside the barrier. We assume $E \ll {\cal V}_{\tbox{wall}}$ 
and therefore ${\cal V}_{\tbox{wall}}$ should have no significance. 
The limit $f\rightarrow\infty$ corresponds to hard walls.
For deformation field $\hat{\mbf{V}}(\mbf{s})$, defined such that 
$\hat{\mbf{V}}\delta x$ is the displacement of a wall element, 
one obtains
\begin{eqnarray}  
\frac{\partial {\cal H}}{\partial x} \ = \ 
- (\mbf{n}(\mbf{s}) {\cdot} \hat{\mbf{V}}(\mbf{s})) 
\ {\cal V}_{\tbox{wall}} \ \delta(z)
\end{eqnarray}
The orientation of each wall element 
is represented by the unit vector $\mbf{n}(\mbf{s})$.    
The logarithmic derivative of the wavefunction along 
the boundary is $\varphi(\mbf{s})/\psi(\mbf{s})$ where 
$\varphi(\mbf{s})=\mbf{n}{\cdot}\nabla\psi$. 
For $z>0$ the wavefunction $\psi(Q)$ is a decaying 
exponential. Hence the logarithmic derivative of 
the wavefunction on the boundary should be equal 
to $-\sqrt{2m{\cal V}_{\tbox{wall}}}/\hbar$.  Consequently 
one obtains the following expression for the 
matrix elements that are associated with the deformation: 
\begin{eqnarray}   \label{ea_j2}
\left(\frac{\partial {\cal H}}{\partial x}\right)_{nm} 
\ = \
-\frac{\hbar^2}{2m}\oint \varphi_n(\mbf{s})\varphi_m(\mbf{s}) 
\ (\mbf{n}{\cdot}\hat{\mbf{V}}) d\mbf{s}
\end{eqnarray}
For uncorrelated chaotic eigenstates we get by 
squaring (\ref{ea_j2}) and using (\ref{ea_i3}), 
the following expression:
\begin{eqnarray} \nonumber 
\hspace*{-2cm}
\left\langle\left|\left(
\frac{\partial {\cal H}}{\partial x}
\right)_{nm}\right|^2 \right\rangle
\ \ = \ \ 
\left(\frac{1}{\mboxs{Volume}}\right)^2 \times
\left(\frac{(\hbar k)^2}{m}\right)^2 \times
\\ \nonumber 
\hspace*{-1cm}
\times \oint\oint 
\mbox{Sinc}(k_n|\mbf{s}_2{-}\mbf{s}_1|) \
\mbox{Sinc}(k_m|\mbf{s}_2{-}\mbf{s}_1|) \ 
(\mbf{n}{\cdot}\hat{\mbf{V}}(\mbf{s}_1)) d\mbf{s}_1 \ 
(\mbf{n}{\cdot}\hat{\mbf{V}}(\mbf{s}_2)) d\mbf{s}_2
\end{eqnarray}
If De-Broglie wavelength is a 
small scale, then the difference $|k_n-k_m|$ 
will have no significance in the calculation. 
Soft walls are essential in order 
to have finite bandwidth (see next appendix).
A classical look-alike 
result for $\nu_{\tbox{E}}=\tilde{C}_{\tbox{E}}(0)$ 
is obtained by setting $k_n \sim k_m \sim k$  and 
then multiplying by $2\pi\hbar/\Delta$. 
Assuming that $\lambda_{\tbox{E}}$ is a small scale,  
it is possible to approximate the $(\mbox{Sinc})^2$ by 
a delta function (see (\ref{ea_e7})) and consequently 
the semiclassical estimate that follows from (\ref{e_29})
for individual matrix elements is recovered: 
\begin{eqnarray} \label{ea_j3}  
\hspace*{-2cm}
\left|\left(
\frac{\partial {\cal H}}{\partial x}
\right)_{nm}\right| \ \ \approx \ \ 
\frac{1}{\mboxs{Volume}} 
\ \frac{(\hbar k)^2}{m} 
\ \sqrt{ 
\frac{2\langle|\cos(\theta)|^3\rangle}{\Omega_d}
\ \mbox{Area} \ \lambda_{\tbox{E}}^{d{-}1} } 
\end{eqnarray}
Compared with non-chaotic eigenstate,  
for which $\mbox{Sinc} \mapsto 1$, there is 
a factor $1/\sqrt{{\cal N}}$, where 
${\cal N}= (\mbox{Area} / \lambda_{\tbox{E}}^{d{-}1})$ 
is the number of correlated regions on the surface 
of the cavity.

\section{Finite bandwidth due to soft walls}    
\label{a_softwall}

Equation (\ref{ea_j3}) is valid for hard walls. 
For soft walls  
$|({\partial {\cal H}}/{\partial x})_{nm}|^2$
should be multiplied by a suppression factor 
$c_{\tbox{soft}} \le 1$. This suppression 
factor can be estimated by considering a 
plane wave incident upon the wall \cite{koonin}. 
The result is 
\begin{eqnarray}   
c_{\tbox{soft}} \ = \ \left|
\frac{1}{\varepsilon_n-\varepsilon_m}\left(
\frac{\sin\phi_n\cos\phi_m}{\sqrt{\varepsilon_n}}-
\frac{\sin\phi_m\cos\phi_n}{\sqrt{\varepsilon_m}}
\right)\right|^2
\end{eqnarray}
where $\varepsilon=(k/(2mf/\hbar^{\tbox{2}})^{\tbox{1/3}})^2$ 
is the scaled energy of the eigenstate. The 
reflection phase-shift is obtained via 
$\tan(\phi)=({\cal A}/{\cal A}')\sqrt{\varepsilon}$. 
We define ${\cal A}=Airy(-\varepsilon)$, and 
${\cal A}'$ and  ${\cal A}''$ are the respective 
first and second derivatives. If $\phi_n$ is very 
close to $\phi_m$ then we can make the approximation
\begin{eqnarray}   
c_{\tbox{soft}} \ \approx \ \left[
\frac{({\cal A}')^2-({\cal A}{\cal A}'')}
{({\cal A}')^2+\varepsilon({\cal A})^2} \right]^2
\end{eqnarray}
In the hard wall limit $\varepsilon\rightarrow 0$, 
and ${\cal A}''\rightarrow 0$, consequently this 
factor will be equal unity as expected.  If 
$\varepsilon_n$ and $\varepsilon_m$ are not  
very close, then $\phi_n$ and $\phi_m$ can be treated
as uncorrelated and we obtain
\begin{eqnarray}   
c_{\tbox{soft}} \ \approx \ 
\frac{1}{2\varepsilon}
\ \frac{1}{(\varepsilon_n-\varepsilon_m)^2} 
\ = \ 
2\left(\frac{\hbar}{\tau_{\tbox{cl}}}\right)^2
\frac{1}{(E_n-E_m)^2}
\end{eqnarray}
The bandwidth can be estimated by finding 
the difference $E_n-E_m$ 
for which the last expression becomes of order unity.
This gives the result 
$\Delta_b=\hbar/\tau_{\tbox{cl}}$ where 
$\tau_{\tbox{cl}}$ is the collision time 
with the wall. This is in agreement with 
our semiclassical expectations.  We learn 
that outside the band, the matrix 
element have a power-law decay.  
This result looks of semiclassical nature 
since it adds an $\hbar$ independent factor to 
$C_{\tbox{E}}(\omega)$. A $1/\omega^2$ tail of $C_{\tbox{E}}(\omega)$  
should be associated with a discontinuity of the 
first derivative of $C_{\tbox{E}}(\tau)$ at $\tau=0$. 
See App.\ref{a_CE}. 

\newpage

\end{document}